\documentclass[a4paper,11pt]{article}
\usepackage{jinstpub}

\usepackage[utf8]{inputenc}
\usepackage[T1]{fontenc}

\usepackage{lineno}

\usepackage{graphicx}
\usepackage{xcolor}
\usepackage{rotating}
\usepackage{upgreek}
\usepackage{dcolumn}
\usepackage{bbm}
\usepackage{textcomp}
\usepackage{gensymb}
\usepackage{wasysym}
\usepackage{amsmath,amsthm}
\usepackage{enumerate}
\usepackage{setspace}
\usepackage{comment}
\usepackage{lineno}
\usepackage{mathrsfs}
\usepackage{booktabs}
\usepackage{isotope}
\usepackage[graphicx]{realboxes}
\usepackage{xspace}
\usepackage{enumitem}

\hypersetup{
    colorlinks=true,
    linkcolor=red,
    citecolor=blue,
    urlcolor=blue
}

\setlist[enumerate]{itemsep=0mm}

\newcommand{\theia}{\textsc{Theia}}
\newcommand{\eos}{\textsc{Eos}}

\makeatletter
\renewcommand\thesection{\arabic{section}}

\renewcommand\p@subsection{}
\renewcommand\p@subsubsection{}
\makeatother

\title{The \textsc{Eos} detector: a demonstrator of hybrid optical detection technology}
\author[a,b]{S.~Arora,}
\author[a,b]{M.~Askins,}
\author[c]{A.~J.~Bacon,}
\author[a,b]{Z.~Bagdasarian,}
\author[d,e]{A.~Baldoni,}
\author[f]{L.~Bartoszek,}
\author[g]{M.~Bergevin,}
\author[a,b]{Y.~Bezawada,}
\author[h]{E.~Blucher,}
\author[f]{J.~Boissevain,}
\author[b]{R.~Bonventre,}
\author[a,b]{E.~J.~Callaghan,}
\author[d]{D.~F.~Cowen,} 
\author[d]{K.~DeHolton,} 
\author[i]{M.~Diwan,}
\author[c]{M.~Dubnowski,}
\author[a,b]{P.~Englezos,}
\author[a,b]{S.~Gadamsetty,}
\author[j]{C.~Grant,}
\author[c]{B.~Harris,}
\author[a,b]{M.~R.~Hebert,}
\author[j]{S.~Jeon,}
\author[a,b]{T.~Kaptanoglu,} 
\author[j]{A.~Katt,}
\author[c]{J.~R.~Klein,}
\author[c]{T.~Kroupov\'a,} 
\author[a,b]{L.~Lebanowski,}
\author[a,b]{S.~Lynch,} 
\author[k]{A.~Mastbaum,}
\author[c]{C.~Mauger,}
\author[c]{G.~Mayers,} 
\author[l]{M.~Miller,}
\author[l]{J.~Nachtman,}
\author[c]{S.~Naugle,}
\author[m]{J. Newby,}
\author[c]{M.~Newcomer,}
\author[c]{A.~Nikolica,} 
\author[a,b]{G.~D.~Orebi~Gann,}
\author[n]{A.~Phipps,}
\author[a,b,1]{L.~Pickard,}
\author[c]{R.~C.~Pitelka,}
\author[o]{L.~Ren,}
\author[a,b]{A.~Rincon,} 
\author[i]{R.~Rosero,}
\author[a,b]{N.~Rowe,}
\author[a,b]{H.~J.~Ryoo,}
\author[j]{J.~Ryshkewitch,}
\author[b]{J.~Saba,} 
\author[a,b]{S.~Schoppmann,}
\author[c]{J.~Shen,}
\author[a,b]{M.~Smiley,}
\author[a,b,j]{H.~Song,}
\author[p,q]{H.~Steiger,}
\author[r]{B. Tam,}
\author[l,s]{E.~Tiras,}
\author[t]{W.~H.~To,}
\author[u]{M.~R.~Vagins,}
\author[f]{R.~Van~Berg,}
\author[b]{J.~Wallig,}
\author[d]{G.~Wendel,}
\author[l]{M.~Wetstein,}
\author[p]{M.~Wurm,}
\author[a,b]{G.~Yang,}
\author[i]{M.~Yeh,}
\author[o]{E.~D.~Zimmerman,}
\author[k]{A.~Zummo,}

\affiliation[a]{Physics Department, University of California at Berkeley, Berkeley, CA 94720-7300}
\affiliation[b]{Lawrence Berkeley National Laboratory, 1 Cyclotron Road, Berkeley, CA 94720-8153, USA}
\affiliation[c]{Department of Physics and Astronomy, University of Pennsylvania, Philadelphia, PA 19104-6396}
\affiliation[d]{Department of Physics, Pennsylvania State University, University Park, PA 16802, USA}
\affiliation[e]{United States Military Academy, Department of Physics and Nuclear Engineering, West Point, NY 10996}
\affiliation[f]{Bartoszek Engineering, Aurora, IL 60506, USA}
\affiliation[g]{Lawrence Livermore National Laboratory, Livermore, CA 94550, USA}
\affiliation[h]{The Enrico Fermi Institute and Department of Physics, The University of Chicago, Chicago, IL 60637, USA}
\affiliation[i]{Brookhaven National Laboratory, Upton, New York 11973, USA}
\affiliation[j]{Boston University, Department of Physics, Boston, MA 02215, USA}
\affiliation[k]{Department of Physics and Astronomy, Rutgers, The State University of New Jersey, 136 Frelinghuysen Road, Piscataway, NJ 08854-8019 USA}
\affiliation[l]{Department of Physics and Astronomy, The University of Iowa, Iowa City, IA 52242, USA}
\affiliation[m]{Oak Ridge National Laboratory, 1 Bethel Valley Road, Oak Ridge, TN 37830}
\affiliation[n]{California State University, East Bay, Department of Physics, 25800 Carlos Bee Blvd, Hayward, CA 94542}
\affiliation[o]{University of Colorado at Boulder, Department of Physics, Boulder, CO 80309, USA}
\affiliation[p]{Johannes Gutenberg-Universit{\"a}t, Institute of Physics and EC PRISMA$^{++}$, Mainz, 55099 Mainz, Germany}
\affiliation[q]{Physics Department, Technische Universit{\"a}t M{\"u}nchen, 85748 Garching, Germany}
\affiliation[r]{Department of Physics, University of Oxford, Parks Rd, Oxford OX1 3PU, United Kingdom}
\affiliation[s]{Department of Physics, Erciyes University, 38030, Kayseri, Turkey}
\affiliation[t]{California State University, Stanislaus, Department of Physics, Turlock, CA 95382, USA}
\affiliation[u]{Department of Physics and Astronomy, University of California, Irvine, Irvine, CA 92697-4575, USA}

\note{Corresponding author.}
\emailAdd{leonjamespickard@berkeley.edu}

\abstract{\eos\ is an R\&D testbed for hybrid detector technologies, featuring state-of-the-art sub-ns photosensors, the first implementation of dichroicons in a large-scale demonstrator, and the deployment of novel detection media such as water-based liquid scintillator (WbLS). By separating Cherenkov and scintillation light, \eos\ leverages the benefits of both to explore the potential of next-generation neutrino technologies. An extensive radioactive source calibration program enables the characterization of position, direction, and energy reconstruction performance of a variety of target materials. Furthermore, \eos\ will provide data to refine optical models and inform the development and simulation of future neutrino experiments. This paper describes the as-built design and data-taking plan of \eos, outlining its scientific motivations and role in the development of future detector technologies.}
\begin{document}
\linenumbers
\pagenumbering{roman}
\maketitle
\pagenumbering{roman}
\setcounter{page}{2}
\clearpage
\pagenumbering{arabic}
\nolinenumbers
\section{Introduction}
Neutrinos hold the key to many unsolved questions in physics. Being electrically neutral and without color charge, neutrinos interact only via the weak force and gravity. Their non-zero masses and mixing provide the opportunity to answer many unknown questions about the Universe, such as the level of CP violation in the leptonic sector and whether this is sufficient to explain the matter-antimatter asymmetry~\cite{T2KCP,LeptonicCP,NOVACP}. Neutrinos are also a valuable tool in exploring physics beyond the Standard Model, such as whether they are Majorana in nature~\cite{Majorana}. In addition, precision measurements of neutrino oscillations are essential to characterize the structure of the leptonic sector and are of phenomenological interest: Is their mass ordering normal or inverted~\cite{NeutrinoMixing}? In what octant does $\theta_{23}$ lie~\cite{Nu6,PDG}? Furthermore, neutrinos are an excellent probe for astrophysical phenomena, by testing our understanding of supernova explosion mechanics~\cite{SN1987A,Bethe,SNTheory}, measuring the magnitude and spectrum of the diffuse supernova neutrino background (DSNB)~\cite{DSNB}, and determining the solar neutrino flux and spectra to constrain solar metallicity~\cite{Solar}. In a terrestrial setting, they can be used to evaluate levels of radioactive isotopes within the Earth's crust and mantle~\cite{GeoKamLAND,GeoBorexino,SNO+geo} and also have the potential to be used in nuclear nonproliferation efforts by monitoring their production, or the absence thereof, at weapons test sites~\cite{FissionMonitoring} or in the vicinity of nuclear facilities~\cite{ReactorMonitoring}.

Due to the difficulty of detecting neutrinos---a consequence of their small interaction cross sections, typically ranging from $\sim10^{-44}$ to $10^{-38}$\,cm$^2$~\cite{Zeller}---large target masses or intense neutrino sources are required. In spite of this challenge, there has been much success, primarily using Cherenkov or scintillation target media. Cherenkov detectors such as Super-Kamiokande~\cite{SuperK} and the upcoming Hyper-Kamiokande~\cite{HyperK} benefit from the excellent optical clarity achievable using water, allowing for detector masses of 50 and 260\,kTons, respectively. Furthermore, the directionality provided by Cherenkov emission has proven an invaluable tool in background reduction~\cite{SuperKAtmospheric,SuperKOscillation,SuperKSolar}. Conversely, liquid scintillator detectors such as SNO+~\cite{SNO+Det}, Borexino~\cite{BorexinoDet}, and KamLAND~\cite{KamlandDet} provide high light yields, low energy thresholds, and pulse shape discrimination, all of which have proven crucial for probing neutrino properties~\cite{SNO+Reactor,kamland,BorexinoNature}.

Next-generation neutrino detectors are poised to benefit greatly from advanced technologies that leverage the combination of Cherenkov and scintillation signals~\cite{ASDC}. With the emergence of novel target media, such as water-based liquid scintillators (WbLS)~\cite{YehWbLS} and slow scintillators~\cite{Biller:2020uoi,Guo:2018kcp}, the development of fast sub-ns photosensors, and the inception of dichroicons as a method to perform spectral sorting, the utilization of both signals could enable detector capabilities not achievable with either technique alone. In addition to the benefits of each signal individually, harnessing both simultaneously will offer further advantages - such as using the Cherenkov/scintillation ratio to further reduce backgrounds via particle identification. Employing the latest technologies could enable the realization of a 100 kTon-scale highly-capable detector, such as \theia~\cite{Askins:2019oqj}, with excellent energy resolution, directionality, particle identification capabilities, and exceptional background rejection.

An active program of R\&D over the past decade has established the foundations for next-generation hybrid neutrino detectors. Bench-top testing has demonstrated the separation of Cherenkov and scintillation signatures ~\cite{CHESS2016}, measured light-yields ~\cite{WbLSLightYield,CHESS2020}, and determined timing profiles~\cite{Onken:2020pnv}. Dichroicons, to allow for spectral sorting, have gone from a concept to a deployable design~\cite{dichroicon_paper}. Next-generation fast photosensors are now readily available~\cite{ham_datasheet_r14688,hamamatsu_8in}, with a transit time spread (TTS) of a few 100\,ps or better. In recent years, the scaling of these technologies has accelerated; the 1\,ton WbLS detector at Brookhaven National Lab, to test the optical stability of this novel medium, is operational~\cite{BNL1Ton}. The 30\,ton WbLS experiment, also at Brookhaven, to test transparency over time, has also recently been constructed and is operational~\cite{BNL30ton}. The BUTTON Experiment at Boulby Mine in the UK is under construction and will provide a test of WbLS in a low-background environment~\cite{Button}. Finally, the ANNIE experiment has recently published data from the deployment of a 366\,L WbLS
volume, demonstrating the detection of neutrinos from the Booster Neutrino Beam at FNAL~\cite{ANNIEWbLS}.

\eos~\cite{EosDesign} is complementary to these other efforts, offering high-precision characterization of the performance of these novel technologies in an integrated testbed~\cite{EosWater}, with the ability to study events from sub-MeV to GeV scales. This paper describes the as-built \eos\ detector, along with the suite of sources designed to calibrate the detector and to demonstrate the capabilities of hybrid technology.

\section{Detector goals}

Prior development has produced a range of novel technologies with viability for deployment in next-generation neutrino experiments:

\begin{itemize}
    \item New detection media that provide tunable Cherenkov and scintillation light contributions while allowing the scintillation emission spectrum, light yield, and time profile to be tailored.
    \item The latest sub-ns photomultiplier tubes capable of distinguishing scintillation from Cherenkov light via timing, while enhancing reconstruction and pulse shape discrimination.
    \item Spectral sorting using dichroicons~\cite{Kaptanoglu:2019gtg}.
\end{itemize} 

\eos\ integrates these technologies within a single detector, bridging the gap between component-level R\&D and deployment in next-generation physics experiments. With the installation of dichroicons, fast PMTs, and high photocathode coverage, \eos\ has the capability to utilize the combined Cherenkov/scintillation signatures by separating them topologically, through timing, and by wavelength. The goals of \eos\ are to measure how performance depends on the choice of target material, photon detectors, detector configuration, and readout options. To this end, a suite of sources has been developed to demonstrate energy, direction, and position reconstruction from sub-MeV to GeV energy scales. These results will be used to validate Monte Carlo predictions and understand the scalability of these technologies to kTon-scale and beyond.

In addition to testing WbLS, a variety of other scintillator formulations will be deployed. These include linear alkylbenzene (LAB) which enables the study of scintillators with a range of emission profiles through the addition of varying concentrations of 2,5-diphenyloxazole (PPO), allowing the investigation of how hybrid Cherenkov/scintillation performance changes as light yield increases and the decay time shortens. 
Furthermore, the deployment of a novel, ultra-fast scintillator will constitute the first direct attempt at topological imaging in a working detector.

The goals of \eos\ dictated the final design. To effectively distinguish Cherenkov and scintillation light temporally, PMTs with sub-ns timing were required. Furthermore, the electronics were designed to digitize PMT signals with a timing resolution below 1\,ns and a wide dynamic range. To enable testing of spectral photon sorting, an array of 12 dichroicons was designed with dichroic filters optimized to separate the longer-wavelength Cherenkov emission from the shorter-wavelength scintillation emission. To maximize light collection, the PMTs and dichroicons were arranged to achieve the highest photocathode coverage. The requirement that \eos\ be a flexible testbed for a range of target media necessitated all essential scintillation-handling components be constructed from 316 stainless steel (SS), Teflon, Viton, polypropylene, and acrylic. Finally, since optical modeling is of central importance, the calibration deployment system was engineered to be versatile, enabling the deployment of an array of sources with a vertical precision of 2 \,mm and azimuthal rotation capability.

\section{The \eos\ detector}

\begin{figure}[!h]
    \begin{center}
    \includegraphics[width=0.56\textwidth,trim=19cm 5cm 20.8cm 3cm, clip]{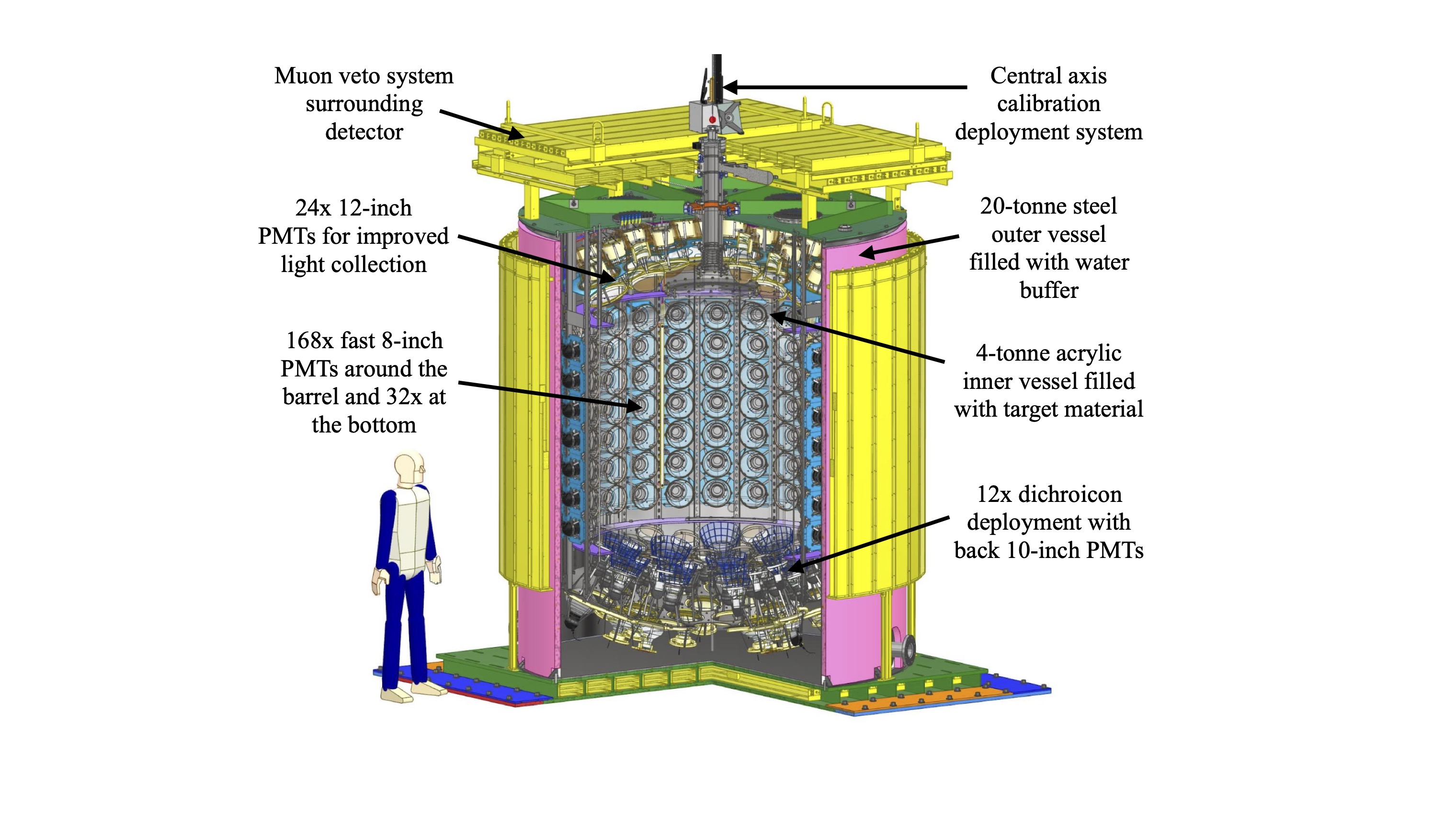}
    \includegraphics[width=0.43\textwidth,trim=0cm 1cm 0cm 0cm, clip]{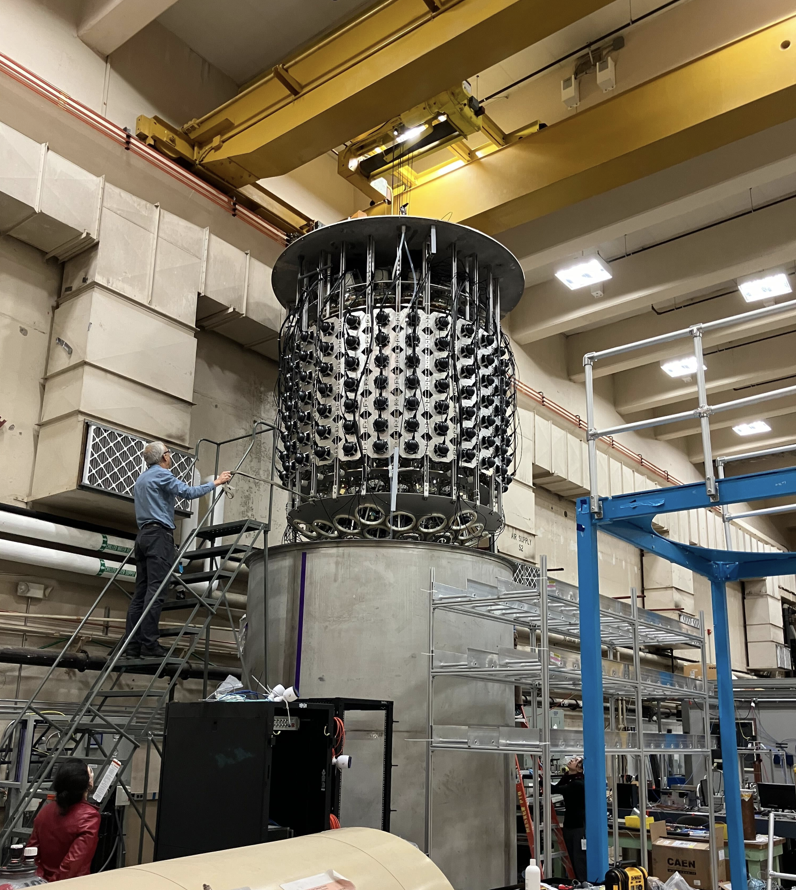}
    \end{center}
    \caption{(Left) Design of \eos\  showing the different detector components; (Right) Detector assembly during IV and PSUP insertion into the OV.}
    \label{f:eos}
\end{figure}

\eos , located at UC Berkeley, consists of an acrylic inner vessel (IV) with a capacity of 4\,tonnes when filled with water. The IV is surrounded by a total of 241 photomultiplier tubes (PMTs) that are secured in place by the PMT support structure (PSUP). Of these PMTs, 12 are equipped with dichroicons and installed on the lower PMT array. The detector assembly is enclosed in a 20-tonne 304 SS outer vessel (OV), which is filled with ultrapure water. A calibration source deployment mechanism is attached to the center of the OV lid, allowing deployment of low-energy radioactive and optical sources along the central axis. Surrounding the OV is the muon veto system, which consists of 68 individual muon veto paddles. Fig.~\ref{f:eos} shows the design rendering and the as-built detector.

\subsection{\eos\ detector components}

The components of \eos\ were designed with the following requirements:
\begin{itemize}
    \item All components in contact with the planned liquids shall be materially compatible to prevent degradation.
    \item The IV shall be optically transparent across the spectral sensitivity range of the PMTs.
    \item The entire assembly, once constructed, shall be movable as a single unit.
    \item The PMTs shall be arranged to maximize photocathode coverage.
    \item The design shall be modular to allow future upgrades.
    \item The entire structure shall be sufficiently rigid to withstand seismic events.
\end{itemize}

\subsubsection{The outer vessel}

The OV tank is a right circular cylinder, 3.36\,m tall and 2.84\,m diameter, with a capacity of approximately 20\,tonnes of water. Constructed from 0.64\,cm thick passivated 304 SS to minimize corrosion, it houses the detector insert (as discussed in Sections~\ref{s:PMT_Support_Structure} and~\ref{s:IV}) and is filled with ultrapure deionized water, which undergoes continuous recirculation and purification (as discussed in Section~\ref{s:OV_Fluid_Handling}).

The tank is bolted to a 3.05 $\times$ 3.05 $\times$ 0.17\,m steel base, which is anchored to the floor. In addition to complying with ASCE/SEI 7-10 code and seismic requirements, the base was fabricated with 11 channels in which the lower muon veto panels are installed (explained in Section~\ref{s:muon}).

\begin{figure}[!h]
    \begin{center}
    \includegraphics[width=1.0\textwidth, trim=0cm 10.3cm 0cm 9cm, clip]{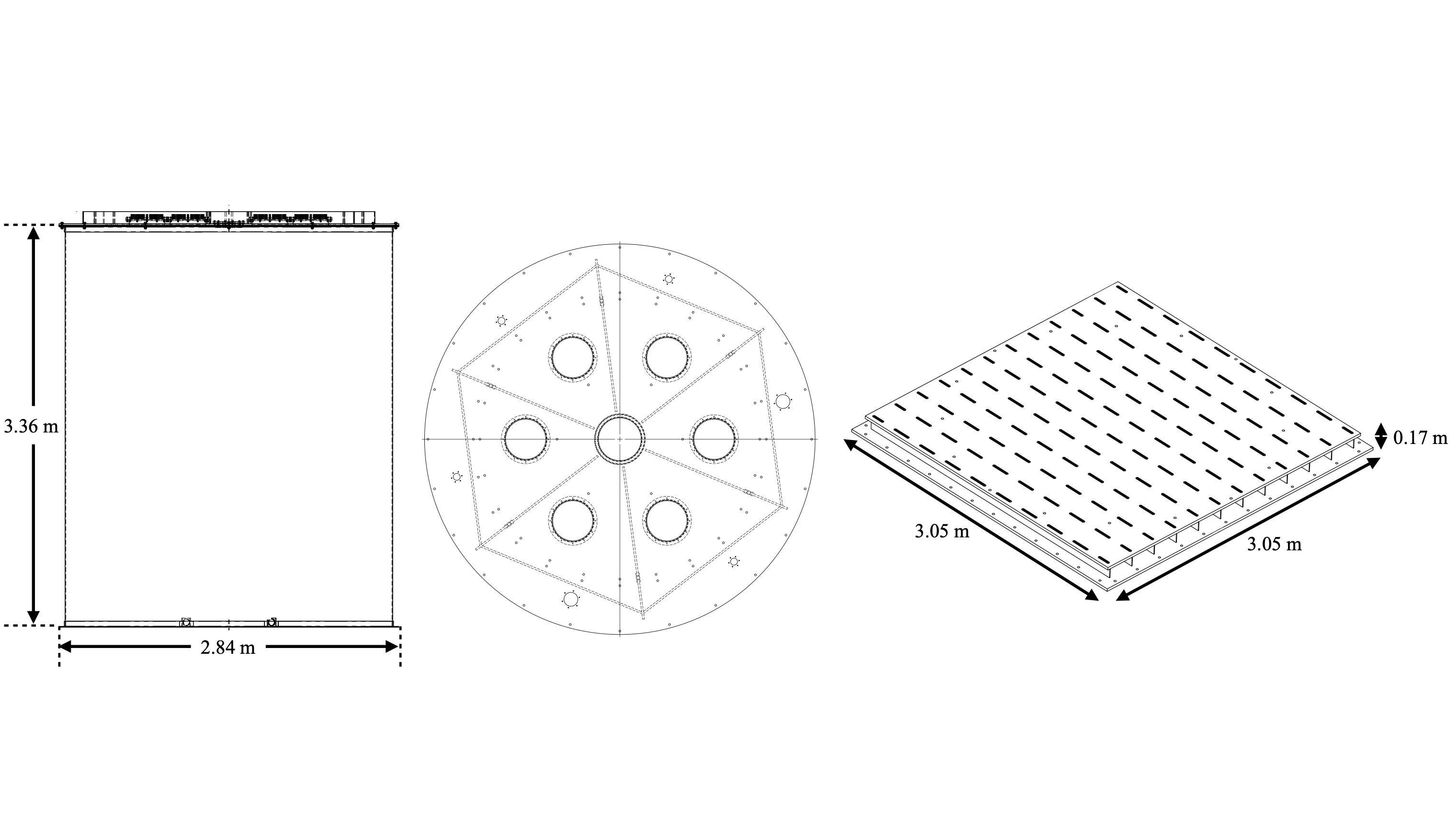}
    \end{center}
    \caption{Side view and cross sectional dimensions of the OV tank (left), top view of the OV lid showing the hexagonal weldment, PMT cable ports, and the central port for source deployments (center), and perspective view of the detector base illustrating the integrated channels used to house the lower muon veto panels (right).}
    \label{f:OV}
\end{figure}

The 304 SS circular lid is 1.27\,cm thick and bolted to the tank. An elastomer gasket with an inner diameter of 2.74\,m, an outer diameter of 2.84\,m, and a thickness of 0.64\,cm was installed between the tank and the lid. The design of \eos\ was such that the detector insert, described in Sections~\ref{s:PMT_Support_Structure} and~\ref{s:IV}, was assembled from the underside of the lid, allowing it to be deployed in the OV using a bridge crane attached to four pick points located on a raised hexagonal weldment. There are six ports on the lid, located 2.44\,m from the central axis, each spaced at 60$^\circ$ intervals. Two ports have a diameter of 10\,cm, and four have a diameter of 5\,cm. These ports are used for fluid handling lines, the cover gas system, and provide the ability to deploy sources within the OV. At a radius of 1.37\,m from the central axis, also spaced at 60$^\circ$ intervals, there are six cable feed-through ports with a diameter of 29\,cm. Each port is sealed with a flange containing a concentric array of 48 bulkhead fittings through which individual cables (from the PMTs, temperature sensors, and level sensors) pass. At the center, there is a 30\,cm diameter port used for the deployment of radioactive sources along the detector central axis (explained in Section~\ref{s:Calibration_deployment_mechanism}). A HVA 72210-0606R gate valve~\cite{gatevalve}, actuated by compressed gas, is mounted above this central port to provide IV access during calibration source deployments (see Section~\ref{s:Calibration_program}). Design drawings of the OV are shown in Fig.~\ref{f:OV}.

\subsubsection{The PMT support structure} \label{s:PMT_Support_Structure}

\begin{figure}[!h]
    \begin{center}
    \includegraphics[width=1.0\textwidth, trim=0cm 5cm 0cm 5cm, clip]{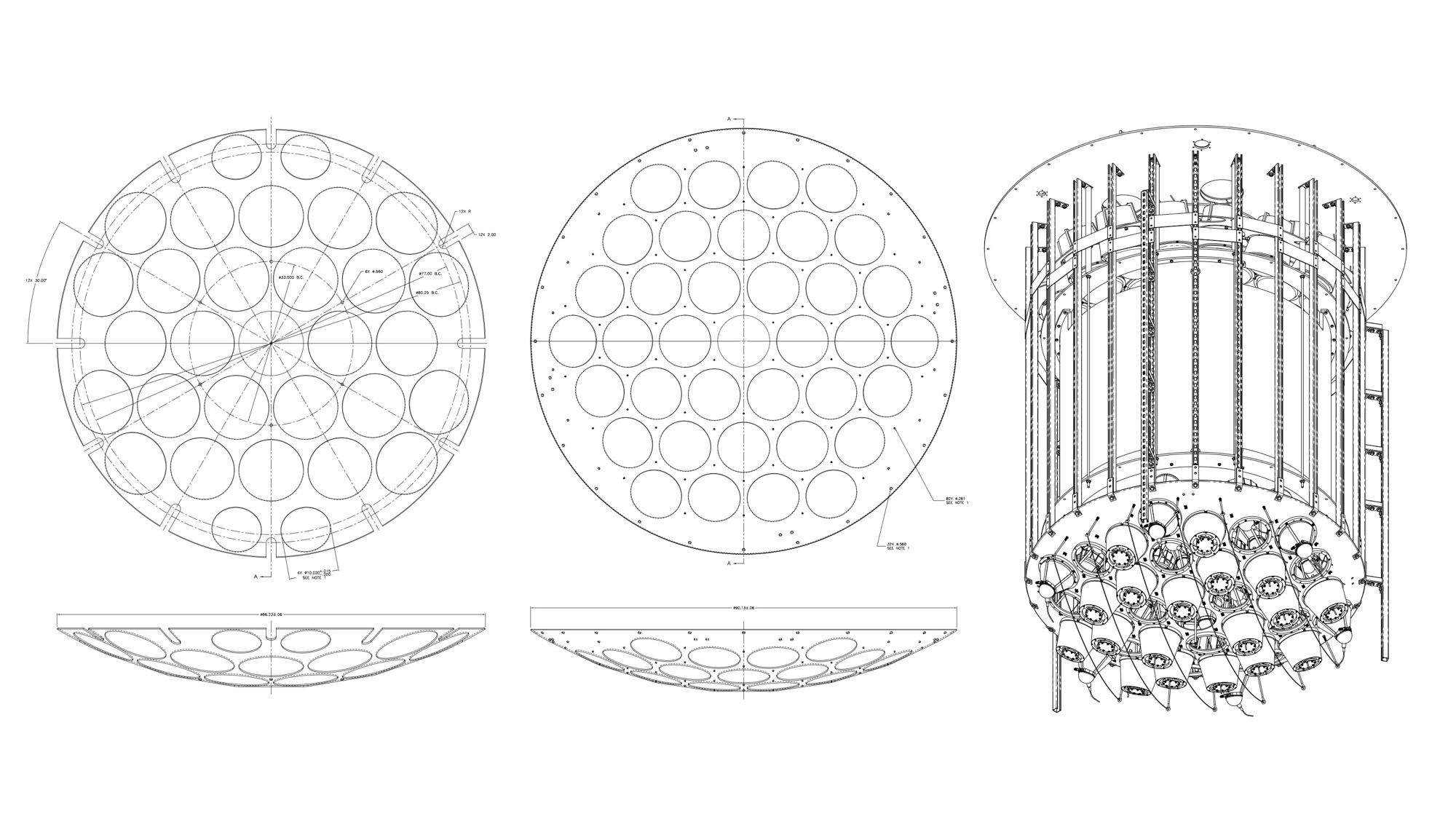}
    \end{center}
    \caption{Top and side view of the upper (left) and lower dish (center). Perspective view of the barrel support structure (right).}
    \label{f:PSUP}
\end{figure}

The PMT support structure, shown in Fig.~\ref{f:PSUP}, is constructed from passivated, electropolished 304 SS and consists of three distinct segments: the upper dish, the lower dish, and the barrel supports. The upper dish is a 0.64-cm-thick, 2.19-m-diameter spherical cap with a 2.14-m spherical radius. It has 27 hexagonally packed holes, for maximum photocathode coverage, each with a diameter of 33\,cm. The central hole is purposely left vacant to allow for the central-axis deployment of calibration sources. The remaining 26 are for PMT installation. A further four holes with a diameter of 25\,cm are located toward the cap edge, providing the option to install additional PMTs to further improve photocathode coverage. The lower dish is a 0.64-cm-thick, 2.29-m-diameter cap with a 2.14-m spherical radius. It has 37 hexagonally packed, 28-cm-diameter holes. This design was adopted to allow for the maximum possible photocathode coverage when mounting the lower PMT array. The barrel support structure was constructed using 24 vertical unistrut columns, spaced every 15$^{\circ}$ at a radius of 1.02\,m from the central axis. Between each unistrut pair, a column of seven PMT plates was bolted in place, for a total of 168 barrel PMTs. Schematics of the top dish, bottom dish, and barrel support structure are shown in Fig.~\ref{f:PSUP}. 

\subsubsection{The inner vessel} \label{s:IV}

\begin{figure}[!h]
    \begin{center}
    \includegraphics[width=1.0\textwidth, trim=0cm 6cm 0cm 3cm, clip]{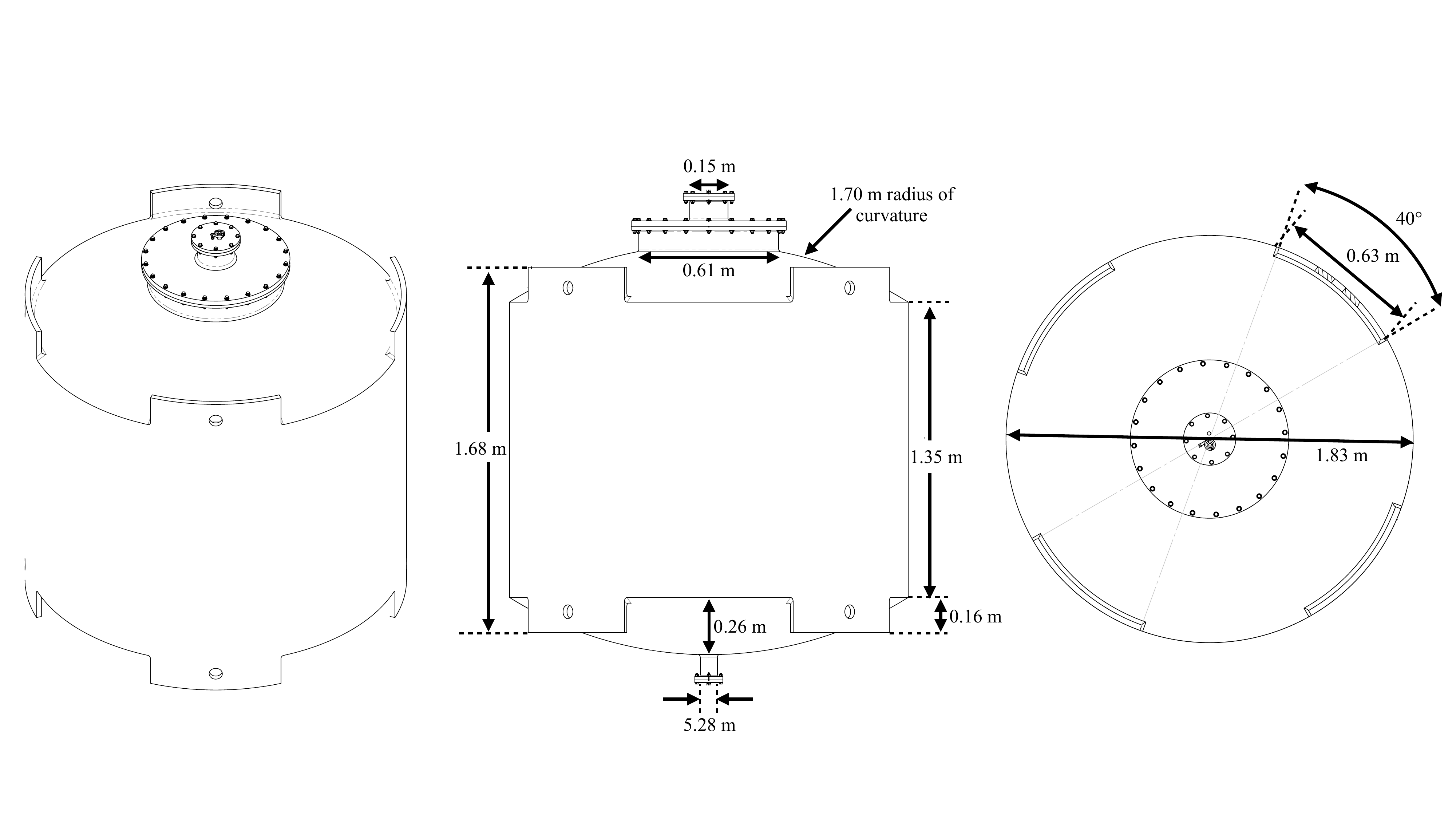}
    \end{center}
    \caption{Projection (left), side view (center) and top view (right) of the \eos\  IV.}
    \label{f:iv}
\end{figure}

\begin{figure}[!h]
    \begin{center}
    \includegraphics[width=0.65\textwidth]{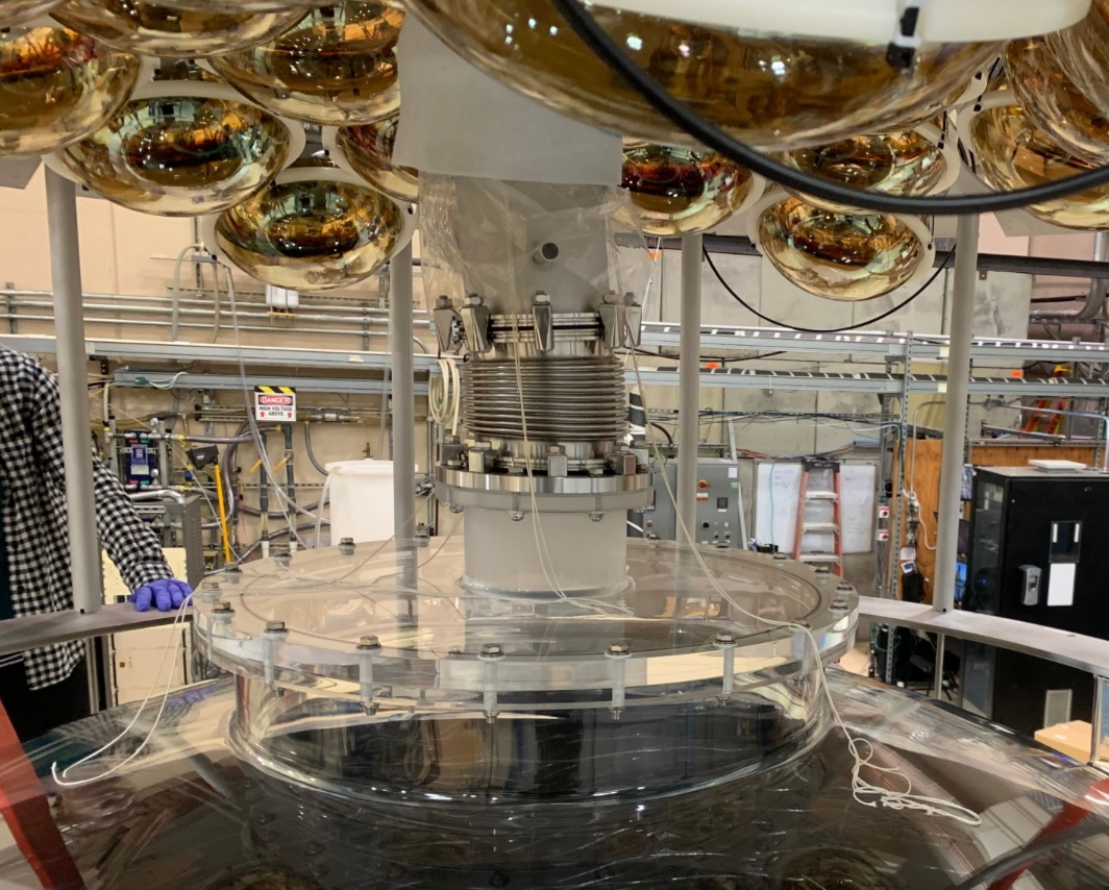}
    \end{center}
    \caption{The neck assembly during installation showing the interface between the acrylic IV, the bellows pipe, and the upper connection tube. The bellows pipe allows for the neck assembly to be attached to the underside of the OV lid, while the connection tube also serves as the injection point of the recirculation system.}
    \label{f:bellows}
\end{figure}

The \eos\ IV is a 2.5-cm-thick UV-transmitting (UVT) acrylic spherocylinder constructed by Reynolds Polymer Technology~\cite{reynolds}, with a target mass of approximately 4 tonnes when filled with water (Fig.~\ref{f:iv}). The cylindrical barrel is 1.35\,m tall with a diameter of 1.83\,m, and each hemispherical cap has a radius of curvature of 1.70\,m. At the top and bottom of the cylindrical section are four stands to act as points of contact for the support structure. The lower cap includes a cylindrical column of 5.28\,cm in diameter to allow connection to the drain line of the fluid handling system. The top cap features a 0.61\,m diameter cylinder, which connects to the 0.16\,m diameter lower portion of the "neck". Above this, a 316 SS bellows pipe and connection tube are attached and sealed with ISO-160-style flange gaskets, forming the upper neck. This assembly serves as both an opening for calibration source deployment and a connection point for the fluid handling return line. The \eos\ neck is shown in Fig.~\ref{f:bellows}.

\subsection{Photon detection} 

Three types of PMTs are deployed in \eos\ (Hamamatsu R14688-100~\cite{ham_datasheet_r14688}, R11780~\cite{r11780_paper}, and R7081-HQE~\cite{ham_datasheet_r7081}). These PMTs and associated optical components were selected and configured with the following requirements:
\begin{itemize}
    \item Sub-ns timing to enable effective Cherenkov/scintillation separation.
    \item High quantum efficiency between 300 and 600\,nm for effective detection of both Cherenkov and scintillation light.
    \item A dichroicon array for testing spectral photon sorting capabilities.
    \item Maximum possible photocathode coverage within detector design constraints.
    \item Watertight sealing for deployment within the OV.
\end{itemize}

\subsubsection{8" barrel PMTs}

\begin{figure*}[h!]
    \centering
    \includegraphics[scale=0.6]{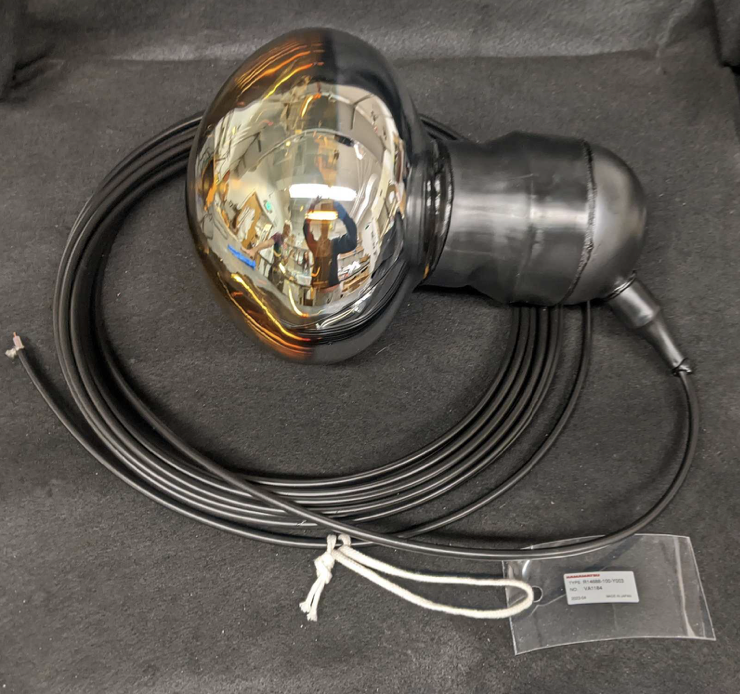}
    \caption{The Hamamatsu R14688-100 8-inch PMT deployed in \eos. 168 are mounted in columns of seven around the barrel, with 36 deployed on the lower dish.}
    \label{f:8in_PMT}
\end{figure*}

204 8-inch R14688-100 Hamamatsu PMTs are installed in \eos\ (Fig.~\ref{f:8in_PMT}). Of these, 168 are positioned in 24 columns of seven around the barrel, with the remaining 36 installed on the bottom dish. These PMTs use a box-and-linear-focused 10-stage dynode structure, are potted by the manufacturer for waterproofing, and are each equipped with a 20\,m waterproof cable. The R14688-100 is a state-of-the-art fast PMT with sub-ns FWHM timing, essential for the improved vertexing and Cherenkov/scintillation separation that \eos\ expects to demonstrate. Furthermore, they have excellent charge resolution, enabling individual photoelectron counting.

Testing prior to installation was performed at Berkeley~\cite{hamamatsu_8in}, demonstrating an average gain of $1 \times 10^{7}$, a peak-to-valley ratio frequently above 4, and a transit time spread of $1.00 \pm 0.08$\,ns (FWHM) for all PMTs. The dark rate for a subset of 14 PMTs was measured to be $2.18 \pm 1.40$\,kHz at 22\,$^{\circ}$C, a rate expected to decrease by a factor of $\sim$5 at the 8\,$^{\circ}$C operating temperature of \eos.

\begin{figure*}[h!]
    \centering
    \includegraphics[scale=0.3]{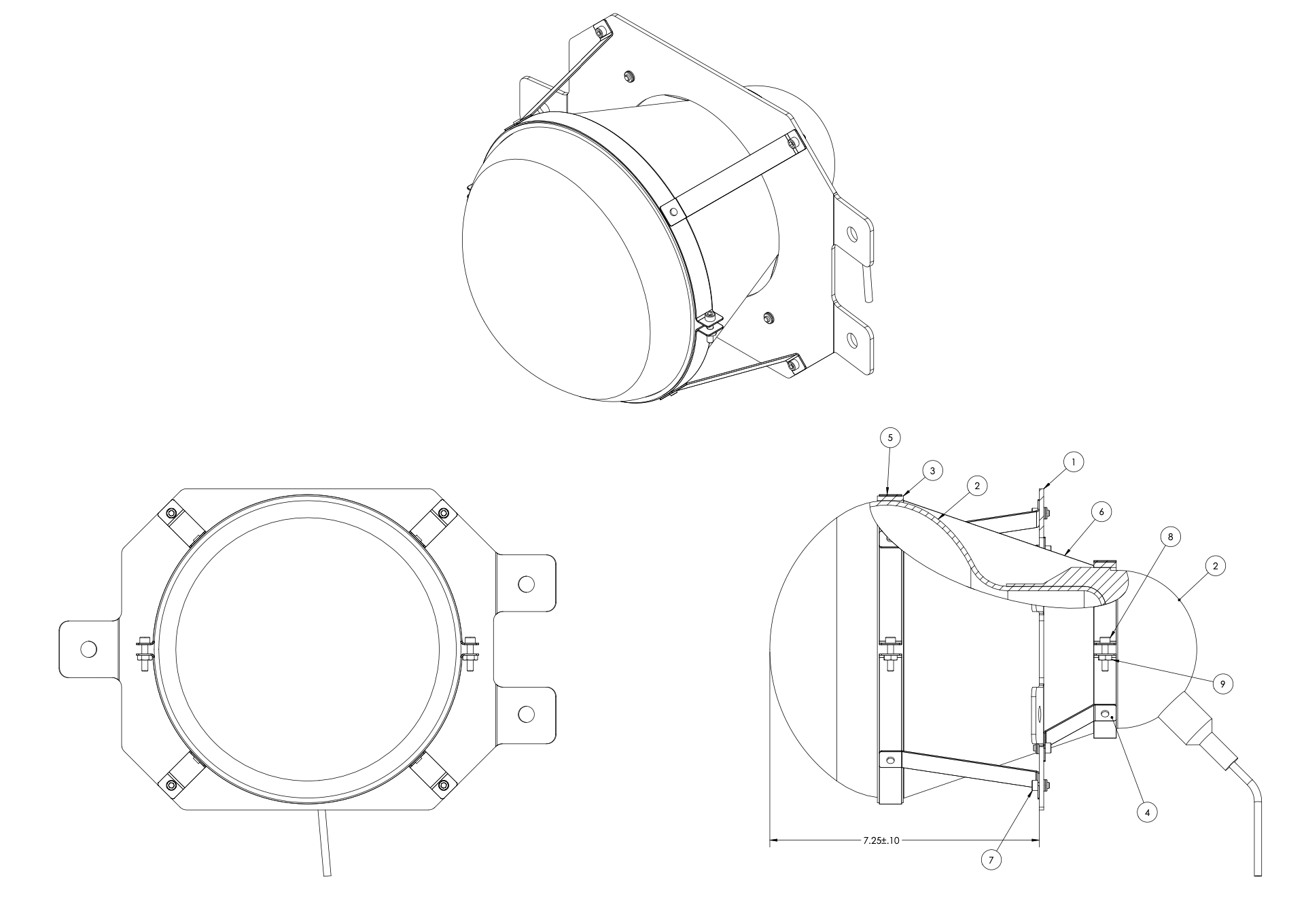}
    \caption{Perspective, front, and side views of the 304 SS plates designed to attach the 8-inch PMTs to the unistrut barrel support structure.}
    \label{f:8in_PMT_plate}
\end{figure*}

\begin{figure*}[h!]
    \centering
    \includegraphics[width=0.65\textwidth]{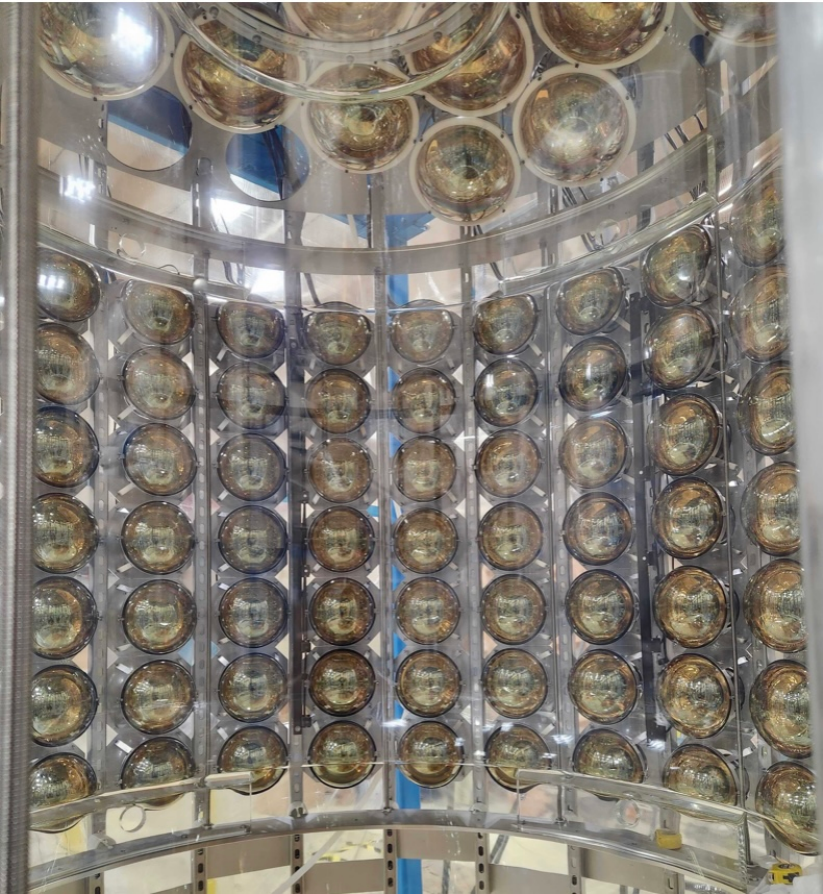}
    \caption{View of the PMT columns after installation on the barrel support structure.}
    \label{f:8in_PMT_columns}
\end{figure*}

To mount the 168 barrel PMTs to the unistrut supports, 316 SS mounting plates were designed, as shown in Fig.~\ref{f:8in_PMT_plate}. The partially constructed barrel array is shown in Fig.~\ref{f:8in_PMT_columns}.

\subsubsection{12" top dish PMTs}

\begin{figure*}[h!]
    \centering
    \includegraphics[width=0.65\textwidth]{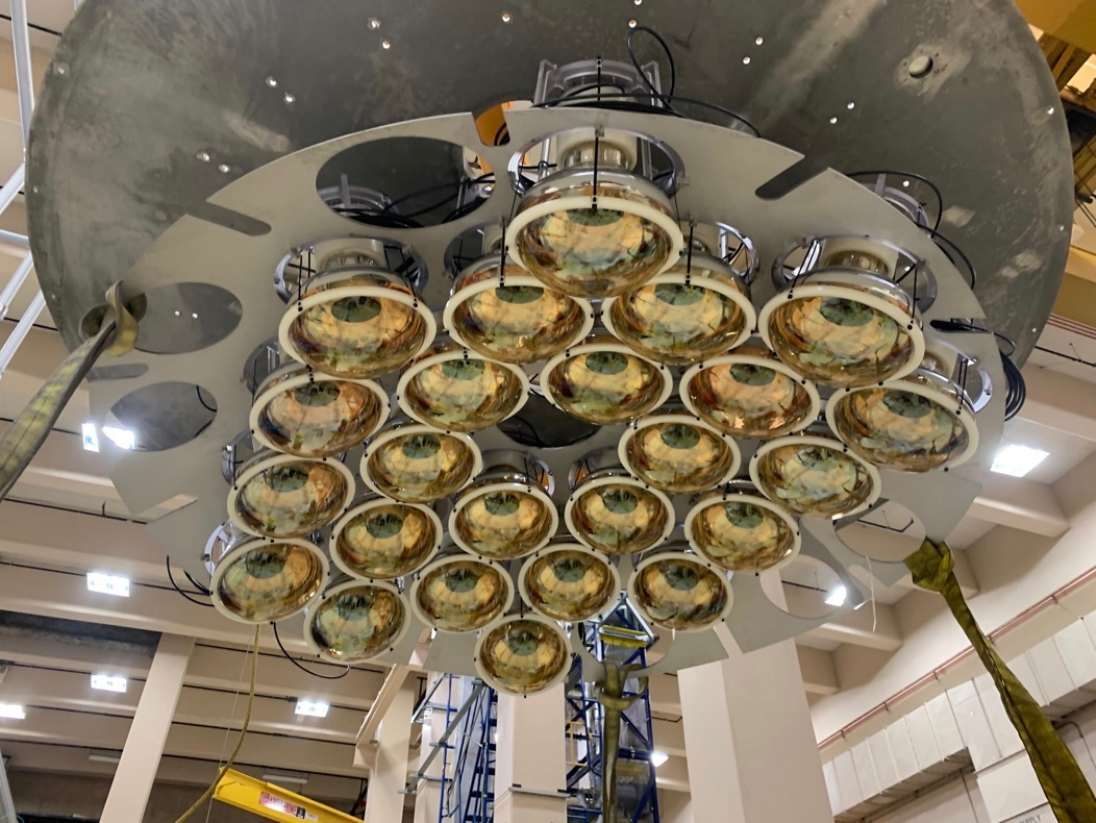}
    \caption{The 24 installed R11780 PMTs mounted to the top dish using polycarbonate brackets.}
    \label{f:top_array}
\end{figure*}

24 R11780 12-inch PMTs, originally developed for the Long Baseline Neutrino Experiment (LBNE) program~\cite{LBNECDR}, are positioned on the upper dish to further increase photocathode coverage. These repurposed PMTs are instrumented with a conformally coated printed circuit board (PCB) and waterproofed for use in \eos. Fig.~\ref{f:top_array} shows the PMTs installed on the upper dish. A technical paper describing the performance of the R11780 PMTs is provided in~\cite{r11780_paper}. 

\subsubsection{Dichroicon array} 
\label{s:dichroicons}
\begin{figure*}[h!]
    \centering
    \includegraphics[width=0.65\textwidth]{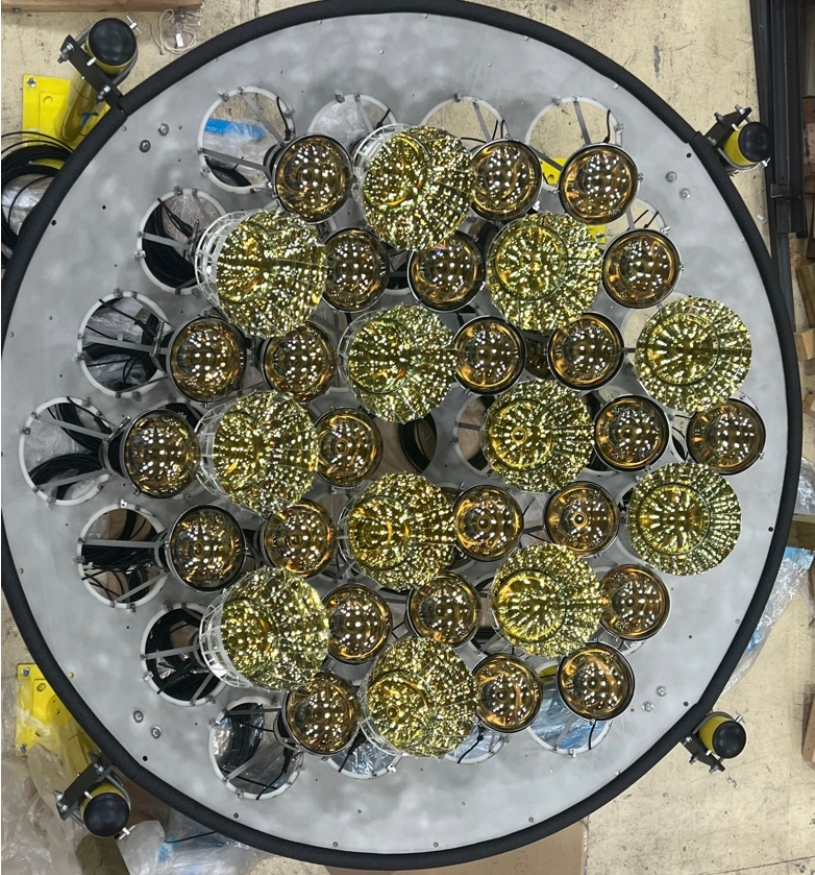}
    \caption{The lower dish with 16 R14688-100 8-inch PMTs and 12 dichroicons installed. On the underside of this dish, behind the dichroicons, 13 10-inch PMTs were mounted.}
    \label{f:bottom_array}
\end{figure*}

 Of the 36 R14688-100 8-inch PMTs mounted above the bottom dish, 12 have dichroicons attached (see Fig.~\ref{f:bottom_array}). The dichroicons were installed to sort photons by wavelength, enabling \eos\ to spectrally differentiate between Cherenkov and scintillation light. Behind the dichroicons, mounted on the underside of the bottom dish, are 13 R7081-HQE 10-inch PMTs.  Bench-top measurements of prototype dichroicon performance have shown promising results with greater than 90$\%$ Cherenkov purity, defined as the fraction of detected photons that are due to Cherenkov emission~\cite{dichroicon_paper}.

\begin{figure*}[h!]
    \centering
    \includegraphics[scale=0.53]{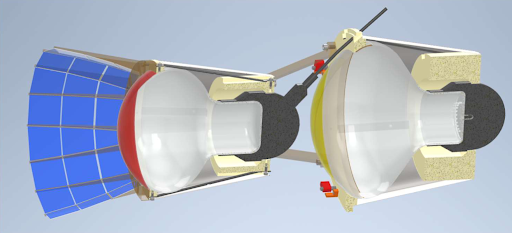}
    \caption{Cross section rendering of the 8-inch dichroicon with the 10-inch PMT mounted behind. In \eos, the 10-inch PMTs are deployed in an offset position relative to the dichroicons to increase photon acceptance; the PMT is shown centered here for illustrative purposes.}
    \label{fig:dichroicon_slice}
\end{figure*}

\begin{figure*}[h!]
    \centering
    \includegraphics[scale=0.14, trim=13cm 9cm 13cm 3cm, clip]{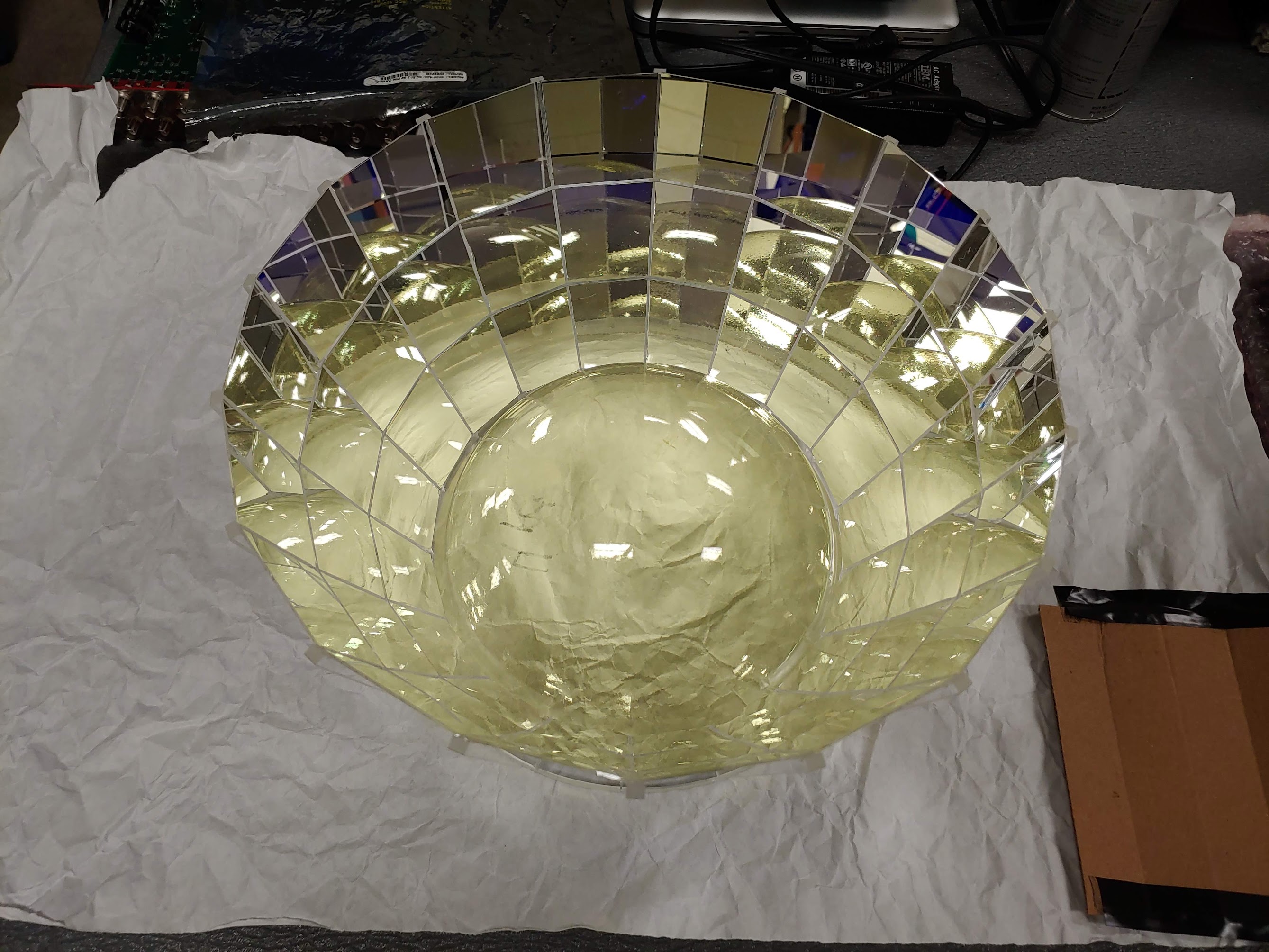}
    \caption{The Winston cone arrangement of dichroic filters with the central molded longpass filter.}
    \label{fig:penn_dichroicon}
\end{figure*}

\begin{figure*}[h!]
    \centering
    \includegraphics[width=0.8\textwidth]{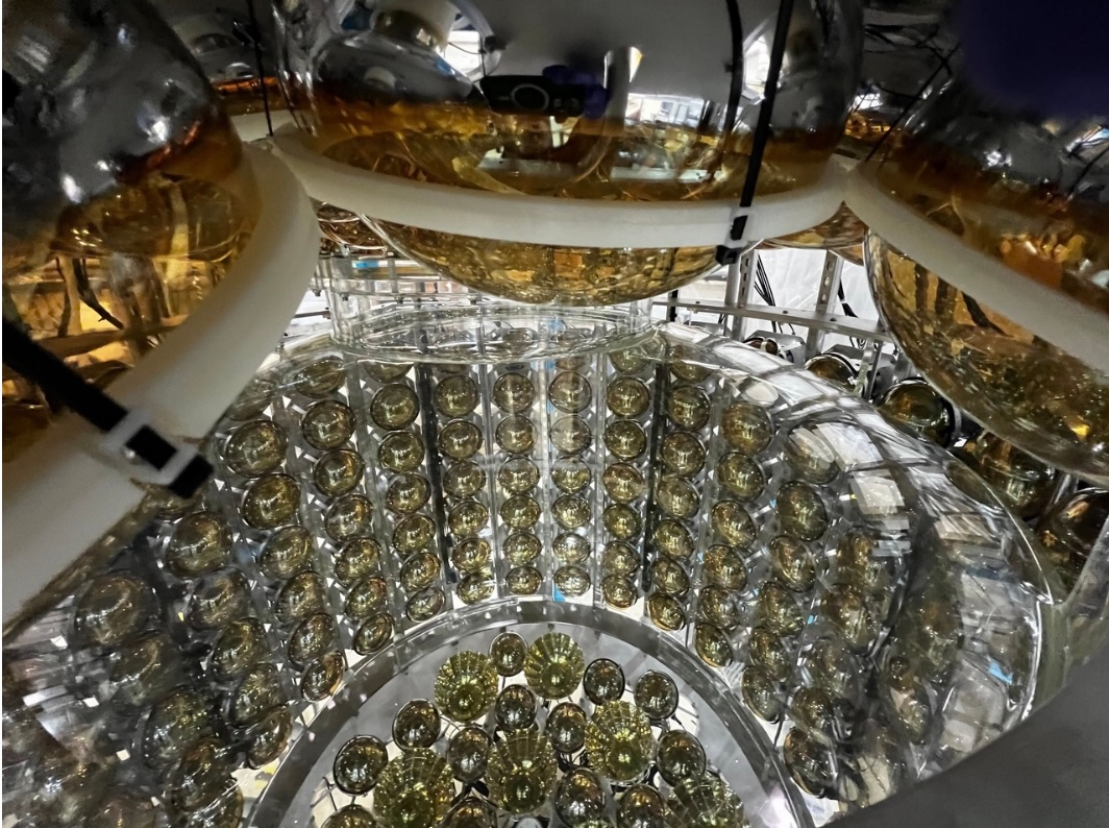}
    \caption{The top, bottom and barrel PMT arrays after installation.}
    \label{f:pmtarray}
\end{figure*}

Photon sorting is accomplished using a Winston cone configured with dichroic filters. By mounting an absorbing long-pass filter over the R14688-100 PMT photocathode and short-pass dichroic filters to define the surface of the Winston cone, the shorter-wavelength scintillation light will pass through the cone, allowing it to be detected by the back array of PMTs, and Cherenkov light will be concentrated toward the aperture of the dichroicon. A cross-sectional rendering of the full dichroicon arrangement is shown in Fig.~\ref{fig:dichroicon_slice}. A Winston cone with the dichroic and long-pass filters attached is shown in Fig.~\ref{fig:penn_dichroicon}. The short-pass dichroic filters are manufactured by Knight Optical~\cite{knight_dichroic_sp}. A cut-on wavelength of 450\,nm was selected, as this is optimal for separating Cherenkov and scintillation photons for the formulations to be deployed in \eos. The individual filters were glued into transparent acrylic support structures to create the desired Winston cone shape. The absorbing long-pass filters, also produced by Knight Optical~\cite{knight_acrylic_lp}, were molded to conform to the hemispherical face of the PMTs. The installed top dish, bottom dish, and barrel array are shown in Fig.~\ref{f:pmtarray}.

\subsection{Fluid handling}

The \eos\ fluid handling system was designed with the following requirements:
\begin{itemize}
    \item Ensure material compatibility of all components in contact with the liquid to prevent optical degradation via leaching.
    \item Attain 18\,$\mathrm{M\Omega\cdot}$cm ultrapure water.
    \item For the OV, cool to $\sim$8\,$^\circ$C to prevent biological growth, reduce the PMT dark rates, and passively chill the IV.
    \item Provide a sufficient flow rate to circulate the volumes to maintain long-term cleanliness.
    \item Enable removal of dust ingress and mitigate biological growth.
    \item Monitor flow rates, resistivity and temperature.
    \item Automatically shut off in the event of a leak.
\end{itemize}

\subsubsection{The outer vessel water system}\label{s:OV_Fluid_Handling}

\begin{figure}[!h]
    \begin{center}
    \includegraphics[width=0.8\textwidth,trim=1cm 2cm 1cm 0cm, clip]{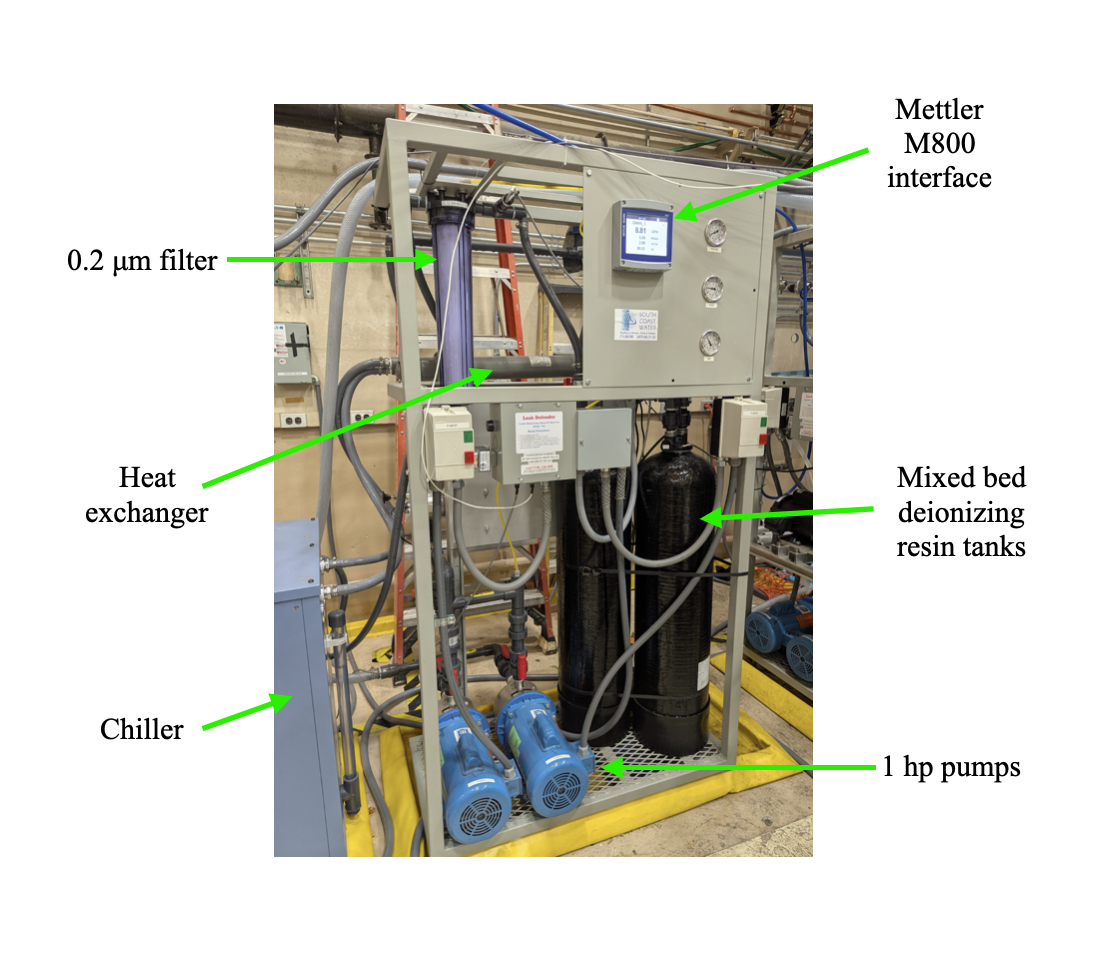}
    \end{center}
    \caption{Front view of the OV fluid handling system with visible components highlighted.}
    \label{f:ov_fluid_handling_system}
\end{figure}

The OV is filled with $\sim$16\,tonnes of ultrapure water that is recirculated using a fluid handling system, shown in Fig.~\ref{f:ov_fluid_handling_system}. The system is designed to prevent biological growth, remove particulates, and maintain a minimum resistivity of 18\,$\mathrm{M\Omega\cdot}$cm, thereby minimizing light attenuation and maximizing detector performance. The OV volume is chilled to $\sim$8$\,^\circ$C to reduce PMT dark rates, limit tank corrosion, and further suppress biological growth. The chilled OV acts as a thermal bath to cool the IV, increasing scintillation light yield by increasing Förster resonant energy transfer~\cite{forster} efficiency.

\begin{figure}[!h]
    \begin{center}
    \includegraphics[width=1.0\textwidth,trim=0cm 3cm 0.3cm 0cm, clip]{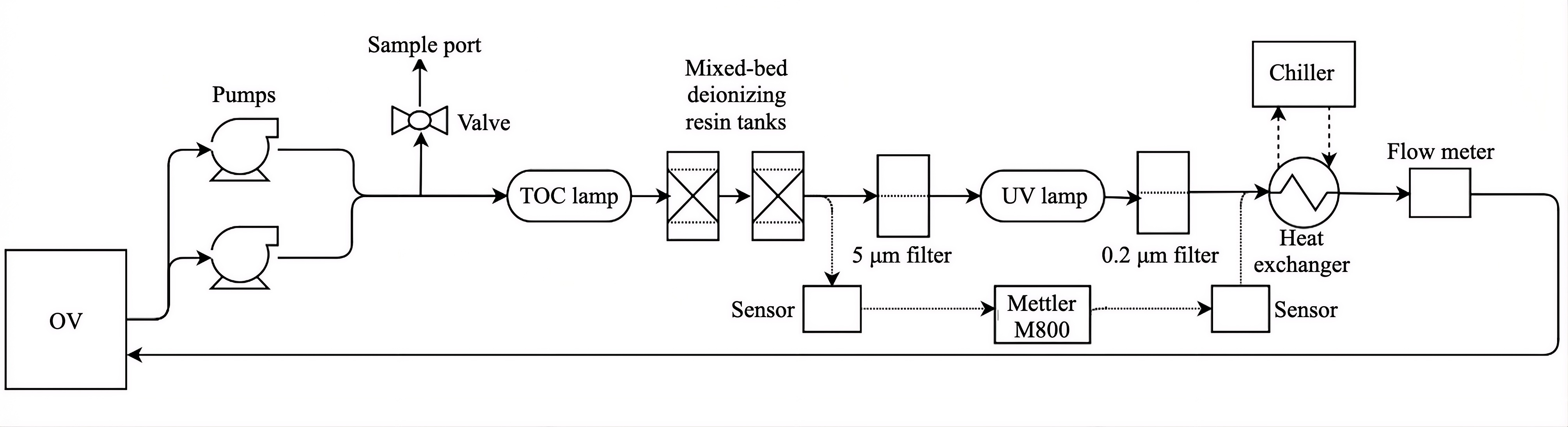}
    \end{center}
    \caption{Diagram showing the flow path and components of the OV water handling system.}
    \label{f:water_handling_system}
\end{figure}

\begin{figure}[!h]
    \begin{center}
\includegraphics[width=0.7\textwidth]{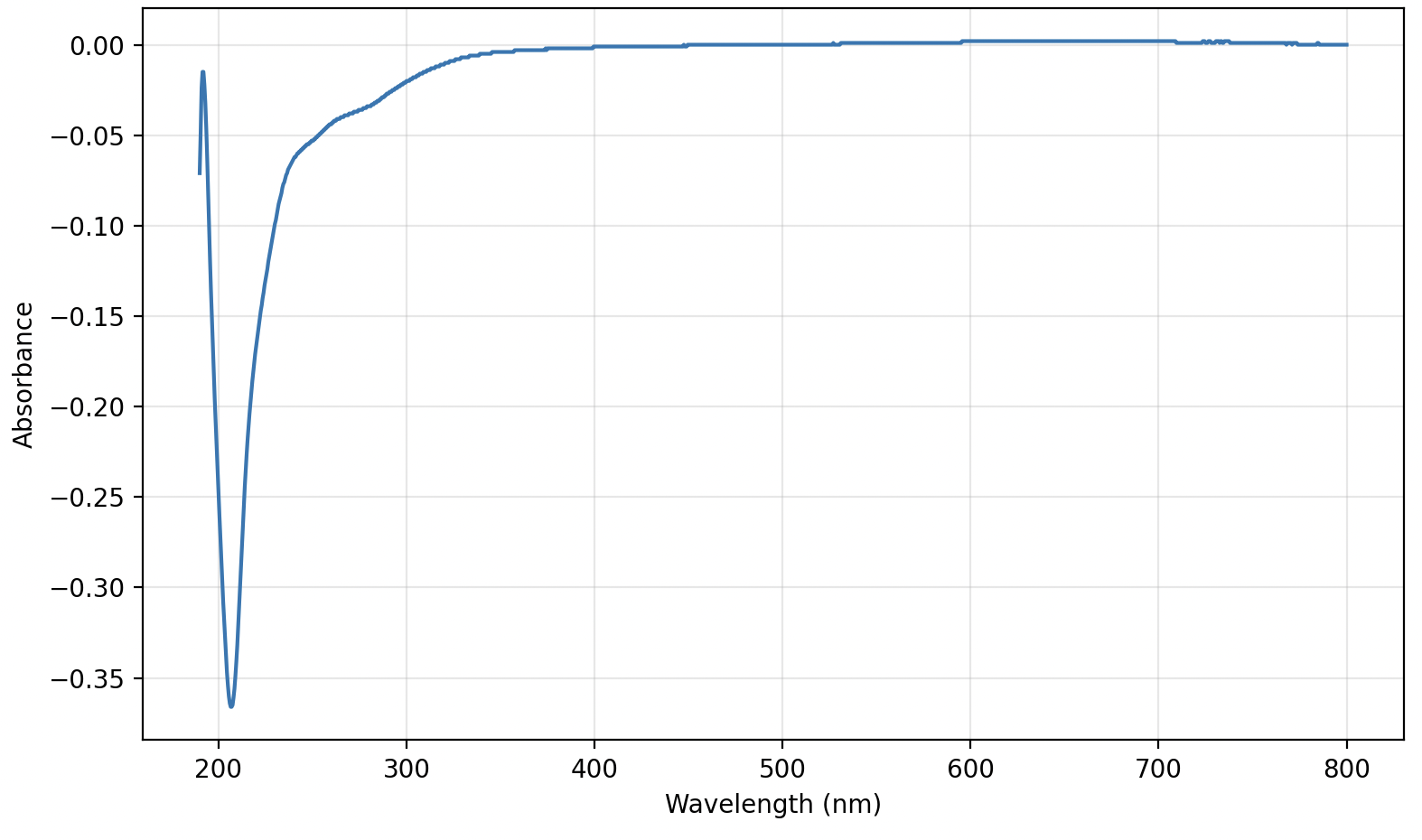}
    \end{center}
    \caption{Ultraviolet-visible absorption spectrum of the purified OV water, demonstrating the performance of the fluid handling system.}
    \label{f:OV_UV_Vis}
\end{figure}

The OV fluid handling system consists of nine main components, shown in Fig.~\ref{f:water_handling_system}, with the following functions:
\begin{enumerate}
    \item Two 1\,hp Goulds Water Technology LB1012TE pumps~\cite{goulds}, installed in parallel, which recirculate the water volume at a rate of $\sim$8\,GPM.
    \item A 187\,nm Sanitron S37C Total Organic Carbon (TOC) lamp~\cite{sanitron} initiates the breakdown of organic material via UV oxidation.
    \item Two 1040 mixed-bed deionization resin canisters remove ionic contaminants.
    \item A 5\,\micro m filter removes larger particulates.
    \item A 254\,nm Sanitron S23A UV lamp~\cite{sanitron} further suppresses biological growth.
    \item A 0.2\,\micro m filter removes smaller particulates.
    \item A heat exchanger and a Thermal Care EQ2A03 chiller unit~\cite{thermal} cool the water during recirculation.
    \item A Mettler M800 interface~\cite{mettler} monitors flow rate, resistivity, temperature, and pH. 
    \item Three Asurity WS-1 water sensors~\cite{ws1} automatically shut off the pumps and isolate the system in the event of a leak.
\end{enumerate}
To ensure optical quality, the absorbance of samples removed from the OV was measured using a Shimadzu UV-2600i Plus ultraviolet-visible spectrophotometer~\cite{shimadzu}. The absorbance spectrum is presented in Fig.~\ref{f:OV_UV_Vis}. No significant absorbance features are observed throughout the wavelength range relevant for PMT sensitivity and detector operation, demonstrating the efficacy of the purification system. Larger negative absorbance values observed at  shorter wavelengths are due to baseline subtraction and instrumental uncertainties and are not physically significant.

\subsubsection{The inner vessel water system}
\begin{figure}[!h]
    \begin{center}
    \includegraphics[width=1.0\textwidth,trim=0cm 1cm 0cm 0cm, clip]{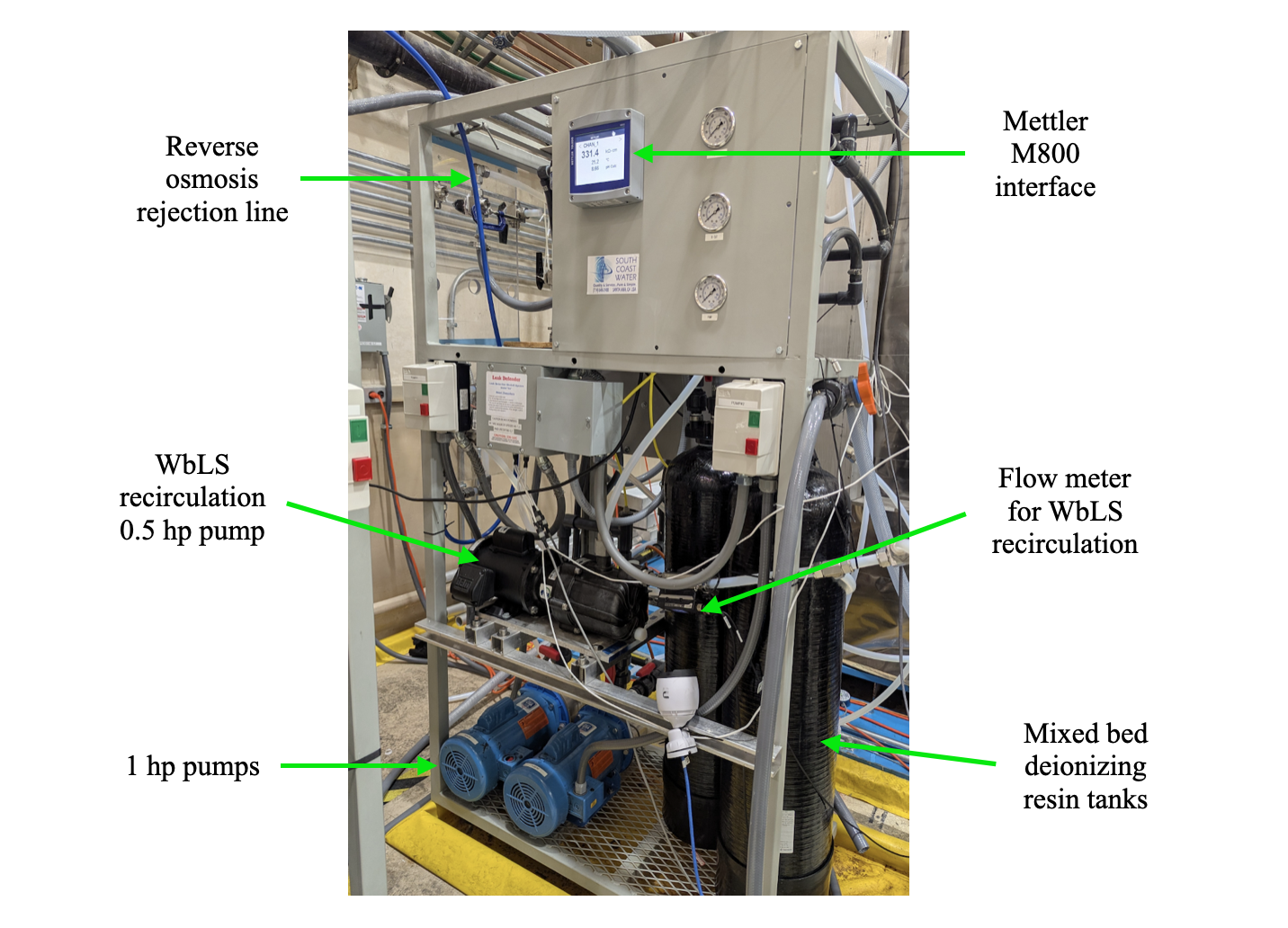}
    \end{center}
    \caption{The IV fluid handling system with visible components highlighted. The output of this stage serves as the input for the RO system, with the RO reject stream feeding back into this stage of the fluid handling loop and the RO product returning to the IV. Note, the WbLS recirculation system was mounted to the IV fluid handling frame, though it was designed to function as an independent system.}
    \label{f:IV_system}
\end{figure}

\begin{figure}[!h]
    \begin{center}
    \includegraphics[width=1.0\textwidth,trim=0.3cm 1cm 1cm 0cm, clip]{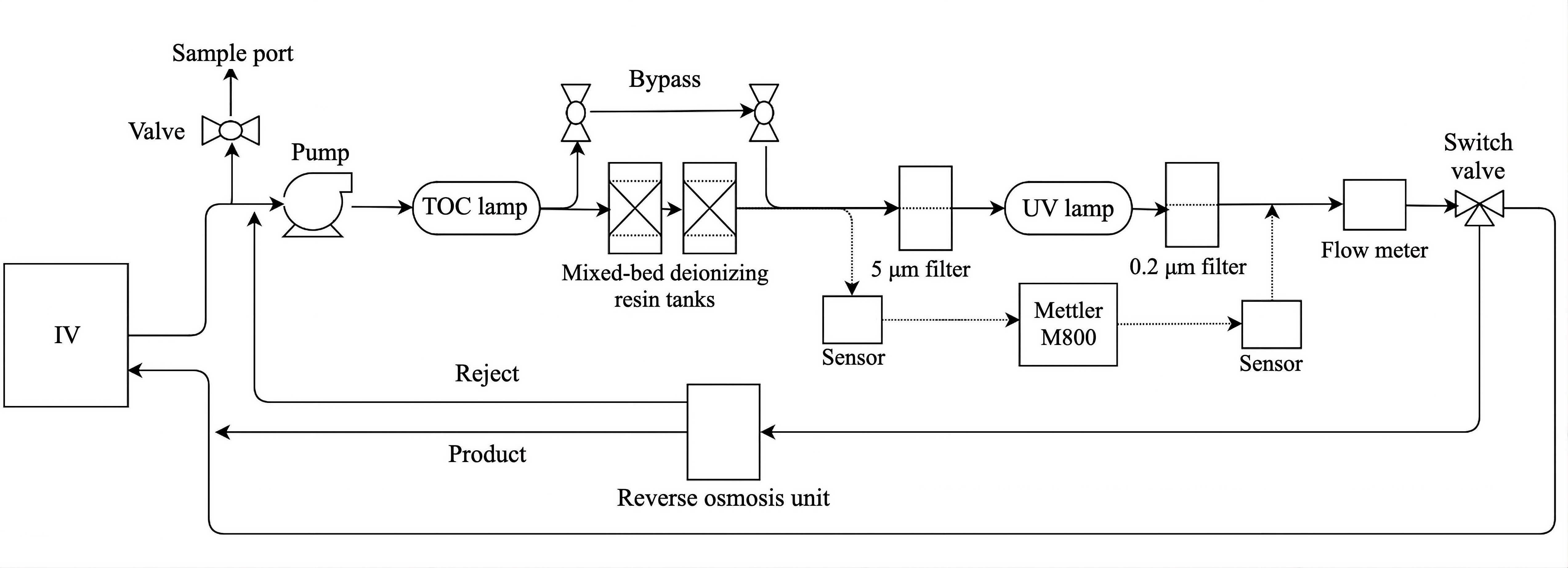}
    \end{center}
    \caption{IV fluid handling P\&ID showing the initial water pre-treatment prior to  passing to the RO unit. Note, the RO reject is recirculated through the entire system to maintain the total water volume.}
    \label{f:fluid_handling_system}
\end{figure}

The 4-tonne IV volume employs a more complex purification system, shown in Fig.~\ref{f:IV_system}, and is designed for two separate detector phases. For the water phase (Fig.~\ref{f:fluid_handling_system}), the system resembles that used for the OV,  with three major differences:
\begin{enumerate}
\item A lower flow rate of $\sim$2\,GPM is used due to the smaller vessel volume.
\item No heat exchanger is used, as the IV relies on passive chilling from the OV to reach a temperature of $\sim$12\,$^\circ$C.
\item Due to the more stringent purity requirements, a reverse osmosis (RO) system was installed as the final stage of purification. This is essential to remove any plasticizers that can leach into the volume and potentially produce a lower-quality WbLS product upon the addition of the scintillating components. The RO reject line is recirculated through the IV water system to prevent water loss. 
\end{enumerate}

\begin{figure}[!h]
    \begin{center}
    \includegraphics[width=0.7\textwidth]{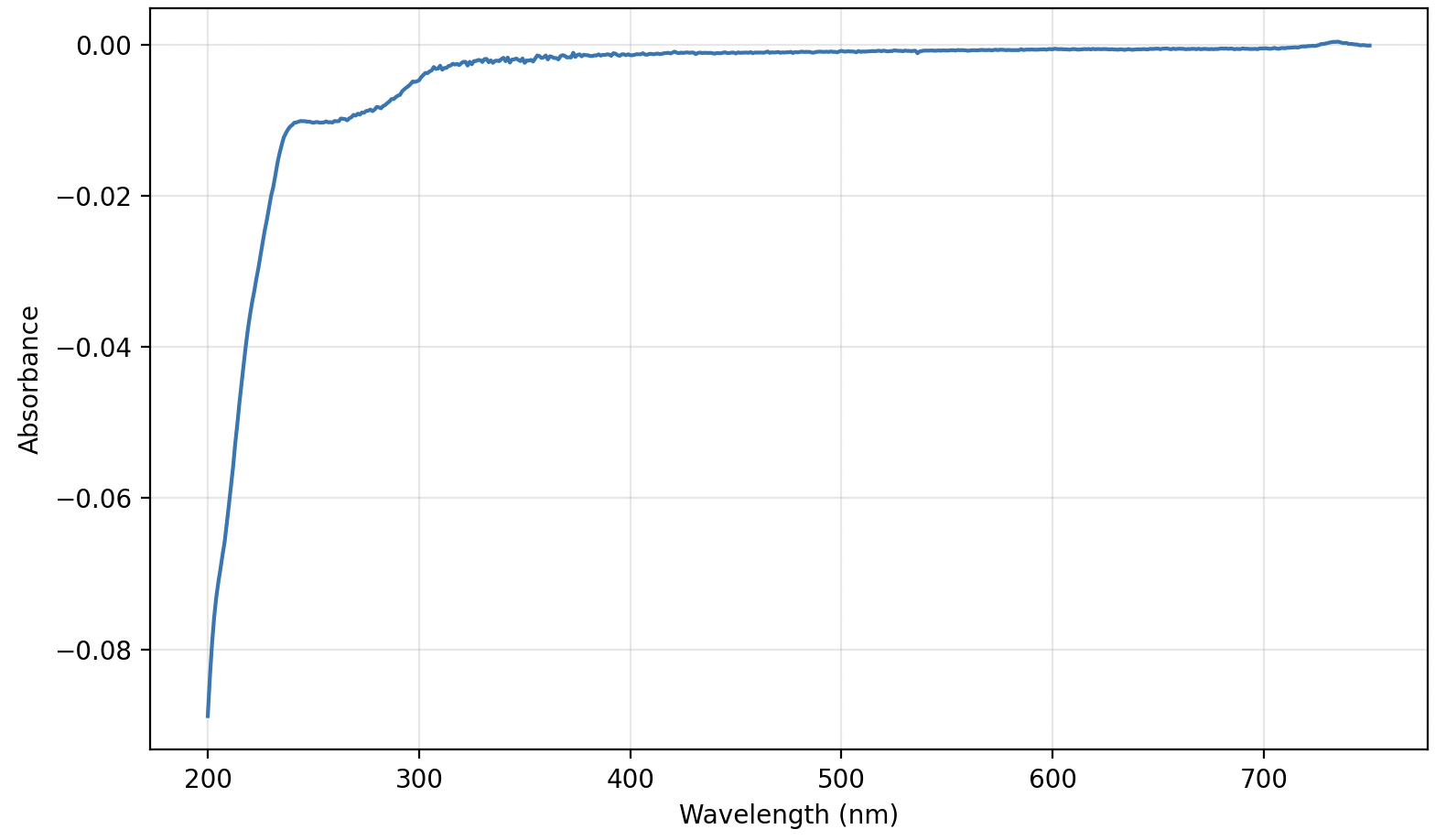}
    \end{center}
    \caption{Ultraviolet-visible absorption spectrum of purified IV water, demonstrating the performance of the recirculation system.}
    \label{f:IVWater_UV_Vis}
\end{figure}

The absorbance of purified IV water was also measured using a Shimadzu UV-1800 ultraviolet-visible spectrophotometer~\cite{shimadzu_uv} to ensure water quality. The absorbance spectra are presented in Fig.~\ref{f:IVWater_UV_Vis}. As with the OV water, no significant absorbance features are observed throughout the wavelength range relevant for detector operation. The larger negative absorbance values at shorter wavelengths are again due to baseline subtraction and instrumental uncertainties. 

\subsubsection{The inner vessel WbLS system}

\begin{figure}[!h]
    \begin{center}
    \includegraphics[width=0.75\textwidth,trim=0cm 0.3cm 0.5cm 0cm, clip]{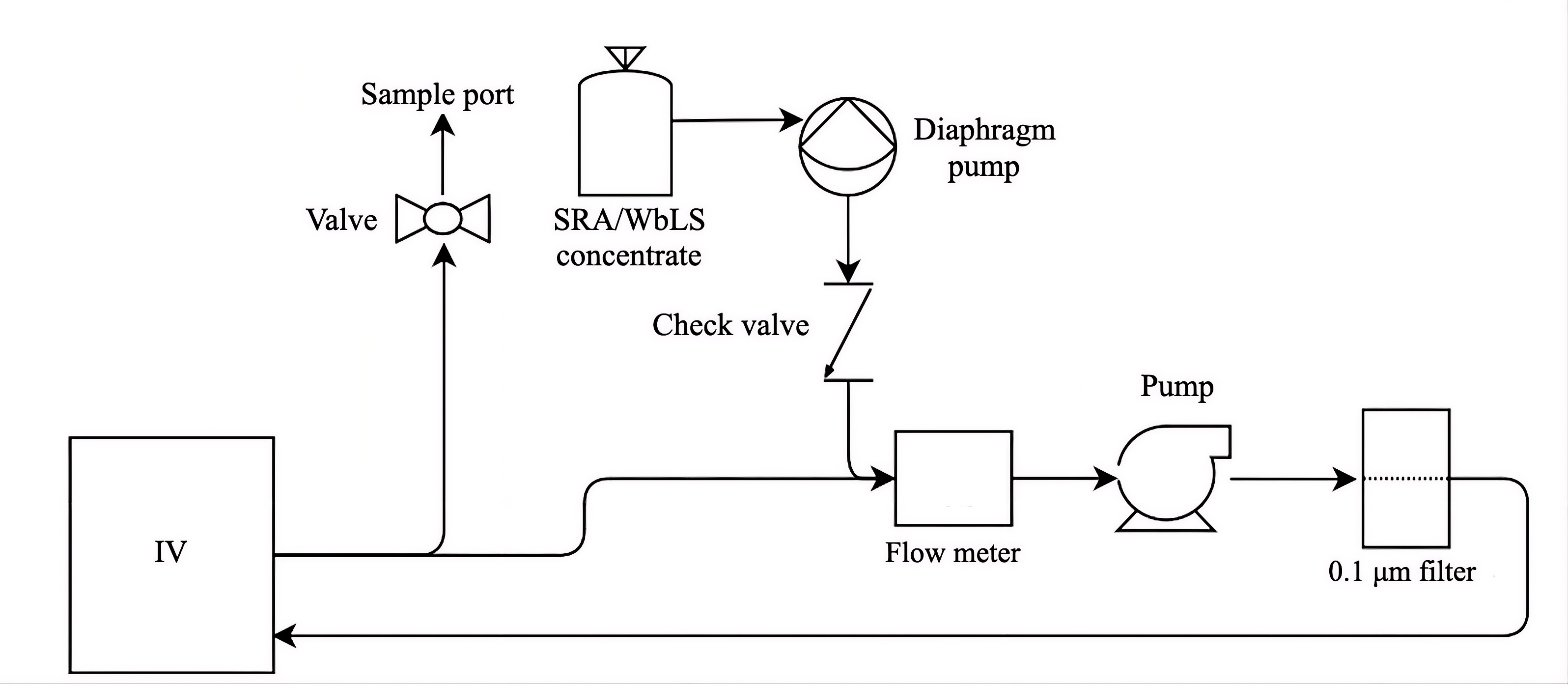}
    \end{center}
    \caption{WbLS flow path highlighting the simplified methodology to maintain optical clarity after the addition of organics. Note, a diaphragm pump is employed to introduce the scattering reduction agent (SRA) and WbLS concentrate, in-line, prior to the recirculation pump. More details of this process are discussed in Section~\ref{s:Deployment}.}
    \label{f:wbls_handling_system}
\end{figure}

As the water system would photodegrade (due to UV irradiation) and eventually remove any injected scintillator (via deionization, filtration and reverse osmosis), a much simpler flow path is used for the WbLS phase, shown in Fig.~\ref{f:wbls_handling_system}. Compatibility with WbLS requires all system elements to be made from 316 SS, Teflon, or polypropylene. The WbLS system has three main components:
\begin{enumerate}
    \item A 0.5\,hp Finish Thompson SP10V-T-6-M297 Teflon-lined pump~\cite{finish} recirculates the IV volume at a rate of $\sim$3\,GPM.
    \item A 25.4\,cm, 0.1\,$\upmu$m polypropylene filter removes particulates and maintains optical clarity.
    \item A separate scintillation line uses a Hi-Techno IX-C060TCN-TF-U diaphragm pump~\cite{iwaki} at a flow rate of $\sim$0.3\,LPM to gradually introduce organic components to the water volume.
\end{enumerate}

\begin{figure}[!h]
    \begin{center}
    \includegraphics[width=0.7\textwidth]{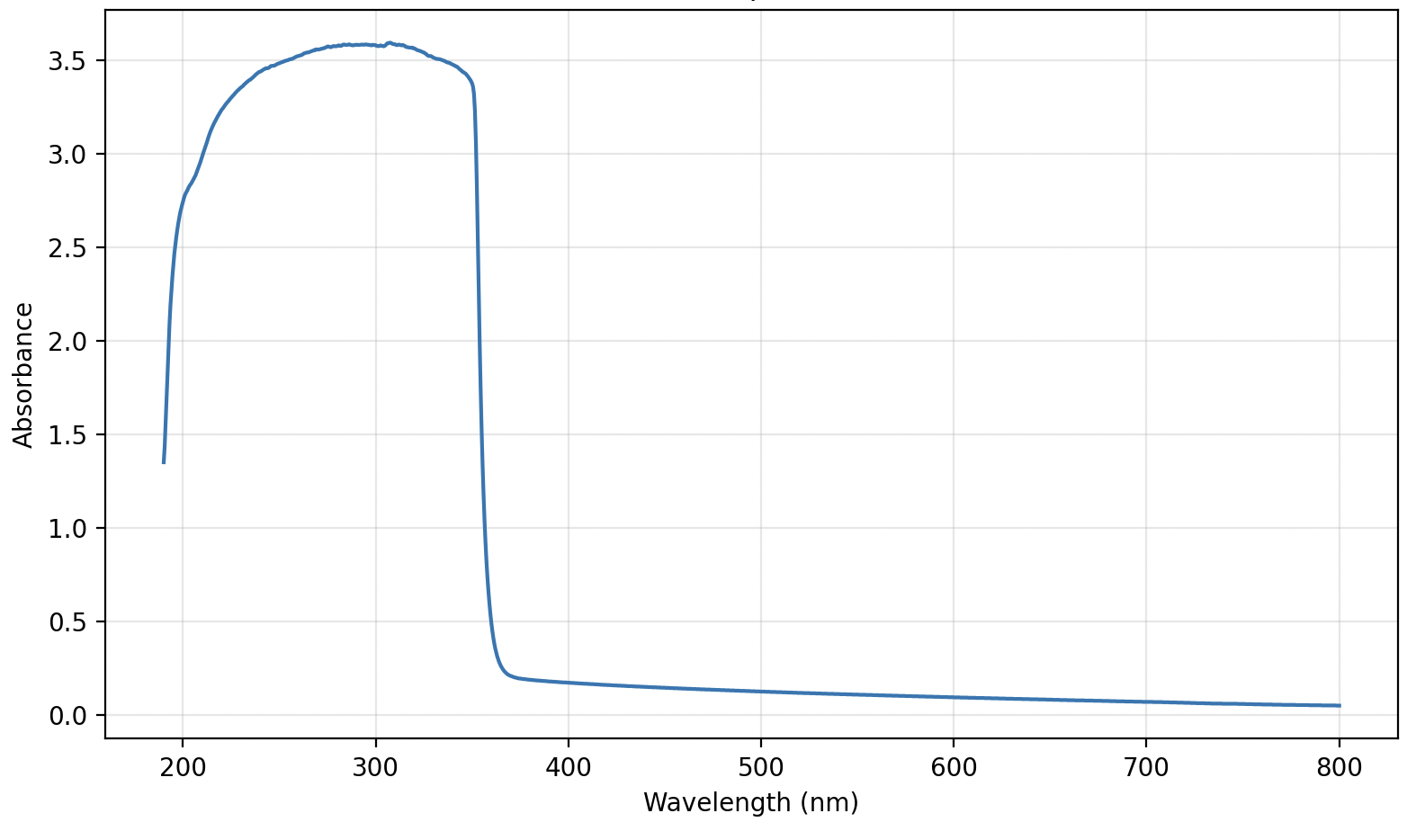}
    \end{center}
    \caption{Ultraviolet-visible absorbance spectrum of an IV sample of 1\% WbLS. The increased absorption below $\sim$365\,nm is expected due to the addition of the scintillation component. Above this wavelength, although the optical transparency is reduced relative to the pure water phase, the final WbLS formulation provided the transparency required for the \eos\ data-taking campaign.}
    \label{f:IVWBLS_UV_Vis}
\end{figure}

Samples from this system were measured using a Shimadzu UV-2600i Plus ultraviolet-visible spectrophotometer. The absorbance spectrum is shown in Fig.~\ref{f:IVWBLS_UV_Vis}. As expected, at wavelengths shorter than $\sim$365\,nm the WbLS is highly absorbing due to the scintillating component. Above this wavelength, the absorbance decreases substantially, providing sufficient optical clarity for the \eos\ data taking campaign.

\subsection{Cover gas system}
\begin{figure}[!h]
    \begin{center}
    \includegraphics[width=0.46\textwidth,trim=0cm 0cm 0cm 0cm, clip]{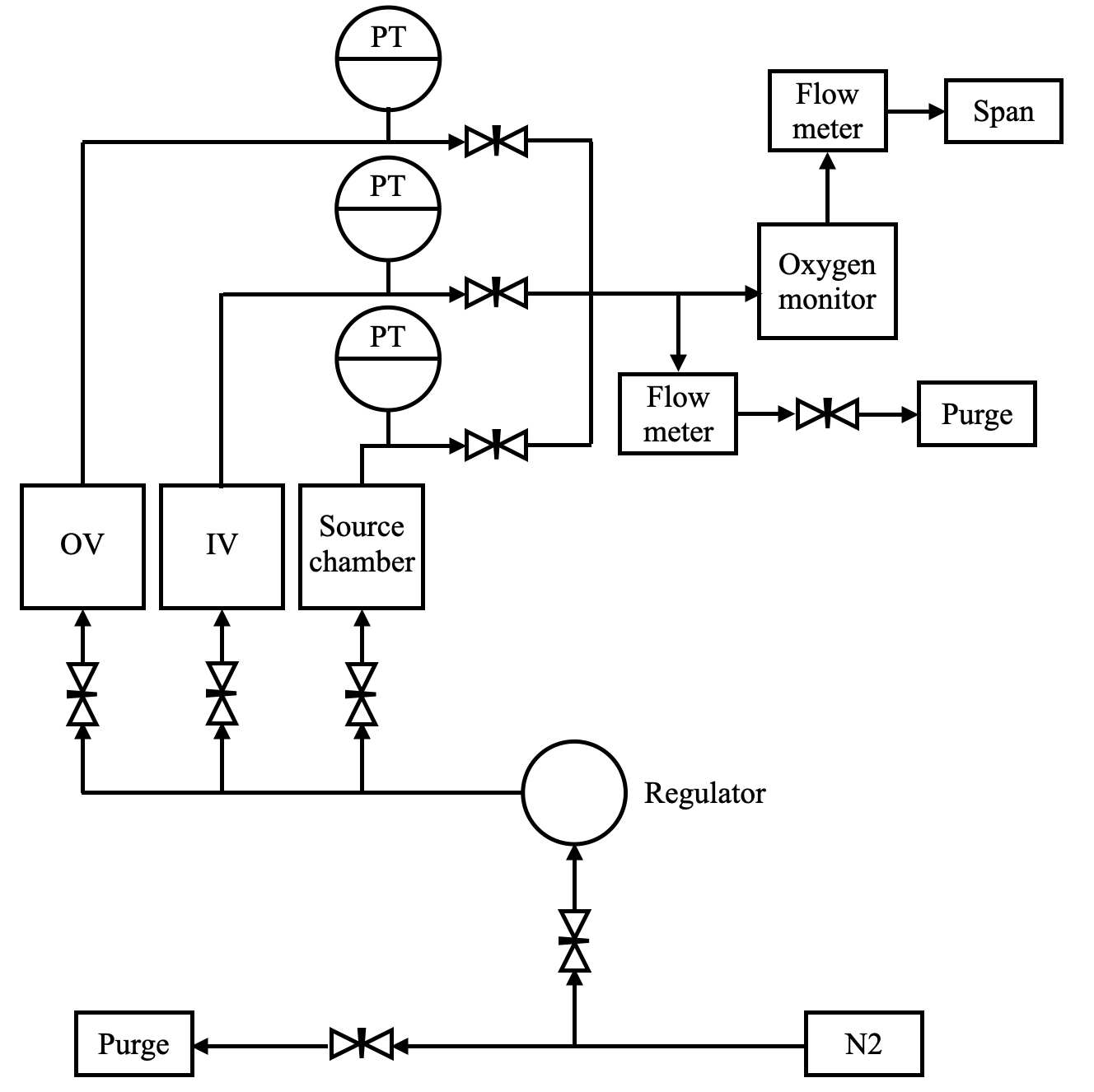}
    \includegraphics[width=0.53\textwidth,trim=0cm 0cm 3cm 0cm, clip]{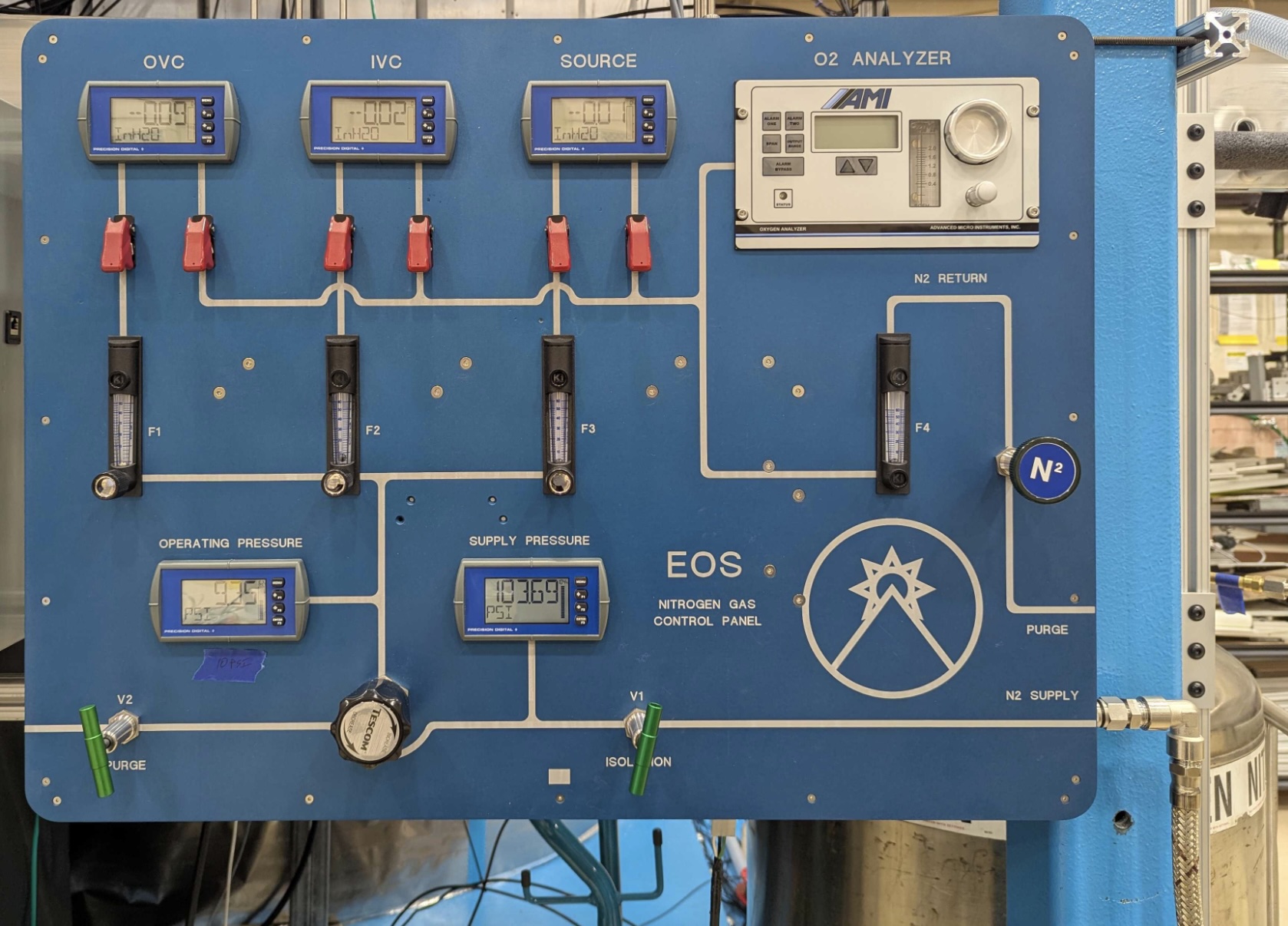}
    \end{center}
    \caption{Flow diagram (left) and control panel (right) for the N$_{2}$ cover gas system. This system allows the operating pressures, flow rates, and O$_{2}$ content to be controlled and monitored.}
    \label{f:covergas}
\end{figure}

The \eos\ cover gas system was designed with the following requirements:
\begin{itemize}
    \item Supply N$_2$ cover gas to prevent O$_2$ ingress into the liquid volumes.
    \item Set and maintain a constant pressure.
    \item Measure system O$_2$ level.
\end{itemize}
To achieve this, \eos\ uses an N$_2$ cover gas system. It supplies individualized flow to the OV, IV, and the calibration source chamber. $\sim$99.998\% pure N$_2$ is supplied from cryogenic boil-off from a 230\,L, 230\,psi Dewar. The purpose of this system is to minimize biological growth and prevent O$_2$ ingress into the detector volumes, both of which would diminish the optical properties of the detection media. Biological growth introduces chromophores scattering sites, and can cause surface fouling, while O$_2$ ingress causes oxidation of the scintillator material. The system O$_2$ concentration is monitored using an AMI 2001RSP Trace Oxygen Analyzer~\cite{ami}. N$_2$ is supplied through 6.4\,mm SS tubing. The pressure for each volume is monitored using Omega PX409 High Accuracy pressure transducers (PT)~\cite{omega} and displayed using Precision Digital PD6604-L2N loop-powered process control meters~\cite{precision_digital_meters}. The gas flow is controlled using Parker Gold Ring Series all-stainless 24\,VDC direct-acting solenoid valves~\cite{parker}. To prevent air ingress into the IV during calibration source deployment, the source chamber can be isolated from the rest of the system, opened to atmosphere, and separately purged, prior to opening the gate valve to the IV. A flow diagram and image of the system interface are shown in Fig.~\ref{f:covergas}.

\subsection{Muon detection} \label{s:muon}

The \eos\ muon veto system (MVS) is designed with the following requirements:
\begin{itemize}
    \item Detect high-energy cosmic rays, primarily muons.
    \item Achieve maximum detector coverage.
    \item Enable the collection of muon datasets for detector calibration and background studies.
\end{itemize}

\begin{figure}[!h]
    \begin{center}
    \includegraphics[width=0.53\textwidth, trim=0cm 1cm 8cm 5cm, clip]{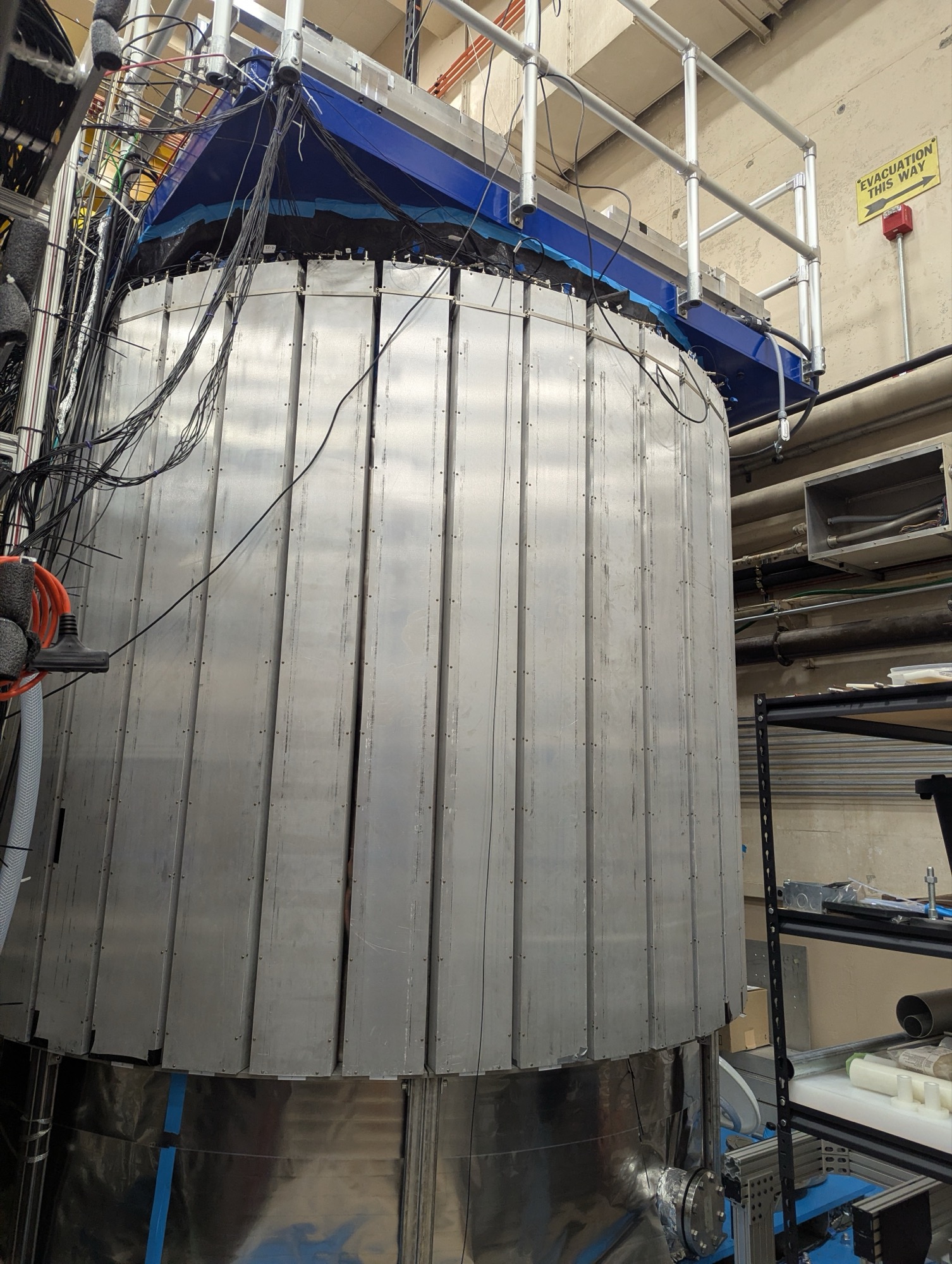}
    \end{center}
    \caption{Installed muon panels surrounding the barrel.}
    \label{f:muonc}
\end{figure}

\begin{figure}[!h]
    \begin{center}
    \includegraphics[width=0.7\textwidth]{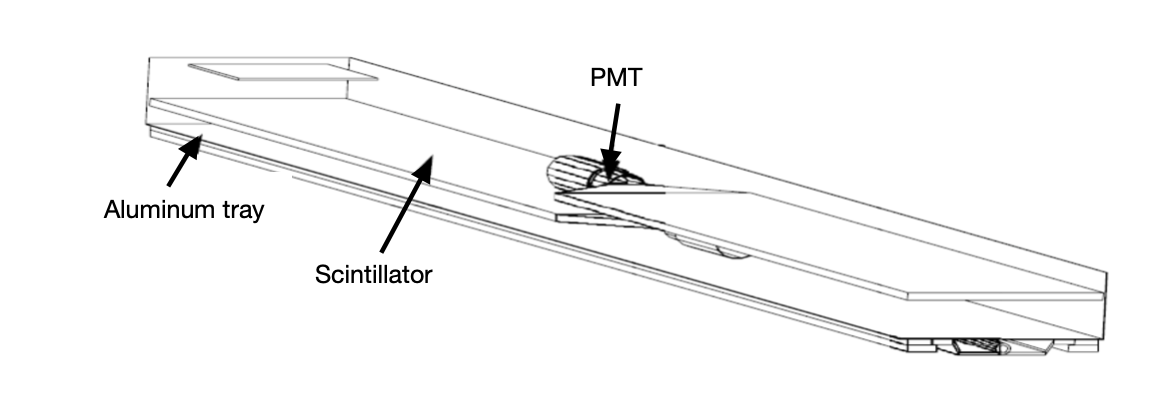}
    \end{center}
    \caption{The paddle and PMT configuration for each of the muon veto panels ~\cite{RHIC}.}
    \label{f:muonb}
\end{figure}

To meet these requirements, the MVS utilizes 68 repurposed scintillator panels from the Solenoidal Tracker at RHIC (STAR) trigger system~\cite{STAR} at the Relativistic Heavy Ion Collider (RHIC)~\cite{RHIC} at Brookhaven National Laboratory. The scintillator panels have dimensions of 241.6\,$\times$\,21.5\,$\times$\,8.7\,cm. Two layers of 10 panels are placed on top of the OV in a crossed configuration. 11 panels are placed within the seismic anchorage underneath, with the remaining 37 surrounding the barrel, as shown in Fig.~\ref{f:muonc}. Each muon panel consists of two paddles composed of BC-408 plastic scintillator, as shown in Fig.~\ref{f:muonb}. One paddle has dimensions of 1\,$\times$\,21\,$\times$\,112.5\,cm$^{3}$, while the other has dimensions 1\,$\times$\,21\,$\times$\,30\,cm$^{3}$. The scintillation light from each paddle is measured individually using Hamamatsu R5946 38\,mm PMTs, for a total of 136 channels. The PMT high voltage (HV) bias is supplied by two CAEN A7030SN boards with 48 HV channels, which are shared by two to three PMTs using Safe High Voltage tees. The data acquisition system (DAQ) is designed for the MVS to be able to veto, trigger, and record hits from charged particles passing through the paddles (see Section~\ref{s:Readout}).

\subsection{Triggering and readout}
\label{s:Readout}

The \eos\ triggering and readout system is designed with the following requirements:
\begin{itemize}
    \item Provide triggered digitization of waveforms from up to 255 PMT channels.
    \item Enable sub-ns timing precision for accurate event reconstruction.
    \item Support a broad dynamic range in signal amplitude.
\end{itemize}

\begin{figure}[!h]
    \begin{center}
    \includegraphics[width=1.0\textwidth,trim=0cm 2cm 0cm 0cm, clip]{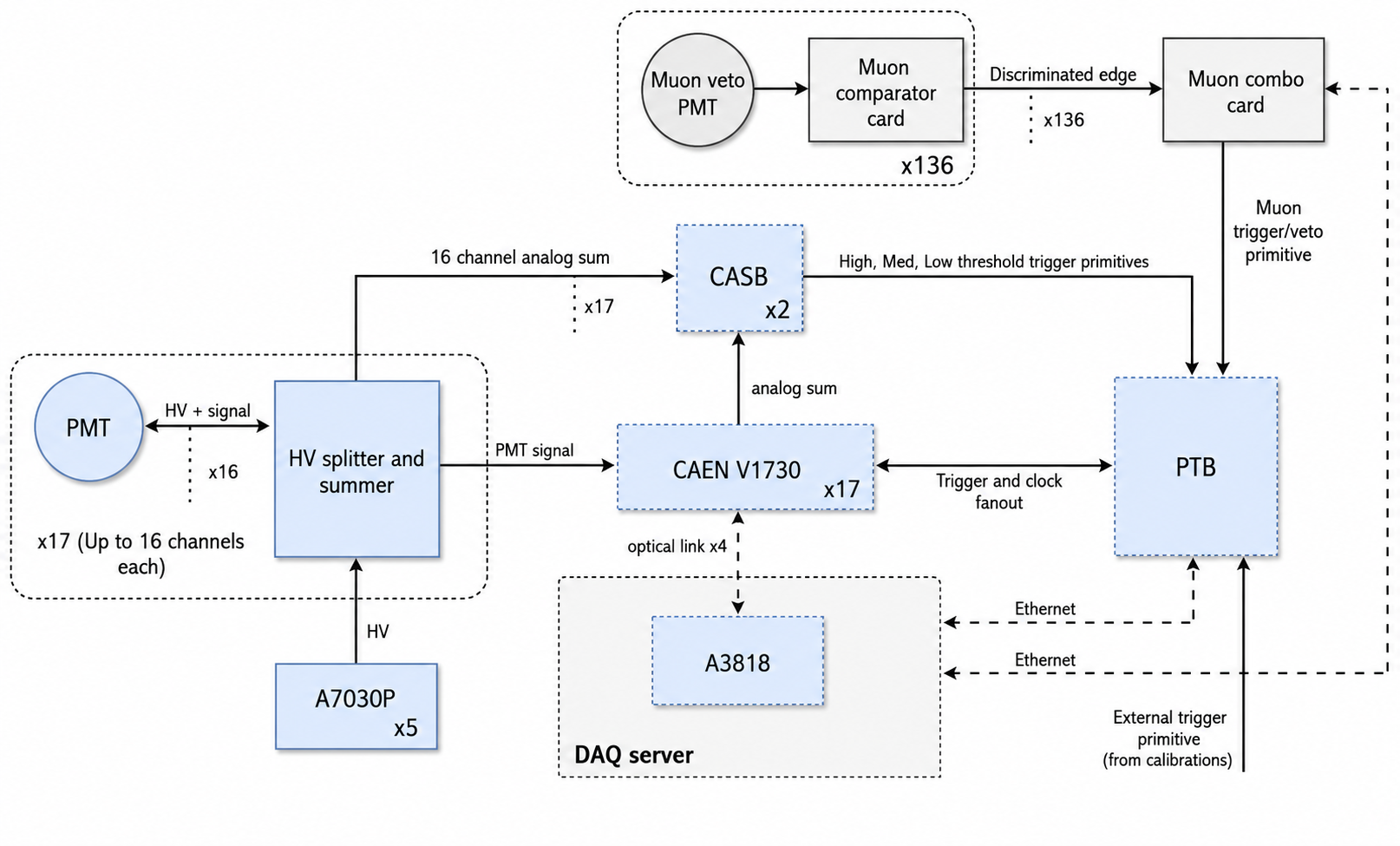}
    \end{center}
    \caption{Block diagram of \eos\  electronics including the high voltage, trigger, digitization, and data readout.}
    \label{f:electronics}
\end{figure}

 Accordingly, the trigger system is flexible and includes programmable logic to support a variety of trigger modes, including calibration sources, cosmic rays, geographic triggering, muon vetoing, and coincidence triggers, with trigger rates up to several kHz. The overall design of the readout and trigger electronics is shown in Fig.~\ref{f:electronics}. The DAQ software is based on the ToolDAQ framework~\cite{tooldaq} and WbLSDAQ~\cite{wblsdaq}.

The PMTs are powered by six CAEN A7030P HV power supply boards, each providing up to 3\,kV and 1\,mA per channel \cite{caenhv}. The HV to each channel is individually controllable and monitored. Each PMT dynode chain is connected by a single cable carrying both signal and positive high voltage to custom ``High Voltage Splitter and Summer'' (HVSS) boards. These boards distribute the high voltage from the A7030Ps and separate the PMT signal, which is then shaped and sent to the digitizers. In addition, the HVSS produces two summed outputs for the trigger: an analog sum of the PMT signals, and a sum of ``NHIT'' signals - fixed width and height square pulses produced for each channel that passes a programmable threshold.

The PMT waveforms are digitized at 500\,MSamples/s using 17 CAEN V1730s modules \cite{caendigitizers}. The pulse shape is then used to extract photon times with sub-ns precision. A common clock and trigger are distributed to all boards from the central trigger board described below. Fixed channel-to-channel timing offsets, arising from small differences in cable length and PMT transit time, are calibrated using a laserball calibration source (see Sec.~\ref{s:laserball}).
Each trigger initiates readout of a programmable number of waveform samples from every digitizer channel. The digitizers are read out in daisy chains over four optical fibers to a CAEN A3818 PCIe controller \cite{caenpci}. The V1730s also provide their own analog NHIT sum output for pairs of channels over a programmable threshold.

The analog sums from the HVSS boards and V1730s modules are sent to custom ``Central Analog Summing Boards'' (CASB). Each CASB receives all the analog sums of one type, performs a global sum, and compares against multiple programmable thresholds. Any threshold crossing produces a trigger primitive that is sent on to the ``Penn Trigger Board'' (PTB), where programmable logic determines whether to generate a system trigger for the digitizers and readout. The CASB features multiple gain paths and active baseline restoration.

The first version of the PTB was developed for the LBNE 35-ton demonstrator~\cite{LBNE}, and subsequent iterations have been used in the ProtoDUNE detector~\cite{DUNE:2020cqd} and the SBND experiment~\cite{doi:10.1146/annurev-nucl-101917-020949}. It uses a Xilinx Zynq-7020 SoC to implement various logical combinations of signals from the CASBs and/or external calibration systems. The PTB generates a ``global trigger'' that is distributed to initiate readout of the PMT waveforms. The programmable logic supports prescaling of input triggers, masking out around a muon veto signal, and low-threshold ``follower-triggers'' for coincidence signals within a time window after a higher-threshold trigger.

The muon veto system contains 136 PMTs, whose anode signals are routed to LM361-based muon comparator boards that convert the pulses into 300\,ns long 0-5\,V logic outputs. These outputs are converted from 0-5\,V to 0-1.8\,V logic levels using the muon combo card. The 1.8\,V logic signals serve as inputs to an Enclustra XU5+PE-3-100 system on module (SOM) \cite{enclustra}, which features a 148-channel input/output (I/O) field-programmable gate array (FPGA). 136 I/Os are dedicated as inputs from the combo card, and the FPGA computes two logic trigger/veto outputs:
``double" - generated when two hits occur in the top array for triggering or vetoing muon decays;
``triple" - generated when three hits occur, with two on the top array and one in either the barrel or bottom array, for triggering or vetoing muon pass-throughs.

The muon FPGA outputs are interfaced directly with the Penn Trigger Board (PTB) with a sub-100\,ns delay, enabling synchronous triggering or vetoing with the \eos\ PMTs. The muon FPGA can also receive one logic input from the PTB to trigger the readout of muon hits.

\subsection{Slow controls}

The \eos\ slow control system was designed with the following requirements:
\begin{itemize}
    \item Record continuous real-time measurements of the detector's environmental conditions.
    \item Monitor the OV water temperature at multiple locations within the volume.
    \item Measure the OV water level.
    \item Provide additional leak detection capabilities.
    \item Enable alerts, alarms, and long-term tracking for detector operations. 
\end{itemize}

To fulfill these requirements, the system includes 15 temperature probes affixed to the barrel PMT support structure within the tank, six wet/dry sensors
located near the base of the tank and liquid handling systems, and two
ultrasonic level sensors mounted to the underside of the tank lid.

\begin{figure}
    \centering
    \includegraphics[width=1.0\linewidth]{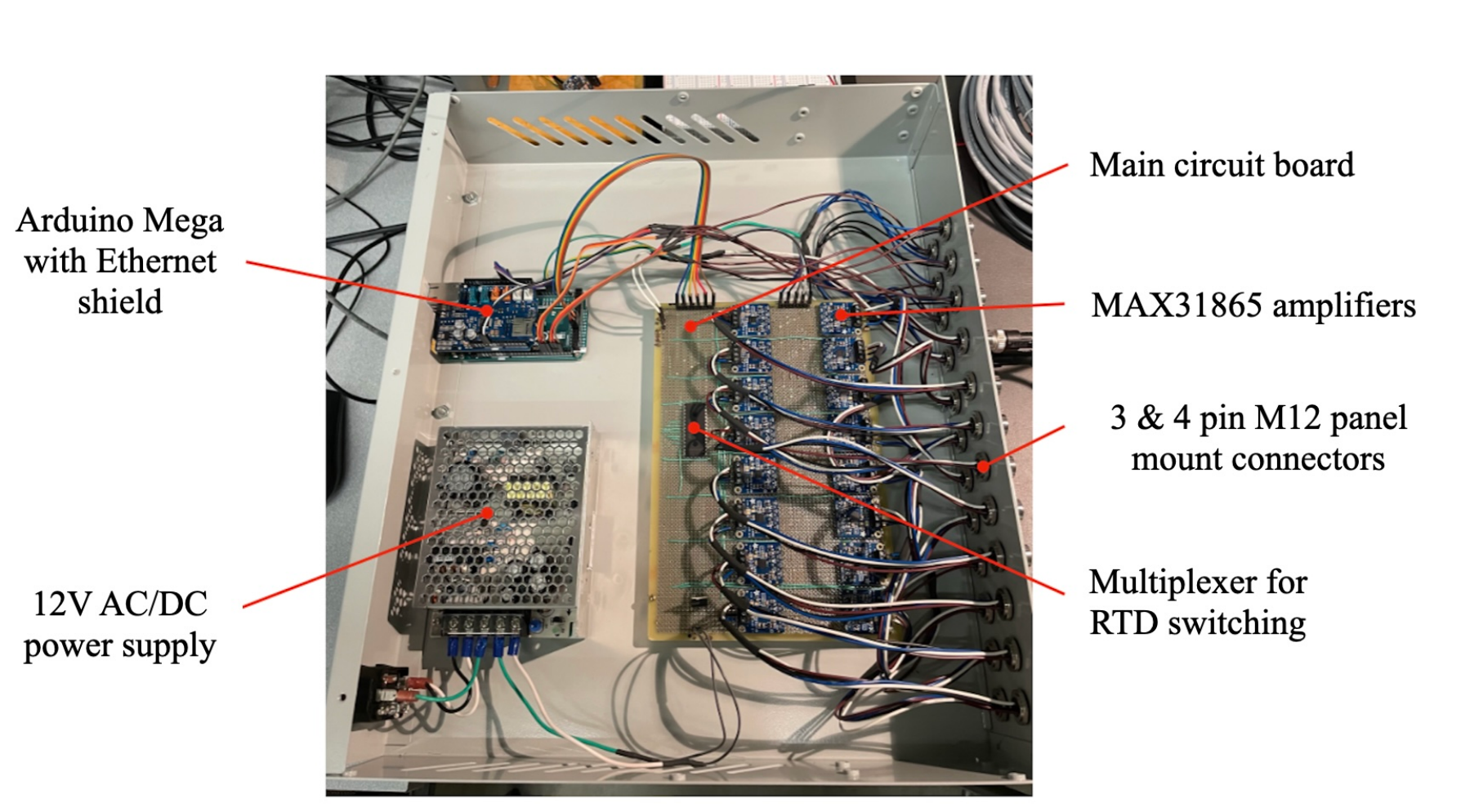}
    \caption{The \eos\ slow controls box, with major components annotated. Sensor inputs
    are received via M12 connectors on the front panel (right) and data is transmitted out via an Ethernet connection to the Arduino at the back (left).}
    \label{fig:eos-sc-box}
\end{figure}

Sensors are connected via M12-terminated waterproof cables to a central slow
controls (SC) box in the \eos\ central DAQ rack, where signals are received by a
custom circuit board. The SC box is shown in Fig.~\ref{fig:eos-sc-box}. The temperature sensors are Omega PR-22-3-100-A-M3 PT100 RTD probes with molded M12
connectors~\cite{pt100}. 
The wet/dry sensors are Adafruit 4965 Water Detection Sensors (WDS)~\cite{wds4965}, which provide a digital signal (high or low) based on the
conductivity across an array of strip lines connected to a transistor biased near saturation. The level sensors are Model \#A02YYUW SEN0311 Ultrasonic Level Sensors (ULS)~\cite{sen0311}, featuring an integrated IP67 waterproof package with UART data readout.

Sensor readout is performed using an Arduino Mega 2560~\cite{arduino2560}. All devices are powered by a TDK Lambda LS75-12 AC/DC power supply providing 12 VDC~\cite{ls75}.

Software running on the Arduino cycles through all sensors with approximately a
1\,Hz frequency, recording measurements that are sent to a data-logging server and stored in a Redis database~\cite{redis}. The server also automatically pushes warnings to a monitoring channel when a water sensor is triggered. The Redis database computes a series of rolling time averages. These metrics are available through a database for data analysis and real-time monitoring through a web application interface.

\subsection{Data flow and monitoring}
\label{sec:data-flow}

The \eos\ data flow and monitoring software was designed with the following requirements:
\begin{itemize}
    \item Produce event-level data from the various \eos\ subsystem fragments.
    \item Index and store these data to disk.
\end{itemize}

Event-level detector data are produced by various \eos\ subsystems (waveform
digitizers, trigger system, muon tagging system, etc.) and aggregated into
complete detector events by matching these fragments using precision
timestamps. This matching is performed by a central Event Builder (EVB) software
instance running on a local data storage server. The EVB listens for incoming connections from detector subsystems. Upon establishing a
connection, the EVB creates a new thread that listens for incoming data packets and applies a packet-type-specific decoding function to extract event data. The EVB
maintains an internal {\tt uthash}~\cite{uthash} hash table indexed by
timestamp and aggregates subsystem data fragments into event objects stored
within it. Once a table entry is complete (i.e., all expected data
components have been received), it is flushed to output. The output is written to disk
in a binary format and transmitted via a TCP network socket. Binary files are
closed when a run terminates or the file size reaches a user-specified limit, at
which time they are converted to a portable HDF5~\cite{doecode_9801} format. The
network connection allows the EVB to dispatch fully constructed events to
downstream monitoring clients.

Live monitoring of data collected by the DAQ is achieved using a monitoring
script that receives and processes event-level data from the EVB via a TCP network
connection. This tool interprets trigger system data to compute the average trigger rates
per trigger type and reduces the full PMT waveform data to a small set of metrics:
the baseline and RMS for each channel, threshold crossing times relative to the trigger,
integrated charge, and channel occupancy (i.e., the fraction of events in which a given
channel has a threshold crossing). These quantities are computed for each waveform
digitizer channel and stored in a persistent Redis database. The monitoring script
is implemented in Python, and to keep up with real-time processing of the
full waveform data at an event rate of a few kHz, performance-critical operations
such as waveform feature extraction are delegated to C code using the Python
{\tt ctypes}~\cite{ctypes} interface.

Real-time monitoring of the \eos\ detector and data acquisition systems is
performed using a custom web-based application. This system is based on the {\tt
minard} application developed by the SNO+ Collaboration
\cite{SNO:2021xpa, minard} and is therefore dubbed {\tt eos-minard}. An
interactive monitoring website provides an interface to run-specific detector
conditions and displays key detector status and performance metrics as
live-updating strip charts, histograms, and in the case of PMT channel-specific
data, visual representations in both the digitizer card/channel space and flat-map projections of the 3D PMT positions.

The monitoring website is built using Python and the Flask WSGI framework~\cite{flask}, and the back-end code can query arbitrary data sources to construct static and live-updating web pages. For live-updating pages, the data are sourced from a central Redis database, where slow-monitoring and aggregated event data are stored as rolling time averages.

A main page displays the history of trigger rates for each trigger type over a
user-specified time range.
This illustrates the stability of different trigger sources and how each
trigger type contributes to the overall detector trigger rate.
A DAQ status page shows a variety of user-selectable, channel-specific metrics,
including the baseline and RMS for each PMT channel, average hit times relative
to the trigger and total charges above threshold, and channel occupancy. These
are displayed in both a ``crate'' view representing the physical layout of
digitizer channels in the readout crate, and in a flat map projection of the PMT
positions. This aids, for example, in identifying whether any unexpected
behavior is clustered in the physical space of the PMTs (e.g., indicating a light
source) or electronics (e.g., indicating a cross-talk effect).
The environmental monitoring page displays real-time data from water
temperature, water level, and leak-detection sensors, with selectable time ranges and histogramming capabilities.

Finally, the monitoring website also includes an interface to the detector and
run-conditions relational databases, providing access to run lists with the state of the detector and DAQ configuration used during data collection, channel threshold settings, and the status and real-time measurements of the
supply voltage and currents from the PMT array high-voltage supply system.

\subsection{Calibration deployment mechanism}

The \eos\  calibration deployment mechanism was designed with the following requirements:
\begin{itemize}
    \item Allow for central-axis deployment inside the IV.
    \item Enable  interchangeable deployment of a suite of sources, as described in Section~\ref{s:Calibration_program}.
    \item Minimize optical shadowing.
    \item Achieve a vertical-axis precision of $\pm $2\,mm.
    \item Traverse the entire vertical range of the IV.
    \item Provide the ability to rotate the source orientation through 360$^\circ$ with a precision of 2$^\circ$.
    \item Maintain a nitrogen-clean environment for source deployment.
    \item Remain light-tight during operation.
\end{itemize}

\begin{figure*}[h!]
    \centering
    \includegraphics[scale=0.45]{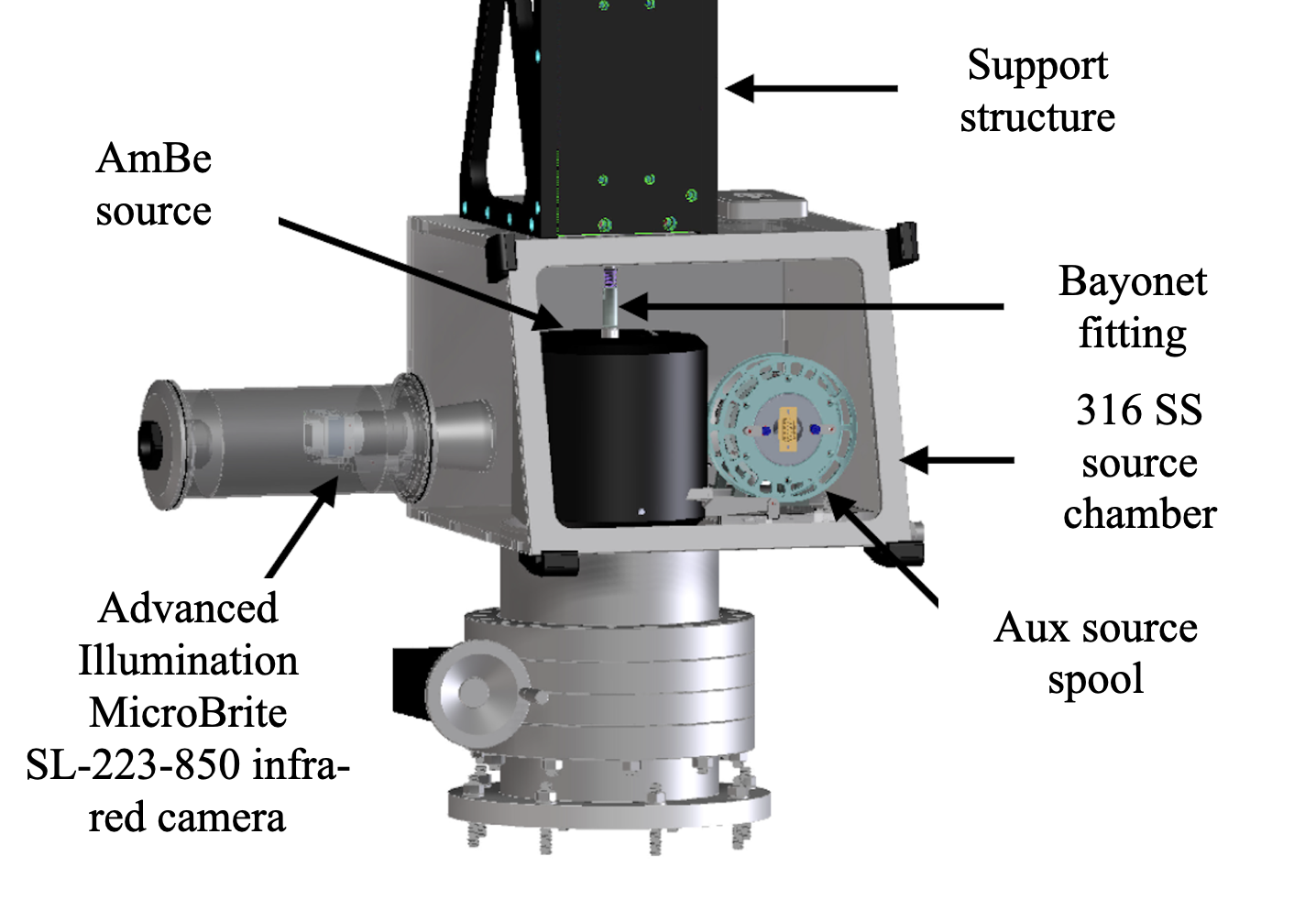}
    \caption{Rendering of the \eos\  calibration source chamber, highlighting the main components. The calibration-rod support structure is bolted to the top of the chamber. The Advanced Illumination MicroBrite camera is mounted on the side. Inside is a spool to allow for sources with aux cables to be deployed. Here, the AmBe source is attached to the calibration rod using a bayonet connector. Note, the front panel to seal the chamber is not shown.}
    \label{fig:sourcechamber}
\end{figure*}

To achieve these criteria, a hermetically sealed source exchange chamber was designed and installed above the detector lid, as shown in Fig.~\ref{fig:sourcechamber}.
Both radioactive and light-emitting sources are attached to a 7.9\,mm diameter 316 SS tube using a bayonet connection, ensuring repeatable source positioning. Installed through the center of the tube is a single-mode Thorlabs S405-XP optical fiber allowing for optical calibration, as discussed in Sec~\ref{s:laserball}. To adjust the vertical location of each source within the detector volume, the rod is controlled using a motor trolley assembly. This assembly employs a linear carriage bearing mounted on a 15\,mm wide linear slide rail and is driven by an Anaheim Automation 23MDSI integrated stepper motor. Travel stop switches are installed at the top of the support structure to prevent the rod from exceeding the vertical design limits of the system.

\begin{figure*}[h!]
    \centering
    \includegraphics[width=0.75\linewidth,trim=2cm 0.4cm 1cm 0cm, clip]{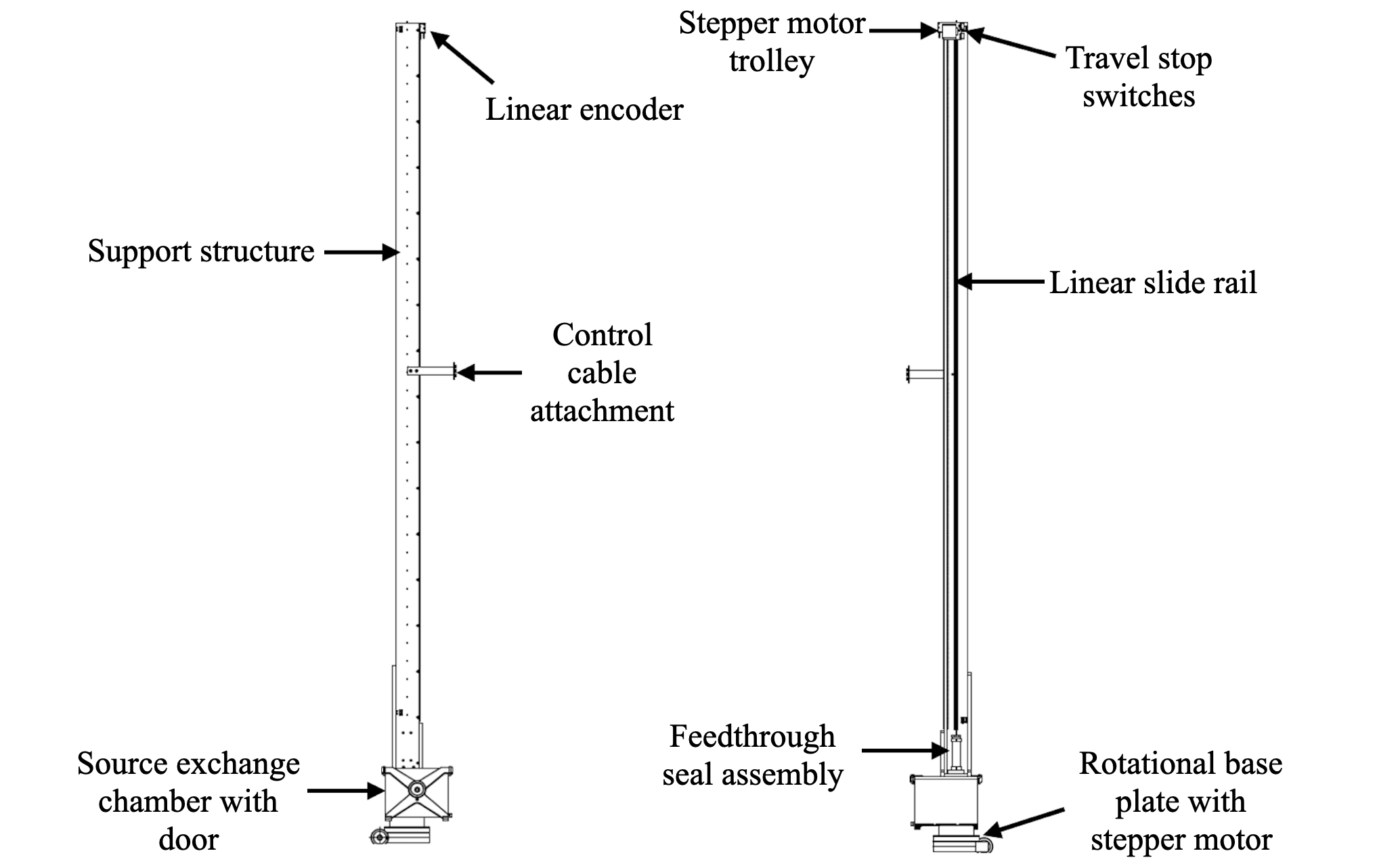}
    \includegraphics[width=0.24\linewidth, trim=9cm 1cm 11.4cm 1cm, clip]{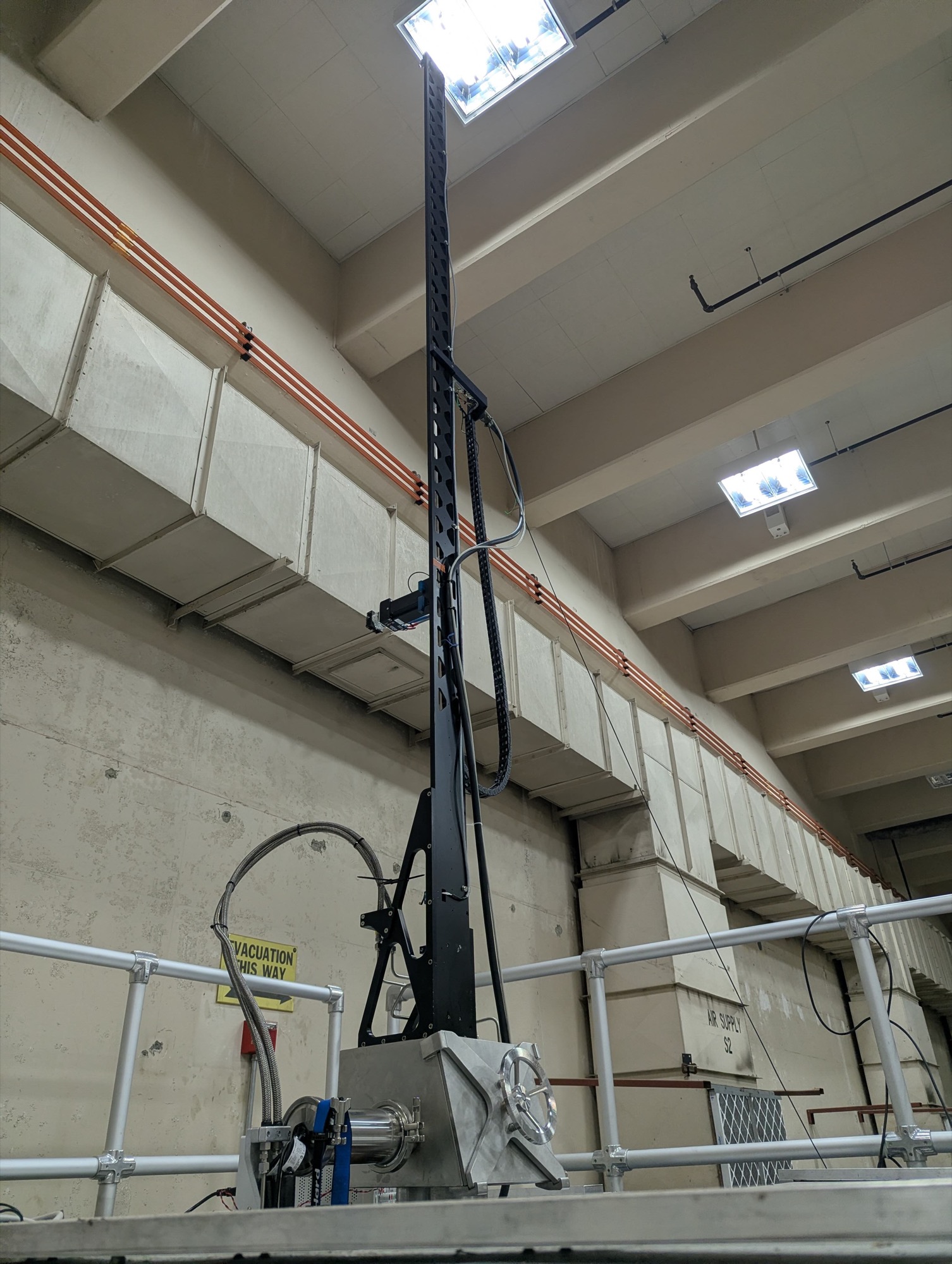}
    \caption{(Left) Front and (center) rear schematic views of the calibration tower with the main components labeled. (Right) Photograph of the deployed calibration tower and source chamber.}
    \label{fig:calibrationtower}
\end{figure*}

The support structure is bolted to the SS source exchange chamber, as shown in Fig.~\ref{fig:calibrationtower}. To allow attachment of sources to the deployment rod, the front panel is removable. Above the source chamber, a feed-through seal assembly is installed. This assembly uses two lip seals to prevent light leaks and air ingress into the chamber, while providing additional support and allowing the rod to move freely through the source chamber. An Advanced Illumination MicroBrite SL-223-850 infrared camera is mounted on the left side of the chamber. This allows viewing of the internal volume when closed as a safety measure to monitor internal components during operation. Inside the chamber, an auxiliary source spool can be attached. This serves as both the automatic cable management system and power connection during deployment of the directional sources, described in Section~\ref{s:Calibration_program}.

The source chamber and calibration tower system are attached to a rotational flange powered by a Thermionics RNN-600 stepper motor, allowing for 360$^\circ$ azimuthal rotation. The entire setup is mounted on top of a gas-actuated gate valve, allowing the source chamber to be isolated when the gate valve is closed and enabling the calibration rod to pass when opened. The rotation, vertical position, and gate valve are controlled using a custom control box.

\label{s:Calibration_deployment_mechanism}
\section{Calibration program}
\label{s:Calibration_program}

The \eos\ calibration program was designed to achieve the following:
\begin{itemize}
    \item PMT timing and charge calibrations.
    \item Measure the light yield of scintillating materials.
    \item Evaluate position and direction reconstruction.
    \item Characterize the optical properties of deployed media.
    \item Determine neutron capture efficiency.
    \item Explore particle identification capabilities.
    \item Calibrate the detector energy scale.
\end{itemize}

Several radioactive and optical sources are used to calibrate detector components and characterize event-reconstruction capabilities, with the development of additional sources ongoing throughout the operation of \eos. A light injection system (LIS) supplies four distinct wavelengths of laser light with tens-of-picoseconds pulse widths to cm-scale light-diffusing balls deployed along the central axis, as well as to optical fibers mounted on the PSUP.
The diffusers are used to calibrate the time response of each PMT channel, while the fibers, which point at PMTs on the opposite side of the detector volume, are designed to enable wavelength-dependent studies of photon scattering and absorption in the target liquid. The calibration deployment system is also used to deploy radioactive sources to study position, direction, and energy reconstruction capabilities. Additionally, they enable calibration of detector efficiency. The primary calibration sources and their physics objectives are summarized in
Table~\ref{tab:calibration_sources}, with detailed descriptions provided in the following sections. 

\begin{table}[htbp]
\centering
\caption{Calibration sources deployed in the Eos detector and their primary physics objectives.}
\label{tab:calibration_sources}

\begin{tabular}{l l p{7cm}}
\toprule
Source & Type & Primary physics use \\
\midrule

Laserball & Optical &
Monitoring of optics, optical calibrations, PMT timing and charge calibration \\

Barrel fibers & Optical &
Monitoring optics of material in internal volume \\

Thorium & Gamma source &
Position reconstruction and energy calibration \\

AmBe & Neutron and gamma source &
Direction reconstruction, energy calibration, and neutron capture detection efficiency \\

PuBe & External gamma source &
Direction reconstruction tests \\

$^{137}$Cs & Gamma source &
Position reconstruction, energy calibration, and scintillation time profile measurement \\

Directional & Beta source &
Direction reconstruction \\

Cherenkov & Beta source &
Optical modeling of pure Cherenkov light \\

\bottomrule
\end{tabular}

\end{table}

\subsection{Light injection system}
The goals of this system are to measure PMT channel timing and charge response using isotropically emitted photons at low intensity such that most PMTs detect either no light or a single photon, and to study optical attenuation in the target liquid. The light sources are four NKT Photonics pulsed diode lasers (374, 408, 442 and 510\,nm) with a 19\,ps full width at half-maximum (FWHM)~\cite{NKT}. 

\subsubsection{Laserball}
\label{s:laserball}
\begin{figure*}[h!]
    \centering
    \includegraphics[angle=90, width=0.59\textwidth, trim=0cm 0cm 0cm 0cm, clip]{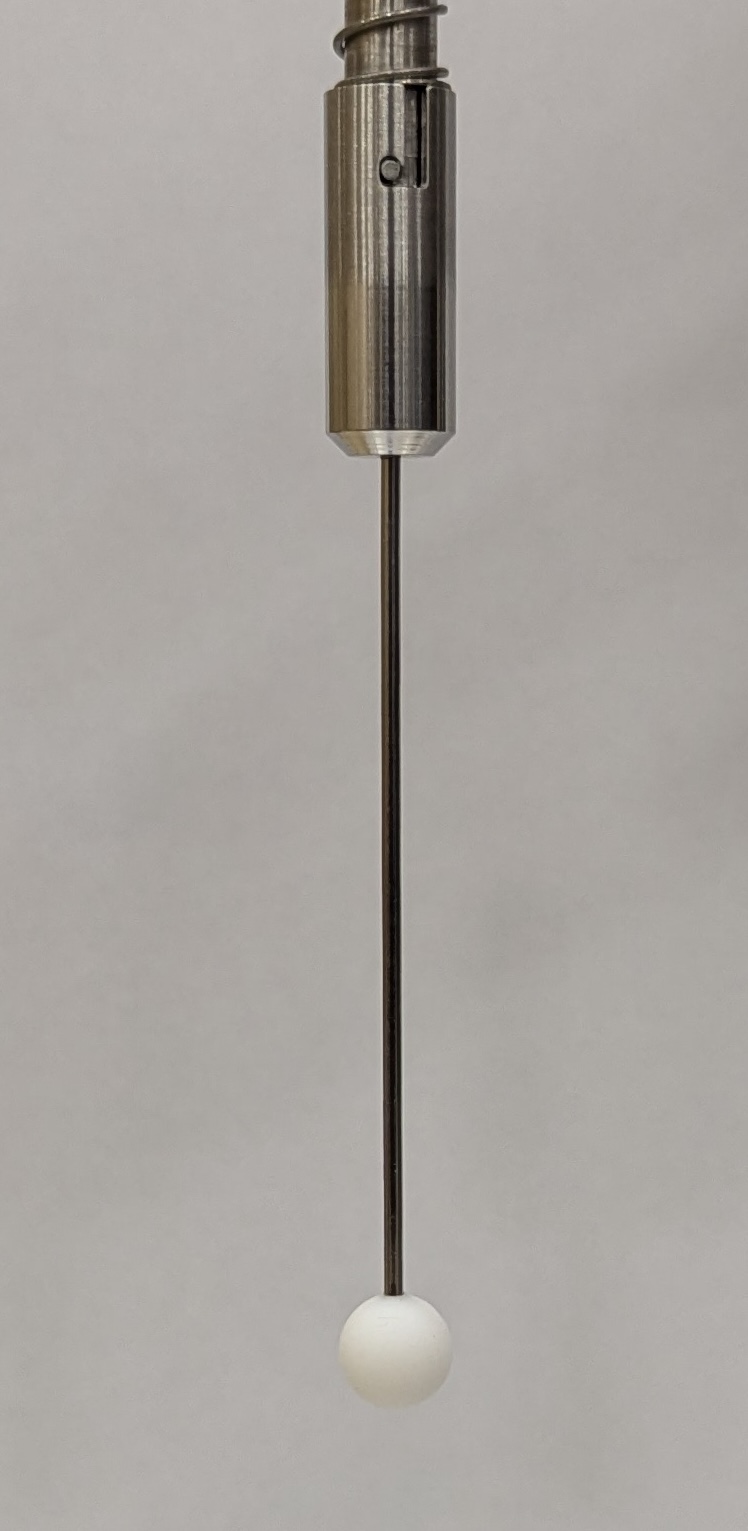}
    \caption{The 9.5\,mm diameter Teflon diffuser ball used to emit near-isotropic light for PMT calibration.}
    \label{fig:laserball}
\end{figure*}

To produce near-isotropic photon emission, a Teflon diffusing ball is deployed at multiple locations along the central axis (see Fig.~\ref{fig:laserball}). Light is supplied via the single-mode fiber running through the deployment rod. Five different diameter Teflon diffusing balls can be deployed (with diameters ranging from 4.8 - 12.7\,mm). This system enables the measurement of PMT timing and charge parameters and is discussed in more detail in~\cite{EosWater}.

\subsubsection{Barrel fibers}
\begin{figure}[!h]
    \begin{center}
    \includegraphics[width=1.0\textwidth, trim=0cm 0cm 0cm 0cm, clip]{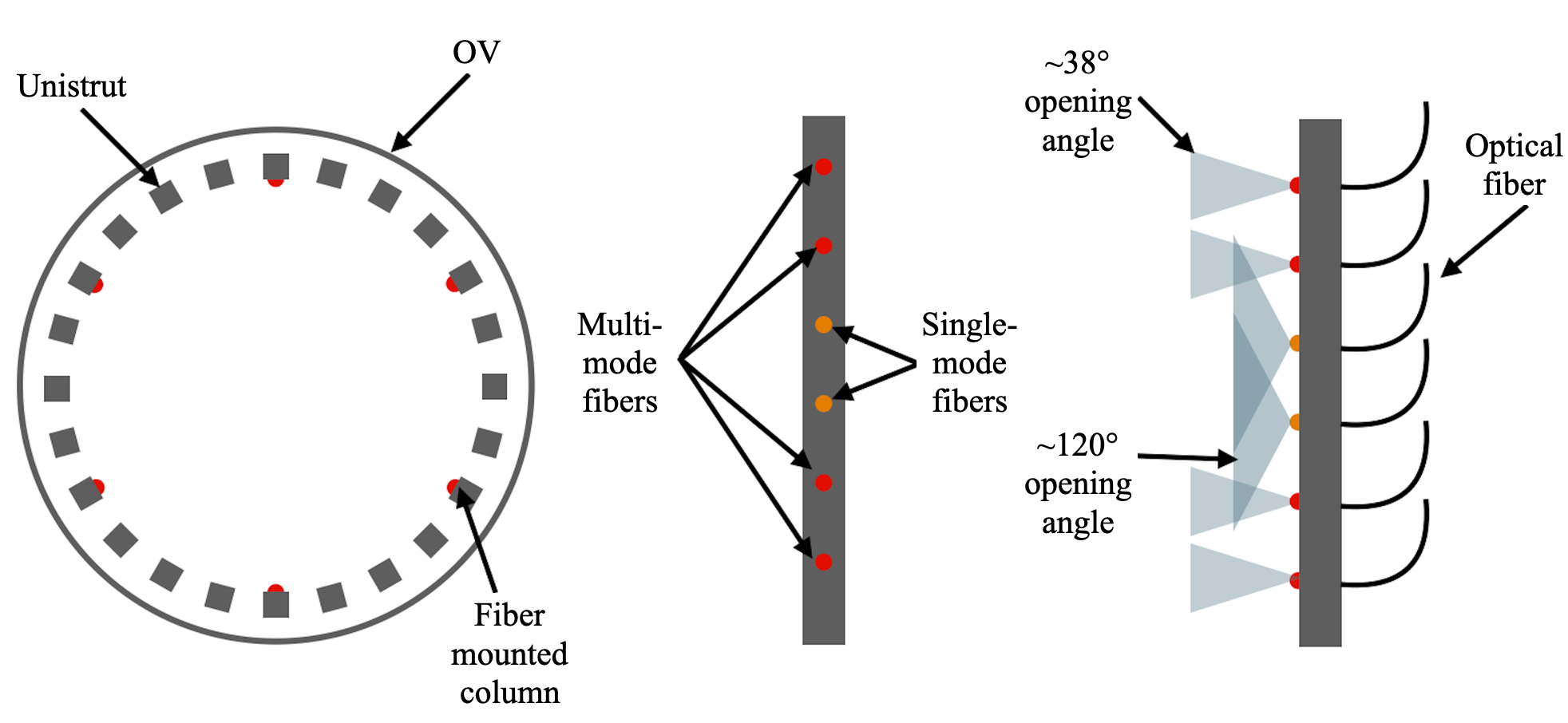}
    \end{center}
    \caption{(Left) Top view showing the barrel fiber arrangement at 60$^{\circ}$ angles around the detector. (Center) Front view of an individual fiber-mounted column, highlighting the locations of the single- and multi-mode fibers. (Right) Profile of an individual column, demonstrating the optical fiber opening angles.}
    \label{f:barrel_fibers}
\end{figure}

A total of 36 barrel-mounted fibers are deployed around the perimeter of the PSUP in six columns separated by 60$^{\circ}$, as shown in Fig.~\ref{f:barrel_fibers}. In each column, two central diffusers are fed by single-mode fibers and have approximately 120$^\circ$ opening angles, with the intention of inducing single photoelectrons to measure PMT time and charge responses without the need to deploy a source within the IV.  
The remaining four diffusers are fed by multi-mode fibers and have approximately 38$^{\circ}$ opening angles, with the intention of studying attenuation in the target liquid. All four diffusers are aimed at the same group of PMTs, centered between the second and third of the seven barrel PMT rows.

\subsection{Thorium sources}
\begin{figure}[!h]
    \begin{center}
    \includegraphics[width=0.71\textwidth, trim=0cm 0cm 0cm 0cm, clip]{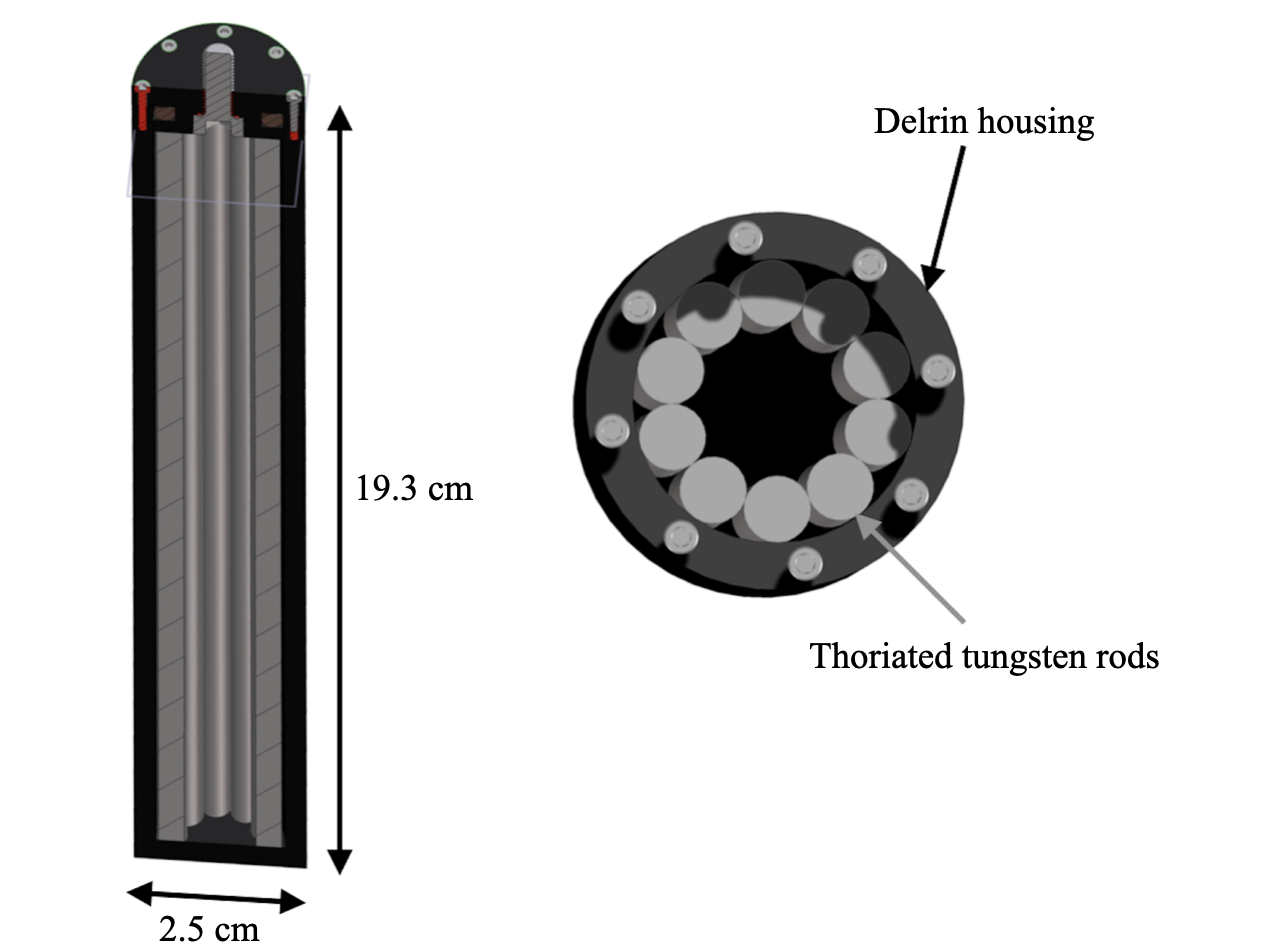}
    \includegraphics[width=0.28\textwidth, trim=6.4cm 0cm 0cm 0cm, clip]{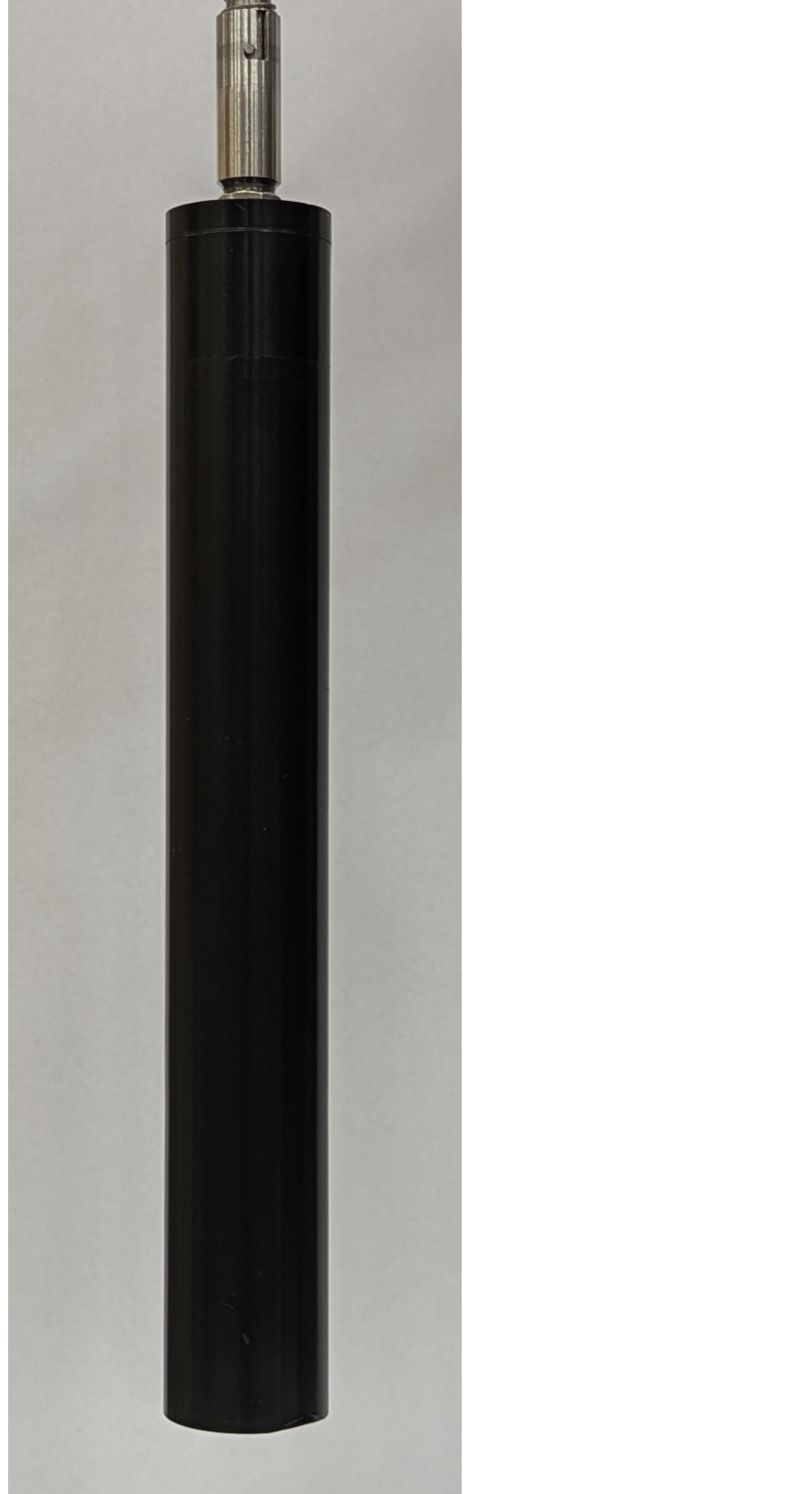}
    \end{center}
    \caption{(Left) Profile-view rendering of the thorium source showing the housing dimensions. (Center) Cross-sectional view illustrating the arrangement of the 10 thoriated tungsten rods within the delrin housing. (Right) Photograph of the assembled source.}
    \label{f:thorium}
\end{figure}

\begin{figure}[!h]
    \begin{center}
    \includegraphics[width=0.47\textwidth, trim=3cm 0.6cm 1cm 0cm, clip]{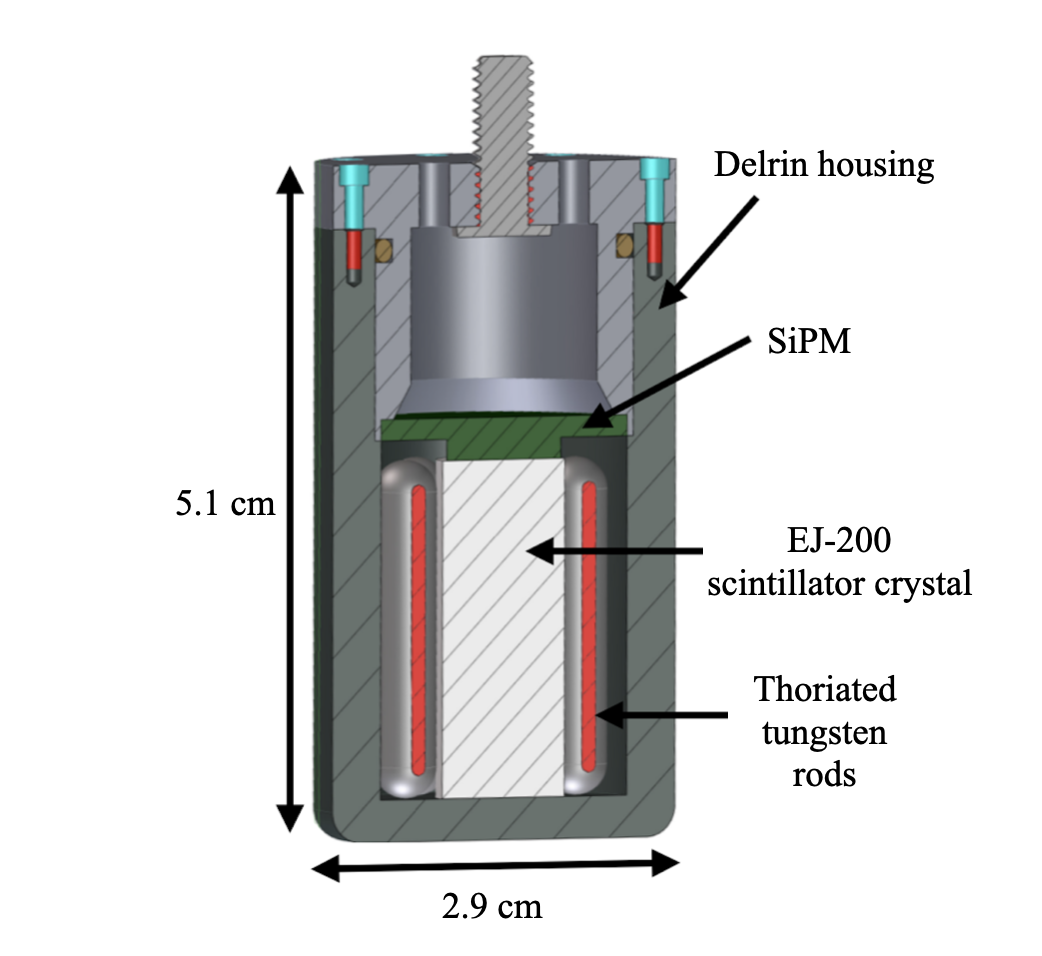}
    \includegraphics[width=0.34\textwidth, trim=1cm 0cm 1.5cm 0cm, clip]{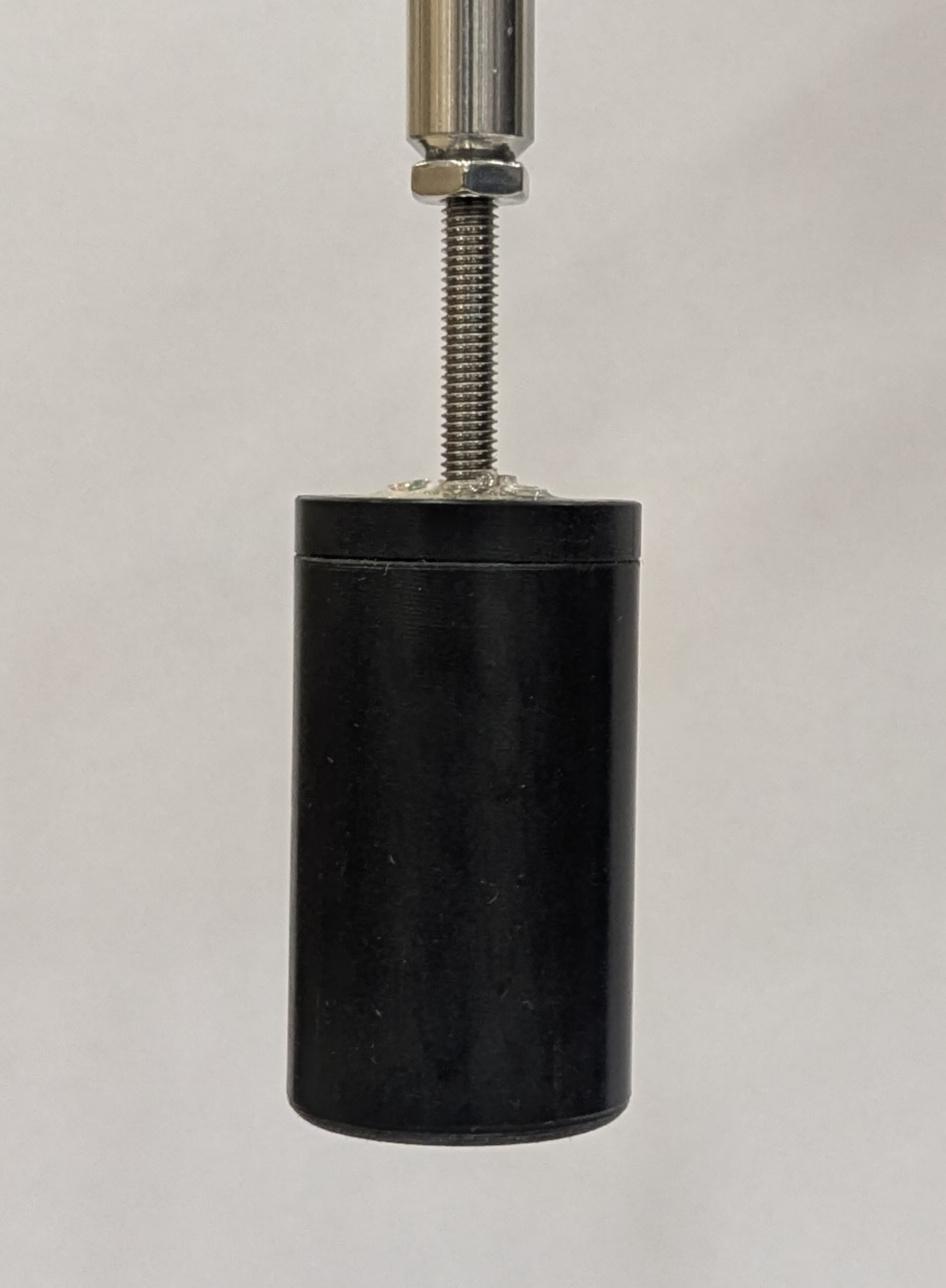}
    \end{center}
    \caption{(Left) Diagram of the internal components of the tagged thorium source, highlighting the placement of the tungsten rods, SiPM, and scintillator crystal. (Right) Photograph of the source.}
    \label{f:tagged_thorium}
\end{figure}

The goals of the thorium sources are to study position and direction reconstruction, as well as the energy scale, using electrons Compton-scattered by the predominantly 2.6\,MeV gammas from the beta decay of $^{208}$Tl in the thorium chain.  
Two sources have been constructed. The first, shown in Fig.~\ref{f:thorium}, is 19.3\,cm tall and 2.5\,cm in diameter. The outer casing is constructed from black delrin and houses a ring of ten 18\,cm tall, 4\,mm diameter, 4\% thoriated tungsten welding rods. The second source is a tagged thorium source shown in Fig.~\ref{f:tagged_thorium}, and uses a black delrin case to house a ring of ten 25.4\,mm long, 4\,mm diameter, 2\% thoriated tungsten rods. The rods surround a 25.4\,mm tall, 10.2\,mm diameter Eljen EJ-200 scintillator crystal. Above this crystal, a Hamamatsu S14160-6050HS silicon photomultiplier (SiPM) is mounted on a PCB to provide power and readout capability. This setup allows for the detection of betas ($\leq$ 2.4\,MeV) in the crystal, while the 2.6\,MeV gamma deposits its energy in the detector. This source is designed as an upgrade to the first, providing a purer dataset and a more point-like gamma emission location.

\subsection{AmBe source}
\begin{figure}[!h]
    \begin{center}
    \includegraphics[width=0.59\textwidth]{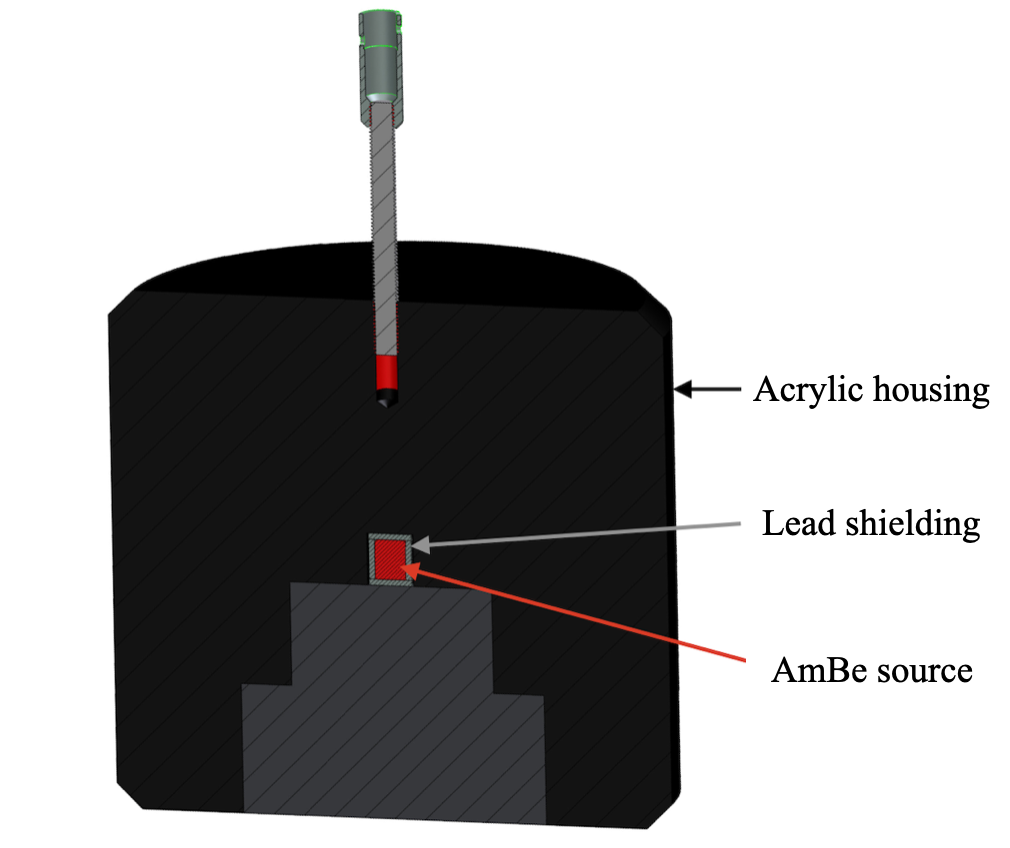}
    \includegraphics[width=0.40\textwidth,trim= 6.4cm 11cm 7cm 9.4cm, clip]{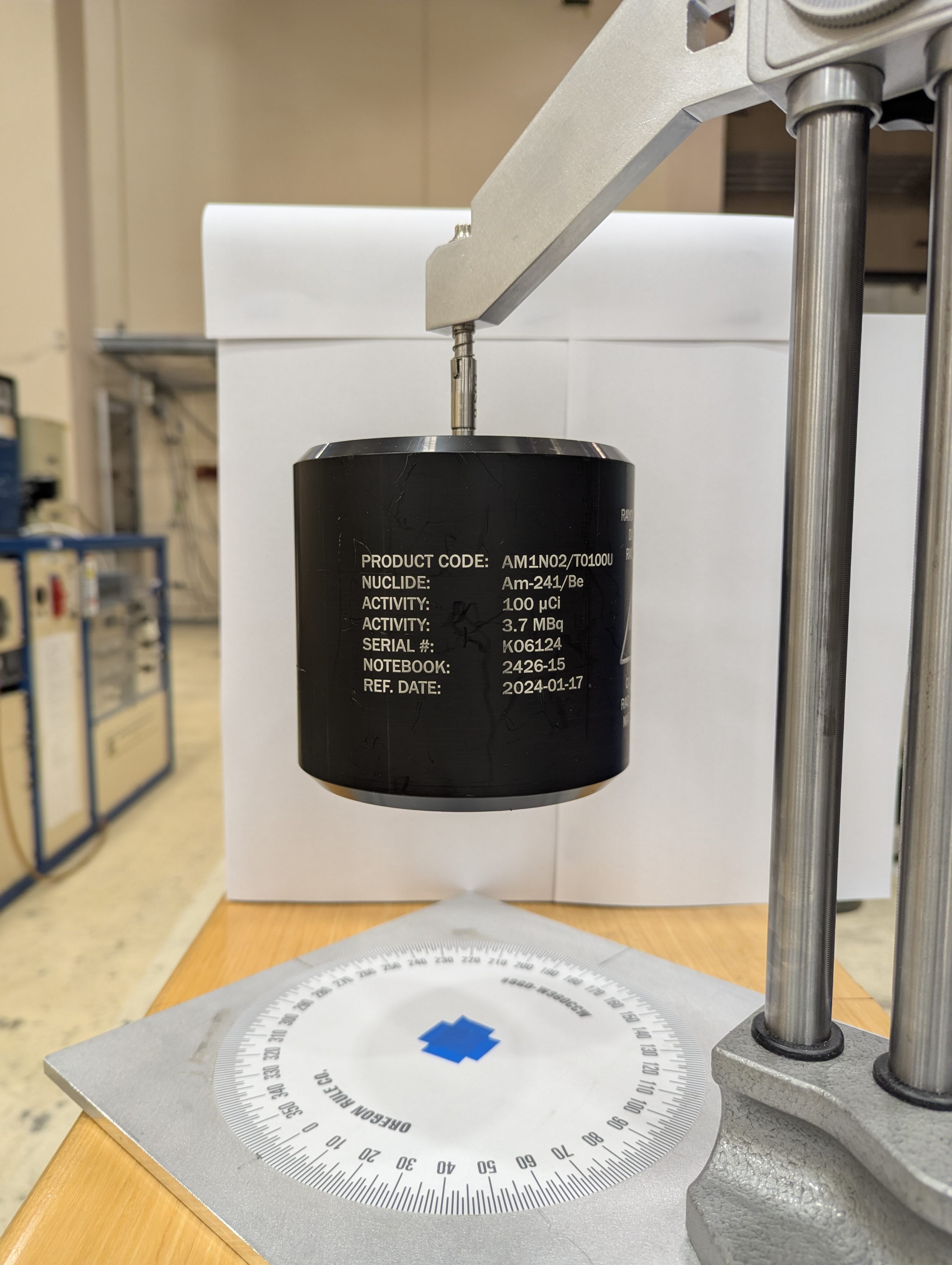}
    \end{center}
    \caption{(Left) Rendering of the AmBe calibration source showing the layered design to attenuate X-ray emission through lead shielding and thermalize emitted neutrons using the acrylic housing. (Right) Photograph of the assembled source housing.}
    \label{f:ambe}
\end{figure}

The AmBe calibration source consists of a 3.7\,MBq active element purchased from Eckert \& Ziegler~\cite{ezag}. The $^{241}$Am decays as follows:
\begin{equation}
    ^{241}\text{Am} \rightarrow{} ^{239}\text{Np} + \alpha + \gamma.
\end{equation}
This can result in the following $\alpha$ capture on $^{9}$Be:
\begin{equation}
\alpha + ^{9}\text{Be} \rightarrow{} ^{12}\text{C}^{*} + \text{n}
\end{equation}
which, for this source, produces a $\sim$880\,Hz neutron fluence. The $^{12}$C$^{*}$ then emits a 4.4\,MeV $\gamma$ with a $\sim$60\% branching ratio:
\begin{equation}
^{12}\text{C}^{*} \rightarrow{} ^{12}\text{C} + \gamma.
\end{equation}
The source is surrounded by 1.5\,mm thick lead shielding, itself encapsulated within a 14\,cm diameter, 13\,cm tall, right-circular cylindrical black acrylic housing, as shown in Fig.~\ref{f:ambe}. The lead shielding provides a high-Z material to attenuate low-energy X-rays emanating from the de-excitation of $^{237}$Np. The acrylic thickness was selected to maximize the volume that would fit through the IV neck. This low-Z material was chosen for its compatibility and ability to thermalize emitted neutrons prior to reaching the target medium, thus minimizing hadronic recoils. Specifically, black acrylic was selected to maximize light absorption and thus minimize the effect of scattered light coincident on the source housing from reaching the PMTs.

The goals of the AmBe source are to study direction reconstruction of the isotropic 4.4\,MeV gammas, provide a high-energy data point for the energy scale, and allow for the measurement of the neutron capture detection efficiency. Furthermore, there is scope, with a reduced acrylic volume, to measure the scintillator emission time for protons using neutron-proton recoils.

\subsection{PuBe source}
\begin{figure*}[h!]
    \centering
    \includegraphics[width=0.65\textwidth]{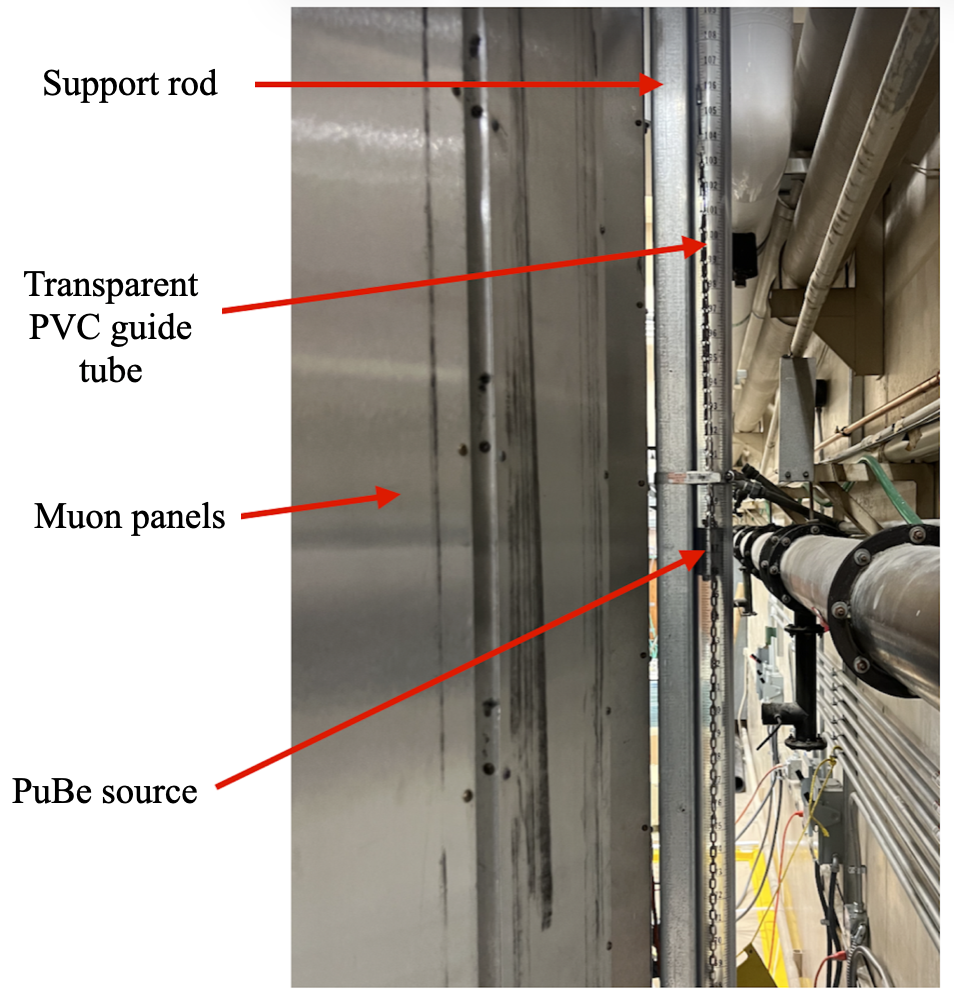}
    \caption{The PuBe source is positioned using a winch system with the radial distance maintained by the deployment tube.}
    \label{fig:pube}
\end{figure*}

The 32\,GBq PuBe source was purchased from Mound Laboratory in 1958.  Its high activity necessitates deployment outside the OV, as depicted in Fig.~\ref{fig:pube}. The decay chain of this source is similar to that of the AmBe, but with the $^{239}$Pu isotope undergoing $\alpha$ emission to produce a neutron and a 4.44\,MeV $\gamma$:
\begin{equation}
^{239}\text{Pu} \rightarrow{} ^{235}\text{U} + \alpha
\end{equation}
\begin{equation}
\alpha + ^{9}\text{Be} \rightarrow{} ^{12}\text{C}^{*} + \text{n}
\end{equation}
\begin{equation}
^{12}\text{C}^{*} \rightarrow{} ^{12}\text{C} + \gamma.
\end{equation}
The source is housed in a right-circular cylindrical container made of tantalum and SS, with a 17\,mm radius and a height of 60\,mm. The source is deployed using a winch to position it at the correct height within the deployment tube (located 1.65\,m from the center of the detector volume).

The goals of this source are threefold. As an external source, it enables testing of reconstruction techniques using 4.4\,MeV gammas originating from a known location outside the detector. It provides an opportunity to test reconstruction techniques without the optical shadowing that is unavoidable with internally deployed sources. Furthermore, due to the high activity, with $1.68\times10^{6}$ neutrons per second emitted, the fluence is large enough that the tagging of external neutrons could be assessed.

\subsection{$^{137}$Cs low-energy gamma source}
The goals of the low-energy gamma source are to study reconstructed position and energy at sub-MeV energies, as well as the scintillation time profile, using Compton-scattered electrons.  A lower-energy gamma ensures a relatively narrow spatial distribution of interactions around the source.  
$^{137}$Cs has a half-life of 30.0 years. 94.6\% of its beta decays result in $^{137\mathrm{m}}$Ba. This emits a 0.662-MeV gamma 90.0\% of the time with a half-life of 153\,s:
\begin{equation}
^{137}\text{Cs} \rightarrow{} ^{137m}\text{Ba} + \text{e}^{-} + \bar{\nu}_{e}
\end{equation}
\begin{equation}
^{137m}\text{Ba} \rightarrow{} ^{137}\text{Ba} + \gamma.
\end{equation}
As a result, the gamma is observed separately from the beta. Furthermore, the betas are easily shielded by a few mm of SS.  

A small cylindrical SS capsule, 3\,mm in diameter and 6\,mm in height, minimizes optical shadowing as well as the amount of energy deposited in the container.  
This source was produced at Spectrum Techniques~\cite{specTech} by pipetting a liquid solution into the SS cylinder and allowing it to dry, leaving the radioactive material on the inner surface. The container was then sealed with epoxy. The container was subsequently coated with Weld-On 40 to further protect the seal. The activity was estimated at 1.1\,kBq. 

\begin{figure*}[h!]
    \centering
    \includegraphics[angle=90, width=0.59\textwidth, trim=  0cm 0cm 0cm 0cm, clip]{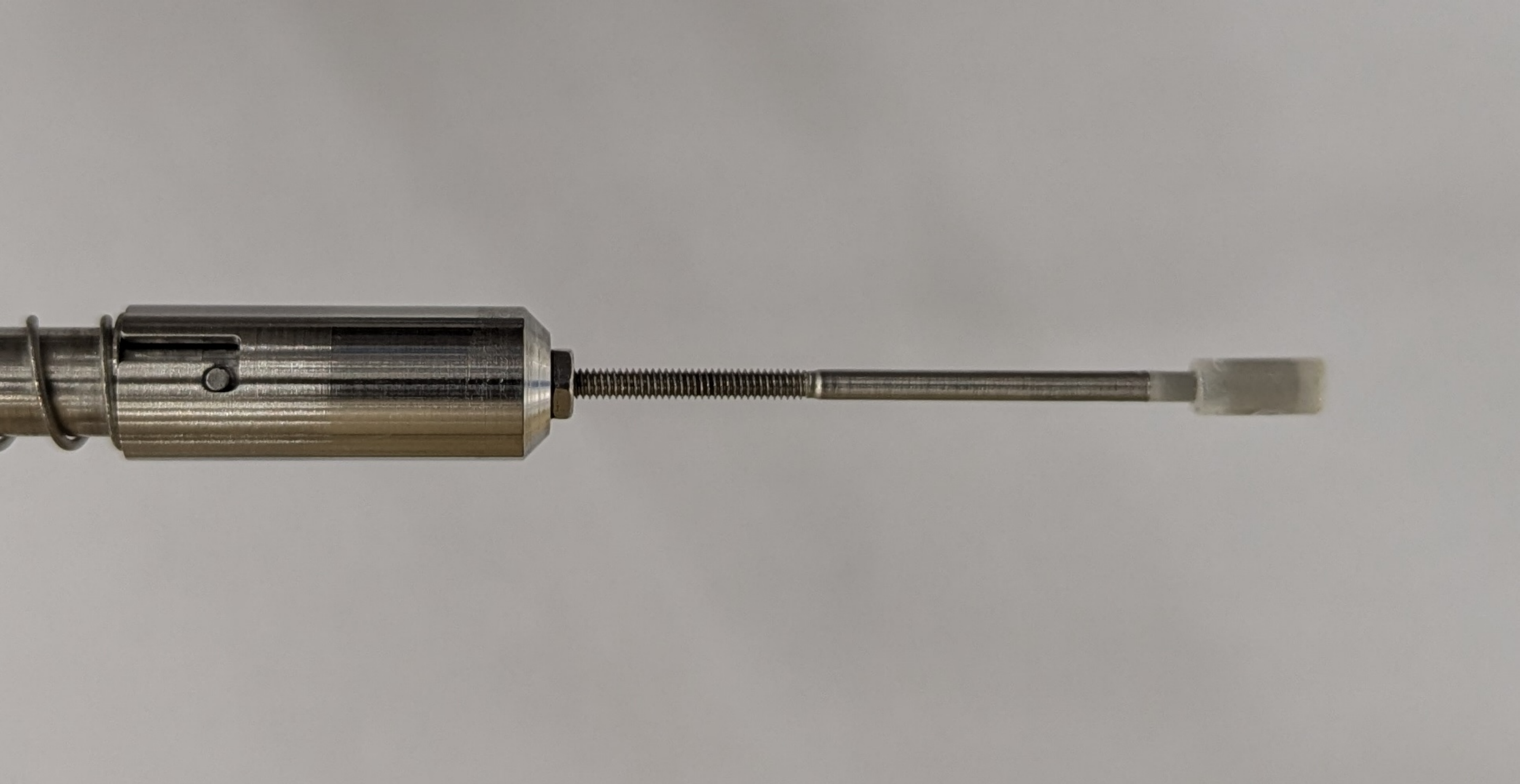}
    \caption{Photograph of the $^{137}$Cs source showing the SS rod and Weld-On 40 encapsulated source container.}
    \label{fig:cs137}
\end{figure*}

\subsection{Directional beta sources}

\begin{figure}[!h]
    \begin{center}
    \includegraphics[width=0.65\textwidth,trim= 0.5cm 2.5cm 1cm 3cm, clip]{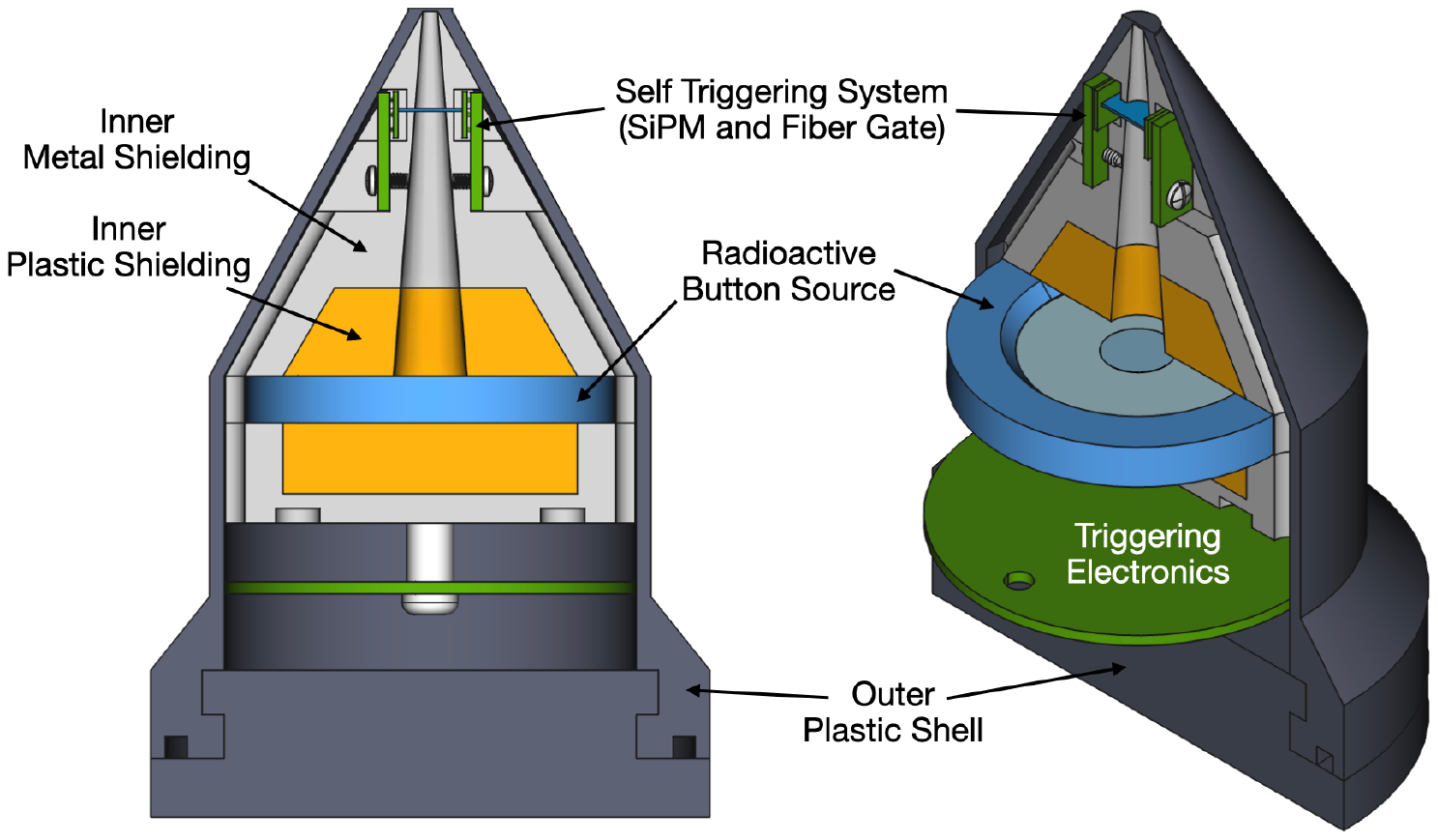}
    \includegraphics[width=0.31\textwidth, trim={0cm 0cm 0cm 0cm}, clip]{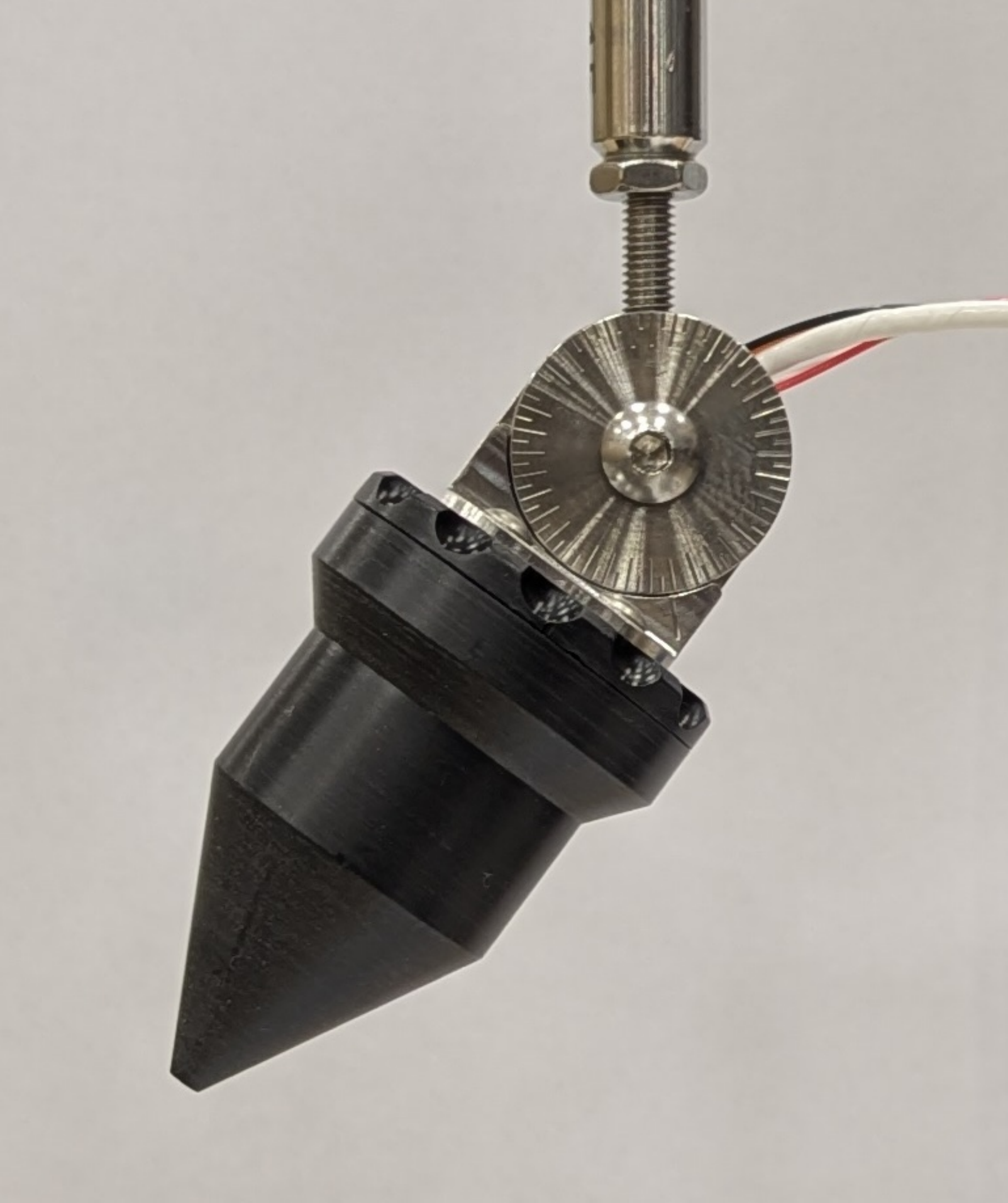}
    \end{center}
    \caption{Rendering showing the components of the directional source (left). A photograph of the 30\,mm source, highlighting the adjustable mechanism used to set the polar angle prior to deployment (right).}
    \label{f:directional}
\end{figure}

\begin{figure}[!h]
    \begin{center}
    \includegraphics[width=1.0\textwidth,trim= 0cm 0.3cm 0cm 0cm, clip]{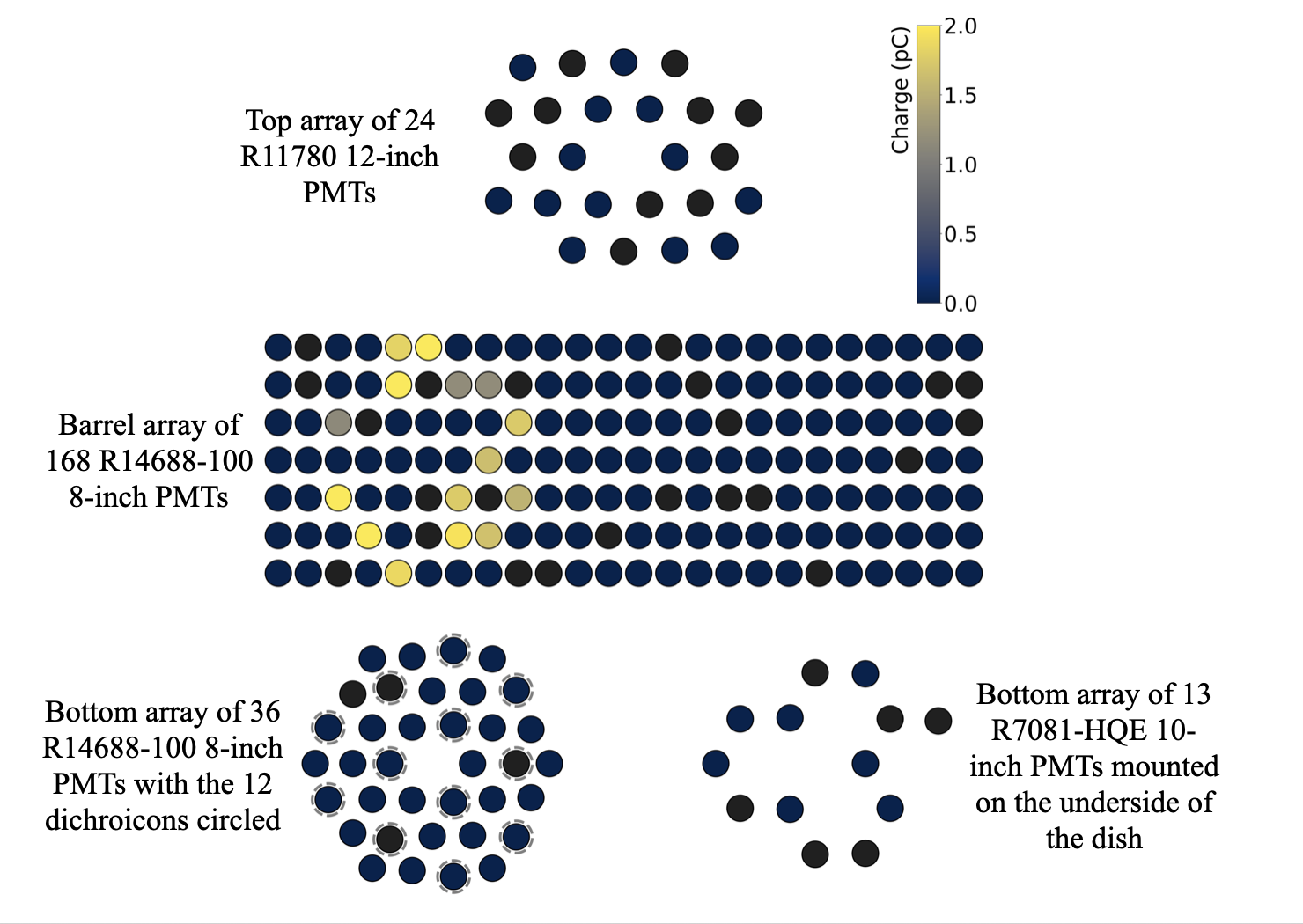}
    \end{center}
    \caption{Event display of a Cherenkov ring produced in water by an electron emitted from the directional source. The color scale indicates the PMT charge. Further details, including the calculation of the PMT charge, are provided in~\cite{EosWater}.}
     \label{f:display}
\end{figure}

The directional source is a novel calibration device designed to study direction reconstruction performance. The source, as shown in Fig.~\ref{f:directional}, is a small capsule housing a radioactive isotope that emits a collimated stream of betas in a known direction and has self-triggering capabilities. A borehole in the shielding material allows betas to exit the source capsule in a fixed direction, while those emitted in other directions are absorbed by the shielding. A rear connector interfaces the source to the calibration arm, allowing it to be oriented at any selected polar angle. Two identical capsules have been constructed for different beta-emitting isotopes: $^{90}$Sr and $^{106}$Ru. The $^{90}$Sr decay is as follows:
\begin{equation}
^{90}\text{Sr} \rightarrow{} ^{90}\text{Y} + \text{e}^{-} + \bar{\nu}_{e}
\end{equation}
\begin{equation}
^{90}\text{Y} \rightarrow{} ^{90}\text{Zr} + \text{e}^{-} + \bar{\nu}_{e}.
\end{equation}
The decay of $^{106}$Ru is as follows:
\begin{equation}
^{106}\text{Ru} \rightarrow{} ^{106}\text{Rh} + \text{e}^{-} + \bar{\nu}_{e}
\end{equation}
\begin{equation}
^{106}\text{Rh} \rightarrow{} ^{106}\text{Pd} + \text{e}^{-} + \bar{\nu}_{e}.
\end{equation}
These isotopes emit betas with a Q-value of 2.28\,MeV and 3.5\,MeV, respectively. $^{106}$Ru emits gammas in 20\% of decays. These 185\,kBq sources were purchased from Eckert \& Ziegler~\cite{ezag}. An example of such an event is shown in Fig.~\ref{f:display}.

The innermost polyoxymethylene (POM) shielding layer reduces Bremsstrahlung radiation, as emitted betas lose energy in this lower-density material before reaching the higher-density brass shield. A conical borehole encompasses the entire 5\,mm active area of the sources and is angled to achieve effective collimation. Based on RAT-PAC2 simulations~\cite{ratpac}, 80\% of the betas exit within 16.10 (20.60) degrees of the pointed direction for $^{90}$Sr ($^{106}$Ru). A hermetically sealed, 1\,mm thick outer POM shell encapsulates the source components. The thickness of the shell is reduced to 0.2\,mm at the borehole exit to minimize energy loss and deflection of the betas.

The self-triggering system consists of a scintillating fiber gate and two Broadcom AFBR-S4N33C013 SiPMs. Optical glue is applied to 0.2\,mm diameter optical fibers to create a thin ribbon at the borehole exit. Betas passing through this ``fiber gate'' produce scintillation light when exiting the capsule. This light is detected by the SiPMs placed on either side of the fiber gate. The SiPM signals are amplified and discriminated to produce a transistor–transistor logic (TTL) signal upon coincidence of both SiPMs. The self-triggering system can tag betas with an efficiency of $\sim$50\%.

As exiting betas travel less than a 1\,cm after exiting the source capsule, self-shadowing effects must be accounted for. Namely, when light is produced from beta emission, a shadow is cast behind the source capsule, providing additional directional information that can potentially bias direction reconstruction results. To thoroughly study this effect, three directional sources with body diameters of 30 mm, 20 mm, and 12 mm have been constructed for each isotope. Further details on the design, characterization, and deployment of these directional sources will be presented in a dedicated publication.

\subsection{Cherenkov sources}
\begin{figure}[!h]
    \begin{center}
    \includegraphics[width=1.0\textwidth]{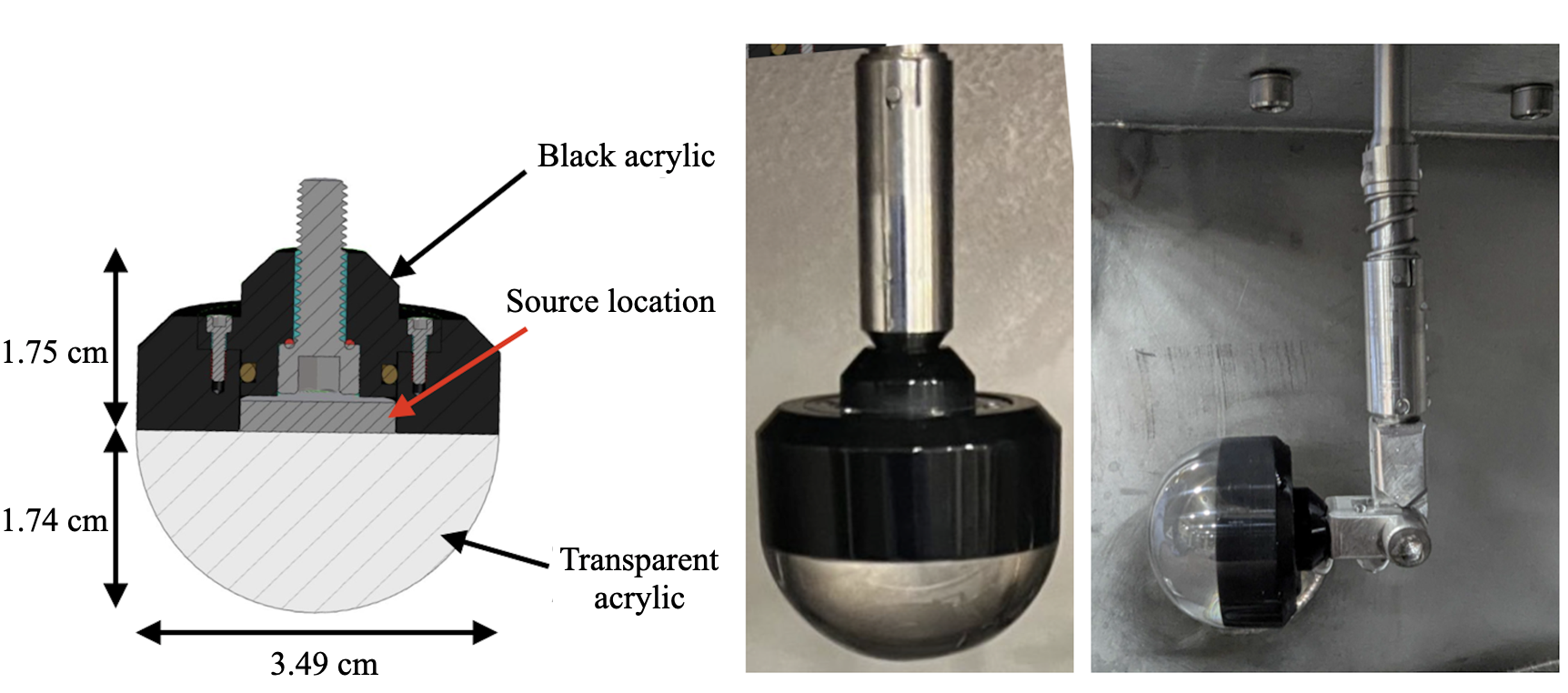}
    \end{center}
    \caption{The dimensions and components of the Cherenkov source (left). The vertical Cherenkov source which is oriented to produce downward going light (center). The elbowed source which allows for Cherenkov light to be pointed at a range of angles (right).}
    \label{f:cherenkov}
\end{figure}

The Cherenkov sources have three main goals: to produce a repeatable wavelength spectrum to analyze optics, to provide quasi-directional light emission, and to enable a pure Cherenkov spectrum when deployed in scintillator. Furthermore, by constructing otherwise identical sources using UVT and UV-absorbing (UVA) acrylic, modeling of absorption and scattering in a scintillator target can be validated with the resulting spectral differences. To achieve this, three Cherenkov source housings were produced, as shown in Fig.~\ref{f:cherenkov}. Each housing was machined to identical dimensions but with three different transparent acrylic hemispheres: one using UVT and two using UVA acrylic. The center of the housing holds a $3.7\times 10^{5}$\,Bq $^{90}$Sr button source from Eckert \& Ziegler~\cite{ezag}. The thickness of the acrylic hemisphere was chosen to fully absorb the emitted betas (Q = 2.3\,MeV), so that a self-contained Cherenkov spectrum is produced. The hemispherical shape was chosen so that emitted Cherenkov photons would be near normal incidence upon exiting the acrylic. 
Two source attachments were designed: a vertical attachment and an elbow attachment, allowing the quasi-directional source to be oriented between $\pi/2 \leq \theta \leq \pi$.

\section{Detector simulation framework}

\eos\ is modeled using the RAT-PAC2 simulation and analysis package~\cite{ratpac}. This framework is based on the original RAT and RAT-PAC packages, developed initially for the Braidwood reactor experiment~\cite{Braidwood} and developed further for other neutrino experiments such as SNO+~\cite{SNO+Det}, MiniCLEAN~\cite{Wang:2017jts}, DEAP~\cite{DEAP} and \theia~\cite{Askins:2019oqj}. RAT-PAC2 is a Geant4-based~\cite{GEANT4:2002zbu} tool that provides a comprehensive simulation package, including physics interactions, production and propagation of optical photons at the single-photon level, and full detector geometry, 3D PMT modeling, charge and timing of PE-level photodetection, and data acquisition, readout, and event building.
RAT-PAC2 also allows for run-by-run parameterization of the detector status. 

\begin{figure}[!h]
    \begin{center}
    \includegraphics[width=0.5\textwidth]{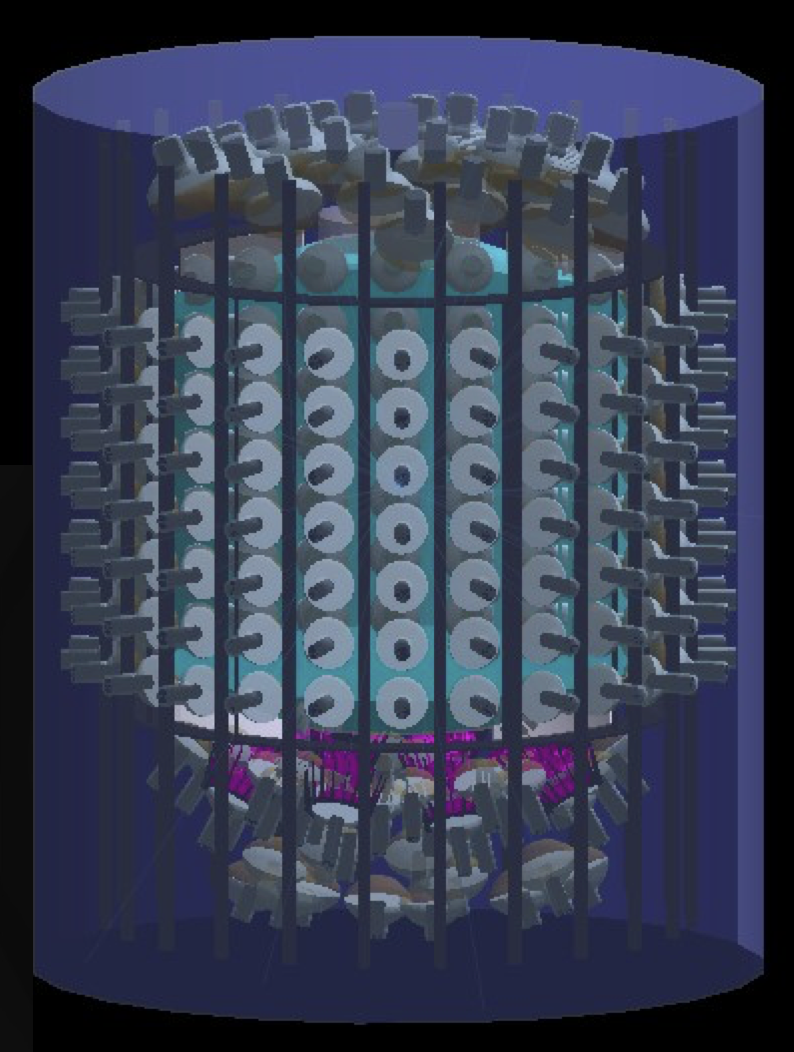}
    \end{center}
    \caption{Visualization of the \eos\ detector geometry in RAT-PAC2.}
    \label{f:ratpac2}
\end{figure}

The simulated \eos\ geometry, as shown in Fig.~\ref{f:ratpac2}, includes all PMT types, dichroicons and support structures. 
Material and optical properties are derived from bench-top measurements and fine-tuned with an extensive in-situ calibration campaign~\cite{EosWater}. During this campaign, the laserball was deployed at 13 positions along the central axis of \eos, using 374, 408, 442 and 515\,nm light. This enables simulation of reflections, refraction, scattering and absorption. The PMT response - quantum efficiency, transit time, pulse-charge, transit time spread and pulse shape - is based on Hamamatsu testing and further refined using measurements performed at UC Berkeley.

The \eos\ event reconstruction suite consists of traditional likelihood, timing, and machine learning approaches. Three time-based methods have been employed for event vertexing. Two directional reconstruction methods using detected photoelectron positions have also been implemented. A machine learning approach has also been developed that simultaneously fits the vertex and direction. An in-depth explanation of these reconstruction methodologies is discussed in~\cite{EosWater}.

\section{Deployment record and plan}
\label{s:Deployment}
After completion of the \eos\ build in early 2024, initial commissioning of the subsystems was undertaken. Subsequently, the various detector fill phases began, with the full suite of calibration sources (Sec.~\ref{s:Calibration_program}) deployed in each phase. In May 2024, the first phase began with the IV and OV filled with water. This enabled comprehensive detector commissioning and characterization using a well-understood medium. At this stage, event reconstruction algorithms were tested and compared with MC simulations to ensure that the detector response was correctly parameterized. The results of this phase are presented in~\cite{EosWater}.

In November 2024, the first WbLS phase began. This involved a two-stage injection process in which a scattering reduction agent (SRA) was added, followed by the scintillating LS concentrate. The SRA is a high-concentration surfactant mixture designed to improve the final optics. However, after injection, the optical properties of the water + SRA solution were significantly degraded compared to pure water, exhibiting increased absorbance and reduced scattering lengths. A campaign was undertaken to understand this effect, which found that the components of WbLS are sensitive to small temperature variations that affect the formation of a stable mixture; a steady injection is imperative to prevent emulsification; and continuous filtration can recover the desired optical properties. Once the optics were recovered, the WbLS concentrate was injected in February 2025.

To characterize a more reproducible WbLS mixture, the initial formulation was drained from the IV, and a second water phase began in April 2025. The purpose of this was to clean the IV of remaining organics and to re-characterize the detector. Data from this phase were consistent with the initial water phase, demonstrating that the detector had no significant residual organic contamination.

A second WbLS phase began in August 2025 with the injection of both SRA and WbLS concentrate. For this injection, several procedural improvements were implemented: the organic materials were shipped from BNL to Berkeley in a temperature-controlled environment; the SRA and WbLS concentrate were injected into the IV at room temperature; and the injection used a diaphragm pump to minimize emulsification. In November 2025, the WbLS concentration was increased to 2\%. The different organic loadings will be characterized for light yield, optical properties, and scintillation timing response to understand the impact these have on reconstruction performance. A detailed discussion of the WbLS deployment and its results will be presented in an upcoming paper.

In the coming months, LAB/PPO scintillator formulations with various fluor loadings, similar to that of SNO+, will  be deployed \cite{Anderson:2020xxb}. This will provide an essential point of reference for detector performance. It will enable testing of Cherenkov/scintillation separation capabilities using ns-timing photosensors and dichroicons, in a well-characterized scintillator. Furthermore, the prospect of deploying slow scintillators for enhanced Cherenkov/scintillation separation, fast scintillators to explore the possibility of topological imaging, and various isotopic loadings, are being considered.

\section{Future upgrade paths}
Although advanced technologies are already implemented in \eos, further upgrades are planned or being considered. The general axes of improvement are timing, photon yield, and further Cherenkov/scintillation separation. 

    Given the relatively small size of \eos, the PMTs subtend a large solid angle, so each channel typically detects more than one photon even for low-energy interactions. The existing CAEN V1730 digitizers sample at 500\,MSamples/s, yielding just a few samples per pulse for \eos's fast PMTs. Advanced analysis techniques perform well at determining pulse arrival times, particularly when well separated. However, the 250\,MHz Nyquist signal bandwidth limits the ability to resolve pulses that pile up, reducing the ability to distinguish a Cherenkov photon from prompt scintillation light recorded on the same channel. Upgrading to faster CAEN digitizers, such as the CAEN V1742~\cite{caendigitizersv1742}, would be expensive and generate a substantially higher data rate. This, in addition to the deadtime for such devices, would ultimately limit the achievable trigger rate.

    An alternative approach is to use {\it analog processing}---the extraction of waveform features using analog techniques. The Analog Photon Processor (APP) ASIC, being developed at the University of Pennsylvania, extracts rising and falling edge times with an anticipated precision of around 25\,ps, provides the integral over threshold of each resolved `pulse packet', and up to two peak heights and times for each packet. A PING/PONG architecture ensures deadtimeless operation. The extracted features are stored in an analog FIFO that can then be read out at rates much lower than those required for waveform digitization---ADCs with $\mu$s conversion times are more than fast enough.
    
Furthermore, an additional 20 R14688-100 PMTs have been procured. These may replace the slower R11780 PMTs and, if beneficial, could also be installed on the lower dish behind the existing dichroicons for improved Cherenkov/scintillation separation. More speculatively, detector performance could be further enhanced by installing additional dichroicons on the barrel or top dish. The extra dichroicons could fit in the interstitial space between the existing barrel PMTs, and concentrate light on the smaller PMTs while ensuring the existing 8'' PMTs are not shadowed to scintillation light.  Such modifications would require minor modifications to the existing support structure, potentially included additional small PMTs between the current 8'' tubes.

\section{Redeployment options}

Due to its advanced technology, flexible configuration, and scale, upon completion of operations at Berkeley, \eos\ could be repurposed for a range of future measurements. One promising avenue is to redeploy to the Spallation Neutron Source (SNS) at Oak Ridge National Laboratory (ORNL). The energy range of neutrinos produced by the SNS is a few tens of MeV~\cite{COHERENT},  providing an exciting opportunity to study neutrino interactions in the supernova energy regime.

Redeploying \eos\ at the SNS opens the possibility of improving the understanding of supernova neutrino phenomenology. Measurements of charged- and neutral-current cross sections on $^{16}$O and $^{12}$C would reduce nuclear cross-section uncertainties, refining interaction models used to predict event rates in large-scale detectors and improving the interpretation of future supernova neutrino signals.  

$^{7}$Li is of interest to the astrophysical neutrino community due to its large $\nu_e$ charged-current cross section, low interaction threshold, and sharply peaked differential cross section, allowing for excellent neutrino energy reconstruction. Doping the \eos\ target medium with $^{7}$Li would enable measurements of the $\nu_e$ charged-current cross section on this isotope at SNS energies. Such measurements would provide an important benchmark for nuclear models and support the development of future Li-doped solar neutrino detectors.

In the precision era of neutrino physics, neutrons constitute a significant source of background. The intense neutron fluence at the SNS provides an ideal environment to develop and validate neutron identification techniques using hybrid technologies. If successful, these methods could significantly improve background rejection and enhance the sensitivity of future rare-event experiments.

A further benefit of deployment at the SNS arises from the flux intensity -- $\sim$10$^{23}$ protons-on-target per year.  \eos\ could be located within 20\,m from the target, resulting in a large geometric acceptance for searches for beyond-the-Standard-Model particles. Sterile neutrino searches can be performed by looking for an increase in the electron (anti-) neutrino interaction rate due to short-baseline oscillations. Furthermore, if light scalar or pseudo-scalar particles are produced through stopped-pion decays, searches for their interaction or decay can be performed. The hybrid Cherenkov and scintillation capabilities of \eos\ could provide powerful discrimination against Standard Model backgrounds through topological event reconstruction, fast timing, and directional and energy information, thereby allowing visible decay signatures such as e$^+$e$^-$ and $\gamma \gamma$ to be identified.

As reactor antineutrinos provide an unshieldable signal, monitoring nuclear reactor operations using antineutrinos is of interest from a non-proliferation standpoint. The sensitivity of mid-field reactor monitoring is largely determined by background rejection and detector mass. In this context, the deployment of \eos\ near a nuclear reactor presents an opportunity to investigate the potential of hybrid detector technologies for reactor antineutrino detection. Through capabilities such as Cherenkov/scintillation separation and novel particle identification techniques for improved background rejection, together with technologies that enable the cost-effective deployment of larger detectors, \eos\ would provide a platform to assess the potential improvements to reactor monitoring offered by next-generation detector concepts.

Furthermore, \eos\ could be deployed in high-energy neutrino beams at FNAL or CERN. This would afford the opportunity to further the long-baseline effort for next-generation technologies. Innovations such as hadronic event reconstruction in scintillator-based media, advanced event identification with fast photosensors, and sub-Cherenkov PID capabilities all provide potential improvements to existing techniques. This superior background rejection, PID, and event reconstruction will greatly enhance the capabilities of next-generation neutrino detectors, such as \theia, to perform both high-precision oscillation measurements and rare-event detection.

\section{Conclusion}

\eos\ is a flexible R\&D testbed for next-generation neutrino detector technology. Utilizing state-of-the-art photosensors, incorporating the first use of dichroicons in a large-scale experiment, and enabling deployment of novel detection media such as WbLS, \eos\ will demonstrate the capabilities of hybrid detector technologies.

For a variety of liquid scintillator and WbLS formulations, position, direction, and energy reconstruction capabilities will be determined. The ability to separate Cherenkov and scintillation signals using large photocathode coverage, photon sorting, and fast timing will be explored. Additionally, \eos\ provides an ideal testbed to refine optical models for these novel detection media.

\eos\ serves as an integrated testbed for evaluating these technologies for use in future experiments. It will help guide the design and construction of larger detectors, such as the \theia\ concept, maximizing their potential to address outstanding questions in neutrino physics.   
\section{Acknowledgements}

Work conducted at Lawrence Berkeley National Laboratory was performed under the auspices of the U.S. Department of Energy under Contract DE-AC02-05CH11231. The work conducted at Lawrence Livermore National Laboratory was supported by the U.S. Department of Energy under contract DE-AC52-07NA27344, release number LLNL-JRNL-2021645. The work conducted at Brookhaven National Laboratory was supported by the U.S. Department of Energy under contract DE-AC02-98CH10886. The work conducted at Oak Ridge National Laboratory was supported by the U.S. Department of Energy under contract number DE-AC05-00OR22725. The work conducted at Oxford was supported by the Royal Society. The project was funded by the U.S. Department of Energy, National Nuclear Security Administration, Office of Defense Nuclear Nonproliferation  Research and Development (DNN R\&D). This material is based upon work supported by the U.S. Department of Energy, Office of Science, Office of High Energy Physics, under Award Number DE-SC0018974.  This work is supported as part of the Growth in Research Opportunities With Traineeship in High energy physics at Minority Serving Institutions (GROWTH-MSI) program funded by the DOE Office of Sciences High Energy and Nuclear Physics, Grant number DE-SC0023725.

\clearpage
\addcontentsline{toc}{section}{References}
\bibliographystyle{JHEP}
\bibliography{bibliography}

@article{CHESS2016,
    author = "Caravaca, J. and Descamps, F. B. and Land, B.J. and Wallig, J. and Yeh, M. and Orebi Gann, G. D.",
    title = "{Experiment to demonstrate separation of Cherenkov and scintillation signals}",
    eprint = "",
    archivePrefix = "arXiv",
    primaryClass = "physics.ins-det",
    doi = "10.1103/PhysRevC.95.055801",
    journal = "Phys. Rev. C",
    volume = "95",
    number = "5",
    pages = "055801",
    year = "2017"
}

@article{CHESS2020,
    author = "Caravaca, J. and Land, B. J. and Yeh, M. and Orebi Gann, G. D.",
    title = "{Characterization of water-based liquid scintillator for Cherenkov and scintillation separation}",
    eprint = "",
    archivePrefix = "arXiv",
    primaryClass = "physics.ins-det",
    doi = "10.1140/epjc/s10052-020-8418-4",
    journal = "Eur. Phys. J. C",
    volume = "80",
    number = "9",
    pages = "867",
    year = "2020"
}

@article{Kaptanoglu:2019gtg,
    author = "Kaptanoglu, Tanner and Luo, Meng and Land, Ben and Bacon, Amanda and Klein, Josh",
    title = "{Spectral Photon Sorting For Large-Scale Cherenkov and Scintillation Detectors}",
    eprint = "",
    archivePrefix = "arXiv",
    primaryClass = "physics.ins-det",
    doi = "10.1103/PhysRevD.101.072002",
    journal = "Phys. Rev. D",
    volume = "101",
    number = "7",
    pages = "072002",
    year = "2020"
}

@article{Biller:2020uoi,
    author = "Biller, Steven D. and Leming, Edward J. and Paton, Josephine L.",
    title = "{Slow Fluors for Highly Effective Separation of Cherenkov Light in Liquid Scintillators}",
    eprint = "",
    archivePrefix = "arXiv",
    primaryClass = "physics.ins-det",
    doi = "10.1016/j.nima.2020.164106",
    journal = "Nucl. Instrum. Meth. A",
    volume = "972",
    pages = "164106",
    year = "2020"
}

@article{Askins:2019oqj,
    author = "Askins, M. and others",
    collaboration = "The Theia",
    title = "{THEIA: an advanced optical neutrino detector}",
    eprint = "",
    archivePrefix = "arXiv",
    primaryClass = "physics.ins-det",
    doi = "10.1140/epjc/s10052-020-7977-8",
    journal = "Eur. Phys. J. C",
    volume = "80",
    number = "5",
    pages = "416",
    year = "2020"
}

@article{Guo:2018kcp,
    author = "Guo, Ziyi and Wang, Zhe",
    editor = "Liu, Zhen-An",
    title = "{Slow Liquid Scintillator for Scintillation and Cherenkov Light Separation}",
    doi = "10.1007/978-981-13-1316-5\_32",
    journal = "Springer Proc. Phys.",
    volume = "213",
    pages = "173--177",
    year = "2018"
}

@article{kamland,
      author         = "Eguchi, K. and others",
      collaboration  = "The KamLAND",
      title          = "{First results from KamLAND: Evidence for reactor
                        anti-neutrino disappearance}",
      journal        = "Phys. Rev. Lett.",
      volume         = "90",
      year           = "2003",
      pages          = "",
      doi            = "10.1103/PhysRevLett.90.021802",
      eprint         = "",
      archivePrefix  = "arXiv",
      primaryClass   = "hep-ex",
      SLACcitation   = "%%CITATION = HEP-EX/0212021;%%"
}

@article{SuperK,
    author = "Fukuda, Y. and others",
    collaboration = "The Super-Kamiokande",
    title = "{The Super-Kamiokande detector}",
    eprint = "",
    archivePrefix = "",
    reportNumber = "",
    doi = "https://doi.org/10.1016/S0168-9002(03)00425-X",
    journal = "Nucl.Instrum.Meth.A.",
    volume = "501",
    pages = "2-3",
    year = "2003"
}

@article{YehWbLS,
author = {Yeh, Minfang and Hans, Sunej and Beriguete, W. and Rosero, R. and Hu, L. and Hahn, R.L. and Diwan, M.V. and Jaffe, D.E. and Kettell, S.H. and Littenberg, L.},
year = {2011},
month = {12},
pages = {51-56},
title = {A new water-based liquid scintillator and potential applications},
volume = {660},
journal = "Nucl. Instrum. Meth. A",
doi = {10.1016/j.nima.2011.08.040}
}

@article{Onken:2020pnv,
author ="Onken, Drew R. and Moretti, Federico and Caravaca, Javier and Yeh, Minfang and Orebi Gann, Gabriel D. and Bourret, Edith D.",
title  ="Time response of water-based liquid scintillator from X-ray excitation",
doi = "https://doi.org/10.1039/D0MA00055H",
journal  ="Mater. Adv.",
year  ="2020",
volume  ="1",
issue  ="1",
pages  ="71-76",
publisher  ="RSC",
}

@misc{ham_datasheet_r7081, author="Hamamatsu",  howpublished="\url{https://www.hamamatsu.com/content/dam/hamamatsu-photonics/sites/documents/99_SALES_LIBRARY/etd/LARGE_AREA_PMT_TPMH1376E.pdf}", title="{Large Area Photocathode Photomultiplier Tubes}", year="2022", note="[Accessed Oct. 27, 2022]"}

@misc{ham_datasheet_r14688, author="Hamamatsu",  howpublished="\url{https://www.hamamatsu.com/content/dam/hamamatsu-photonics/sites/documents/99_SALES_LIBRARY/etd/R14688_R14688-100_TPMH1385E.pdf}", title="{Photomultiplier Tube R14688/R14688-100}", year="2022", note="[Accessed Oct. 27, 2022]"}

@misc{ratpac, author="Stan Seibert and others", howpublished="\url{https://rat.readthedocs.io/en/latest/}", title="{RAT User Guide}", note="[Accessed Oct. 28, 2022]", year="2014"}

@misc{mainz, author="D. Guffanti and others", howpublished="\url{https://indico.desy.de/event/29253/contributions/104889/attachments/66183/81926/gff-MzScintillatorRnD.pdf}", title="{Progress of Water-based Liquid Scintillator Studies in Mainz}", note="[Accessed Oct. 28, 2022]", year="2021"}

@article{GEANT4:2002zbu,
    author = "Agostinelli, S. and others",
    collaboration = "GEANT4",
    title = "{GEANT4--a simulation toolkit}",
    reportNumber = "SLAC-PUB-9350, FERMILAB-PUB-03-339, CERN-IT-2002-003",
    doi = "10.1016/S0168-9002(03)01368-8",
    journal = "Nucl. Instrum. Meth. A",
    volume = "506",
    pages = "250--303",
    year = "2003"
}

@article{Anderson:2020xxb,
    author = "Anderson, M. R. and others",
    collaboration = "The SNO+",
    title = "{Development, characterisation, and deployment of the SNO+ liquid scintillator}",
    eprint = "",
    archivePrefix = "arXiv",
    primaryClass = "physics.ins-det",
    doi = "10.1088/1748-0221/16/05/P05009",
    journal = "JINST",
    volume = "16",
    number = "05",
    pages = "P05009",
    year = "2021"
}

@article{doi:10.1146/annurev-nucl-101917-020949,
author = {Machado, Pedro A.N. and Palamara, Ornella and Schmitz, David W.},
title = {The Short-Baseline Neutrino Program at Fermilab},
journal = {Annual Review of Nuclear and Particle Science},
volume = {69},
number = {1},
pages = {363-387},
year = {2019},
doi = {10.1146/annurev-nucl-101917-020949},

URL = "",
eprint = ""
}

@article{dichroicon_paper,
    author = "Kaptanoglu, Tanner and Luo, Meng and Land, Ben and Bacon, Amanda and Klein, Josh",
    title = "{Spectral Photon Sorting For Large-Scale Cherenkov and
             Scintillation Detectors}",
    eprint = "",
    archivePrefix = "arXiv",
    primaryClass = "physics.ins-det",
    doi = "10.1103/PhysRevD.101.072002",
    journal = "Phys. Rev. D",
    volume = "101",
    number = "7",
    pages = "072002",
    year = "2020",
}

@article{r11780_paper,
    title = "{Characterization of the Hamamatsu R11780 12 in. Photomultiplier
             Tube}",
    journal = {Nucl. Instrum. Meth. A},
    volume = {712},
    pages = {162-173},
    year = {2013},
    issn = {0168-9002},
    doi = {https://doi.org/10.1016/j.nima.2013.02.022},
    url = {https://www.sciencedirect.com/science/article/pii/S0168900213002015},
    author = {J. Brack and B. Delgado and J. Dhooghe and J. Felde and B. Gookin
              and S. Grullon and J.R. Klein and R. Knapik and A. LaTorre and S.
              Seibert and K. Shapiro and R. Svoboda and L. Ware and R. {Van Berg}
              },
}

@misc{knight_acrylic_lp,
    title = {Knight Optical Acrylic Longpass Filters},
    howpublished = {\url{
                    https://www.knightoptical.com/stock/filters/acrylic-colour-filters
                    }},
    author = {{Knight Optical}},
    urldate = {2025-5-16},
    year = 2025,
}

@misc{knight_dichroic_sp,
    title = {Knight Optical 450 nm Dichroic Shortpass Filters},
    howpublished = {\url{
                    https://www.knightoptical.com/stock/shortpass-dichroic-filter-450nm-25mmdia-x-1-1mmthk
                    }},
    author = {{Knight Optical}},
    urldate = {2025-5-16},
    year = 2025,
}

@article{DUNE:2020cqd,
    author = "Abi, B. and others",
    collaboration = "The DUNE",
    title = "{First results on ProtoDUNE-SP liquid argon time projection chamber performance from a beam test at the CERN Neutrino Platform}",
    eprint = "",
    archivePrefix = "arXiv",
    primaryClass = "physics.ins-det",
    reportNumber = "FERMILAB-PUB-20-059-AD-ESH-LBNF-ND-SCD, CERN-EP-2020-125",
    doi = "10.1088/1748-0221/15/12/P12004",
    journal = "JINST",
    volume = "15",
    number = "12",
    pages = "P12004",
    }

@article{SNO:2021xpa,
    author = "Albanese, V. and others",
    collaboration = "The SNO+",
    title = "{The SNO+ experiment}",
    eprint = "2104.11687",
    archivePrefix = "arXiv",
    primaryClass = "physics.ins-det",
    doi = "10.1088/1748-0221/16/08/P08059",
    journal = "JINST",
    volume = "16",
    number = "08",
    pages = "P08059",
    year = "2021"
}

@misc{uthash,
  author = "Hanson, Troy D. and O'Dwyer, Arthur",
  title = {{uthash: a hash table for C structures}},
  howpublished = {\url{https://troydhanson.github.io/uthash/}},
  note = {Accessed: 2024-03-22}
}

@misc{minard,
  author = "",
  collaboration = "The SNO+",
  title = {{GitHub Repository: snoplus/minard}},
  howpublished = {\url{https://github.com/snoplus/minard}},
  note = {Accessed: 2024-04-24}
}

@misc{pt100,
  author = {{Dwyer Omega}},
  title = {{Datasheet: RTD Probes with M12 Molded Connectors}},
  howpublished = {\url{https://assets.dwyeromega.com/pdf/test-and-measurement-equipment/temperature/sensors/rtds/PR-22.pdf}},
  note = {{Accessed: 2023-09-25}}
}

@misc{wds4965,
  author = {{Adafruit}},
  title = {{Datasheet: Adafruit 4965}},
  howpublished = {\url{https://cdn-shop.adafruit.com/product-files/4965/Datasheet.pdf}},
  note = {{Accessed: 2023-09-25}}
}

@misc{sen0311,
  author = {{DFRobot}},
  title = {{Datasheet: A02YYUW Waterproof Ultrasonic Sensor}},
  howpublished = {\url{https://mm.digikey.com/Volume0/opasdata/d220001/medias/docus/201/SEN0311_Web.pdf}},
  note = {{Accessed: 2023-09-25}}
}

@misc{arduino2560,
  author = {{Arduino}},
  title = {{User Manual: Arduino Mega 2560 Rev3}},
  howpublished = {\url{https://docs.arduino.cc/resources/datasheets/A000067-datasheet.pdf}},
  note = {{Accessed: 2023-09-25}}
}

@misc{ls75,
  author = {{TDK Lambda}},
  title = {{Datasheet: LS25-150 Series}},
  howpublished = {\url{https://product.tdk.com/system/files/dam/doc/product/power/switching-power/ac-dc-converter/catalog/ls25-150_e.pdf}},
  note = {{Accessed: 2023-09-25}}
}

@misc{redis,
  author = {{Redis}},
  title = {{Redis: Open Source}},
  howpublished = {\url{https://redis.io/open-source/}},
  note = {{Accessed: 2025-09-05}}
}

@misc{ctypes,
  author = {{Python Software Foundation}},
  title = {{{\tt ctypes} --- A foreign function library for Python}},
  howpublished = {\url{https://docs.python.org/3/library/ctypes.html}},
  note = {{Accessed: 2025-09-05}}
}

@misc{flask,
  author = {{Pallets}},
  title = {{Flask Documentation}},
  howpublished = {\url{https://flask.palletsprojects.com/en/stable/}},
  note = {{Accessed: 2025-09-05}}
}

@misc{doecode_9801,
title = {HDF5-Version 1.12.0},
author = {The HDF Group and Koziol, Quincey},
abstractNote = {The HDF5 library and file format implement the HDF5 data model for storing and managing data. This implementation supports an unlimited variety of datatypes and is designed for flexible and efficient I/O and for high volume and complex data. It is highly portable and extensible, allowing applications to evolve in their use of HDF5. The HDF5 Technology suite includes tools and applications for managing, manipulating, viewing, and analyzing data in the HDF5 format and other implementations.},
doi = {10.11578/dc.20180330.1},
url = {https://doi.org/10.11578/dc.20180330.1},
howpublished = {[Computer Software] \url{https://doi.org/10.11578/dc.20180330.1}},
year = {2020},
month = {2}
}

@article{hamamatsu_8in,
    author = "Kaptanoglu, T. and Rincon, A. and Duce, M. and Kaplan, S. and Koplowitz, J. and  Lynch, S. and Ryoo, H. J. and Orebi Gann, G. D. ",
    title = "{Characterization of the Hamamatsu 8-inch R14688-100 PMT}",
    eprint = "",
    archivePrefix = "arXiv",
    primaryClass = "physics.ins-det",
    doi = "10.1088/1748-0221/19/02/P02032",
    journal = "JINST",
    volume = "19",
    number = "02",
    pages = "P02032",
    year = "2024"
}

@article{STAR,
title = "{The STAR Trigger}",
author = "Bieser, F. S. and others",
journal = "Nucl. Instrum. Meth. A.",
doi = "https://doi.org/10.1016/S0168-9002(02)01974-5",
number = "A",
volume = "499",
place = {Netherlands},
year = {2003},
month = {3}
}

@misc{tooldaq,
  author = {{B. Richards}},
  title = {{The ToolDAQ DAQ Software Framework \& Its Use In The Hyper-K \& ANNIE Detectors}},
  howpublished = {\url{https://doi.org/10.1051/epjconf/201921401022}},
  note = {{}}
}

@misc{wblsdaq,
  author = {{B. J. Land}},
  title = {{WbLSDAQ}},
  howpublished = {\url{http://github.com/BenLand100/WbLSdaq}},
  note = {{}}
}

@article{T2KCP,
    author = "Abe, K. and others",
    collaboration = "The T2K",
    title = "{Constraint on the matter–antimatter symmetry-violating phase in neutrino oscillations}",
    eprint = "",
    archivePrefix = "",
    primaryClass = "",
    doi = "https://doi.org/10.1038/s41586-020-2177-0",
    journal = "Nature",
    volume = "580",
    number = "",
    pages = "339-344",
    year = "2020",
}

@article{LeptonicCP,
    author = "Branco, G. C. and Gonzalez Felipe, R. and Joaquim, F.R. ",
    title = "{Leptonic CP Violation}",
    eprint = "",
    archivePrefix = "",
    primaryClass = "",
    doi = "https://doi.org/10.1103/RevModPhys.84.515",
    journal = "Rev. Mod. Phys.",
    volume = "84",
    number = "515",
    pages = "",
    year = "2012",
}

@article{NOVACP,
    author = "Acero, M. A. and others",
    collaboration = "The NOvA",
    title = "{Improved measurement of neutrino oscillation parameters by the NOvA experiment}",
    eprint = "",
    archivePrefix = "",
    primaryClass = "",
    doi = "https://doi.org/10.1103/PhysRevD.106.032004",
    journal = "Phys. Rev. D.",
    volume = "106",
    number = "032004",
    pages = "",
    year = "2022",
}

@article{NeutrinoMixing,
    author = "Capozzi, F. and others",
    title = "{Neutrino masses and mixing: Entering the era of subpercent precision}",
    eprint = "",
    archivePrefix = "",
    primaryClass = "",
    doi = "https://doi.org/10.1103/PhysRevD.111.093006",
    journal = "Phys. Rev. D.",
    volume = "111",
    number = "093006",
    pages = "",
    year = "2025",
}

@article{Nu6,
    author = "Esteban, I. and others",
    title = "{NuFit-6.0: updated global analysis of three-flavor neutrino oscillations}",
    eprint = "",
    archivePrefix = "",
    primaryClass = "",
    doi = "https://doi.org/10.1007/JHEP12(2024)216",
    journal = "J. High Energ. Phys.",
    volume = "2024",
    number = "216",
    pages = "",
    year = "2024",
}

@article{PDG,
    author = "Navas, S. and others",
    title = "{Review of Particle Physics}",
    eprint = "",
    archivePrefix = "",
    primaryClass = "",
    doi = "https://doi.org/10.1103/PhysRevD.110.030001",
    journal = "Phys. Rev. D.",
    volume = "110",
    number = "030001",
    pages = "",
    year = "2024",
}

@article{Majorana,
    author = "Wilczek, F.",
    title = "{Majorana returns}",
    eprint = "",
    archivePrefix = "",
    primaryClass = "",
    doi = "https://doi.org/10.1038/nphys1380",
    journal = "Nature Phys",
    volume = "5",
    number = "",
    pages = "614–618",
    year = "2009",
}

@article{DSNB,
    author = "Abe, K. and others",
    collaboration = "The Super-Kamiokande",
    title = "{Diffuse supernova neutrino background search at Super-Kamiokande}",
    eprint = "",
    archivePrefix = "",
    primaryClass = "",
    doi = "https://doi.org/10.1103/PhysRevD.104.122002",
    journal = "Phys. Rev. D.",
    volume = "104",
    number = "122002",
    pages = "",
    year = "2021",
}

@article{Solar,
    author = "Vinyoles, N. and others",
    title = "{A New Generation of Standard Solar Models}",
    eprint = "",
    archivePrefix = "",
    primaryClass = "",
    doi = "10.3847/1538-4357/835/2/202",
    journal = "ApJ",
    volume = "835",
    number = "202",
    pages = "",
    year = "2017",
}

@article{SN1987A,
    author = "Hirata, K. and others",
    collaboration = "The Kamiokande",
    title = "{Observation of a Neutrino Burst from the Supernova SN1987A}",
    eprint = "",
    archivePrefix = "",
    primaryClass = "",
    doi = "https://doi.org/10.1103/PhysRevLett.58.1490",
    journal = "Phys. Rev. Lett.",
    volume = "58",
    number = "1490",
    pages = "",
    year = "1987",
}

@article{Bethe,
    author = "Bethe, H. A. and Wilson, J. R.",
    title = "{Revival of a stalled supernova shock by neutrino heating}",
    eprint = "",
    archivePrefix = "",
    primaryClass = "",
    doi = "10.1086/163343",
    journal = "Astrophysical Journal",
    volume = "295",
    number = "",
    pages = "14-23",
    year = "1985",
}

@article{SNTheory,
    author = "Janka, H. T. and others",
    title = "{Theory of core-collapse supernovae}",
    eprint = "",
    archivePrefix = "",
    primaryClass = "",
    doi = "https://doi.org/10.1016/j.physrep.2007.02.002",
    journal = "Physics Reports",
    volume = "442",
    number = "",
    pages = "38-74",
    year = "2007",
}

@article{GeoKamLAND,
    author = "Abe, S. and others",
    collaboration = "The KamLAND",
    title = "{Abundances of Uranium and Thorium Elements in Earth Estimated by Geoneutrino Spectroscopy}",
    eprint = "",
    archivePrefix = "",
    primaryClass = "",
    doi = "https://doi.org/10.1029/2022GL099566",
    journal = "Geophysical
Research Letters",
    volume = "49",
    number = "",
    pages = "",
    year = "2022",
}

@article{GeoBorexino,
    author = "Bellini, G. and others",
    collaboration = "The Borexino",
    title = "{Observation of Geo-Neutrinos}",
    eprint = "",
    archivePrefix = "",
    primaryClass = "",
    doi = "https://doi.org/10.1016/j.physletb.2010.03.051",
    journal = "Phys.Lett.B.",
    volume = "687",
    number = "",
    pages = "299-304",
    year = "2010",
}

@article{FissionMonitoring,
    author = "Carr, R. and others",
    title = "{Sensitivity of seismically cued antineutrino detectors to nuclear explosions}",
    eprint = "",
    archivePrefix = "",
    primaryClass = "",
    doi = "https://doi.org/10.1103/PhysRevApplied.10.024014",
    journal = "Phys. Rev. Applied",
    volume = "10",
    number = "024014",
    pages = "",
    year = "2018",
}

@article{ReactorMonitoring,
    author = "Bernstein, A. and others",
    title = "{Colloquium: Neutrino detectors as tools for nuclear security}",
    eprint = "",
    archivePrefix = "",
    primaryClass = "",
    doi = "https://doi.org/10.1103/RevModPhys.92.011003",
    journal = "Rev. Mod. Phys.",
    volume = "92",
    number = "011003",
    pages = "",
    year = "2020",
}

@article{Zeller,
    author = "Formaggio, J. A. and Zeller, G. P. ",
    title = "{From eV to EeV: Neutrino Cross Sections Across Energy Scales
}",
    eprint = "",
    archivePrefix = "",
    primaryClass = "",
    doi = "https://doi.org/10.1103/RevModPhys.84.1307",
    journal = "Rev. Mod. Phys. ",
    volume = "84",
    number = "1307",
    pages = "",
    year = "2012",
}

@article{HyperK,
    author = "Abe, K. and others",
    collaboration = "The Hyper-Kamiokande",
    title = "{Hyper-Kamiokande Design Report}",
    eprint = "",
    archivePrefix = "",
    primaryClass = "",
    doi = "
https://doi.org/10.48550/arXiv.1805.04163",
    journal = "arXiv: Instrumentation and Detectors",
    volume = "1805.04163",
    number = "",
    pages = "",
    year = "2018",
}

@article{Wang:2017jts,
author = "Wang, J",
    collaboration = "",
    title = "{MiniCLEAN Dark Matter Experiment}",
    eprint = "",
    archivePrefix = "",
    primaryClass = "",
    doi = "
https://doi.org/10.48550/arXiv.1711.02117",
    journal = "arXiv: Instrumentation and Detectors",
    volume = "1711.02117",
    number = "",
    pages = "",
    year = "2017",
}

@article{SuperKAtmospheric,
    author = "Fukuda, Y. and others",
    collaboration = "The Super-Kamiokande",
    title = "{Evidence for Oscillation of Atmospheric Neutrinos}",
    eprint = "",
    archivePrefix = "",
    primaryClass = "",
    doi = "https://doi.org/10.1103/PhysRevLett.81.1562",
    journal = "Phys. Rev. Lett.",
    volume = "81",
    number = "1562",
    pages = "",
    year = "1998",
}

@article{SuperKSolar,
    author = "Cravens, J. P. and others",
    collaboration = "The Super-Kamiokande",
    title = "{Solar neutrino measurements in Super-Kamiokande-II}",
    eprint = "",
    archivePrefix = "",
    primaryClass = "",
    doi = "https://doi.org/10.1103/PhysRevD.78.032002",
    journal = "Phys. Rev. D.",
    volume = "78",
    number = "032002",
    pages = "",
    year = "2008",
}

@article{SuperKOscillation,
    author = "Abe, K. and others",
    collaboration = "The Super-Kamiokande",
    title = "{Atmospheric neutrino oscillation analysis with external constraints in Super-Kamiokande I-IV}",
    eprint = "",
    archivePrefix = "",
    primaryClass = "",
    doi = "https://doi.org/10.1103/PhysRevD.97.072001",
    journal = "Phys. Rev. D.",
    volume = "97",
    number = "072001",
    pages = "",
    year = "2018",
}

@article{SNO+Det,
    author = "Albanese, V. and others",
    collaboration = "The SNO+",
    title = "{The SNO+ Experiment}",
    eprint = "",
    archivePrefix = "",
    primaryClass = "",
    doi = "10.1088/1748-0221/16/08/P08059",
    journal = "JINST",
    volume = "16",
    number = "P08059",
    pages = "",
    year = "2021",
}

@article{BorexinoDet,
    author = "Alimonti, G. and others",
    collaboration = "The Borexino",
    title = "{The Borexino detector at the Laboratori Nazionali del Gran Sasso}",
    eprint = "",
    archivePrefix = "",
    primaryClass = "",
    doi = "
https://doi.org/10.1016/j.nima.2008.11.07",
    journal = "Nucl.Instrum.Meth.A.",
    volume = "600",
    number = "",
    pages = "568-593",
    year = "2009",
}

@article{KamlandDet,
    author = "Suekane, F. and others",
    collaboration = "The KamLAND",
    title = "{An Overview of the KamLAND 1-kiloton Liquid Scintillator}",
    eprint = "",
    archivePrefix = "",
    primaryClass = "",
    doi = "https://doi.org/10.48550/arXiv.physics/0404071",
    journal = "arXiv: Instrumentation and Detectors",
    volume = "0404071",
    number = "",
    pages = "",
    year = "2004",
}

@article{BorexinoNature,
    author = "Agostini, M. and others",
    collaboration = "The Borexino",
    title = "{Comprehensive measurement of pp-chain solar neutrinos}",
    eprint = "",
    archivePrefix = "",
    primaryClass = "",
    doi = "https://doi.org/10.1038/s41586-018-0624-y",
    journal = "Nature",
    volume = "562",
    number = "7728",
    pages = "505-510",
    year = "2018",
}

@article{SNO+Reactor,
    author = "Abreu, M. and others",
    collaboration = "The SNO+",
    title = "{Measurement of reactor antineutrino oscillation at SNO+}",
    eprint = "",
    archivePrefix = "",
    primaryClass = "",
    doi = "https://doi.org/10.1103/gypt-lc9v",
    journal = "Phys. Rev. Lett.",
    volume = "135",
    number = "121801",
    pages = "",
    year = "2025",
}

@article{WbLSLightYield,
    author = "Callaghan, E. J. and others",
    title = "{Measurement of proton light yield of water-based liquid scintillator
}",
    eprint = "",
    archivePrefix = "",
    primaryClass = "",
    doi = "https://doi.org/10.1140/epjc/s10052-023-11242-2",
    journal = "Eur. Phys. J. C.",
    volume = "83",
    number = "134",
    pages = "",
    year = "2023",
}

@article{BNL1Ton,
    author = "Xiang, X. and others",
    title = "{Design, construction, and operation of a 1-ton Water-based Liquid scintillator detector at Brookhaven National Laboratory
}",
    eprint = "",
    archivePrefix = "",
    primaryClass = "",
    doi = "10.1088/1748-0221/19/06/P06033",
    journal = "JINST",
    volume = "19",
    number = "P06033",
    pages = "",
    year = "2024",
}

@article{ANNIEWbLS,
    author = " Ascencio-Sosa, M. and others",
    title = "{Deployment of Water-based Liquid Scintillator in the Accelerator Neutrino Neutron Interaction Experiment
}",
    collaboration = "The ANNIE",
    eprint = "",
    archivePrefix = "",
    primaryClass = "",
    doi = "10.1088/1748-0221/19/05/P05070",
    journal = "JINST",
    volume = "19",
    number = "P05070",
    pages = "",
    year = "2024",
}

@article{ASDC,
    author = " Alonso, J. R. and others",
    title = "{Advanced Scintillator Detector Concept (ASDC): A Concept Paper on the Physics Potential of Water-Based Liquid Scintillator}",
    eprint = "",
    archivePrefix = "",
    primaryClass = "",
    doi = "https://doi.org/10.48550/arXiv.1409.5864",
    journal = "arXiv: Instrumentation and Detectors",
    volume = "1409.5864",
    number = "",
    pages = "",
    year = "2014",}

@misc{gatevalve,
  author = {{72000 Series Standard Cycle Harsh Process Gate Valves}},
  howpublished = {\url{https://www.vacuumvalves.com}},
  note = ""
}

@misc{reynolds,
    author = "{Reynolds Polymer Technology}",
    howpublished = "\url{https://www.reynoldspolymer.com}",
}

@misc{goulds,
  author = {{Goulds Water Technology}},
  howpublished = {\url{https://www.xylem.com/en-us/brand/goulds-water-technology/}},
  note = ""
}

@misc{sanitron,
  author = {{Sanitron Germicidal Ultraviolet Water Purifiers}},
  howpublished = {\url{https://d163axztg8am2h.cloudfront.net/static/doc/3a/d5/09bc85d16b387ac1854dcba91c01.pdf}},
  note = ""
}

@misc{thermal,
  author = {{Accuchiller EQ Series}},
  howpublished = {\url{https://www.thermalcare.com/assets/files/EQ%20Specification%2014.pdf}},
  note = ""
}

@misc{mettler,
  author = {{M800 Multi-channel Transmitter}},
  howpublished = {\url{https://www.mt.com/us/en/home/products/Process-Analytics/transmitter/multi-parameter-digital-transmitter-M800.html?filter[model-family]=M800%20Process&filter[model-family]=M800%20Water}},
  note = ""
}

@misc{ws1,
  author = {{Asurity Flood Detector}},
  howpublished = {\url{https://www.asurityhvacr.com/wet-switch-flood-detector}},
  note = ""
}

@misc{shimadzu,
  author = {{UV-2600i Plus}},
  howpublished = {\url{https://www.shimadzu.com/an/sites/shimadzu.com.an/files/pim/pim_document_file/brochures/25086/c101-e184.pdf}},
  note = ""
}

@misc{shimadzu_uv,
  author = {{Shimadzu}},
  howpublished = {\url{https://www.shimadzu.com}},
  note = ""
}

@misc{finish,
  author = {{Finish Thompson Inc.}},
  howpublished = {\url{https://www.finishthompson.com/products/sp10/}},
  note = ""
}

@misc{iwaki,
  author = {{Iwaki America}},
  howpublished = {\url{https://iwakiamerica.com}},
  note = ""
}

@misc{omega,
  author = {{High Accuracy Pressure Transducers}},
  howpublished = {\url{https://assets.dwyeromega.com/pdf/test-and-measurement-equipment/pressure/pressure-transducers/PX409_Series.pdf}},
  note = ""
}

@misc{ami,
  author = {{AMI Model 2001RS}},
  howpublished = {\url{https://www.amio2.com/oxygen-analyzers/trace-oxygen-analyzers/model-2001rs/#technical-specs}},
  note = ""
}

@misc{precision_digital_meters,
  author = {{PD6600 Loop Leader Loop-Powered Process Meters}},
  howpublished = {\url{https://www.predig.com/sites/default/files/documents/LDS6602.pdf}},
  note = ""
}

@misc{parker,
  author = {{G7 Series Solenoid Valves}},
  howpublished = {\url{https://www.parker.com/content/dam/Parker-com/Literature/Fluid-Control-Division/Catalogs/Parker_FCD_Catalog-G7.pdf}},
  note = ""
}

@misc{caenhv,
  author = {{CAEN A7030 Specifications}},
  howpublished = {\url{https://www.caen.it/products/a7030/}},
  note = ""
}

@misc{caendigitizers,
  author = {{CAEN V1730 Specifications}},
  howpublished = {\url{https://www.caen.it/products/v1730/}},
  note = ""
}

@misc{caendigitizersv1742,
  author = {{CAEN V1742 Specifications}},
  howpublished = {\url{https://www.caen.it/products/v1742/}},
  note = ""
}

@misc{caenpci,
  author = {{CAEN A3818 Specifications}},
  howpublished = {\url{https://www.caen.it/products/a3818/}},
  note = ""
}

@misc{enclustra,
    author = "{Enclustra Mercury Specifications}",
    howpublished = "\url{https://www.enclustra.com/en/products/system-on-chip-modules/mercury-xu5}",
}

@misc{ezag,
    author = {{Eckert \& Ziegler}},
    howpublished = "\url{https://ezag.com}",
}

@misc{specTech,
    author = {{Spectrum Techniques}},
    howpublished = "\url{https://www.spectrumtechniques.com/}",
}

@misc{NKT,
    author = {{NKT Photonics}},
    howpublished = "\url{https://www.nktphotonics.com/products/pulsed-diode-lasers/pilas/}",
}

@article{forster,
    author = " Förster, T",
    title = "{Zwischenmolekulare Energiewanderung und Fluoreszenz}",
    collaboration = "",
    eprint = "",
    archivePrefix = "",
    primaryClass = "",
    doi = "https://doi.org/10.1002/andp.19484370105",
    journal = "Ann. Phys.",
    volume = "437",
    number = "",
    pages = "55-75",
    year = "1948",
}

@article{DEAP,
    author = "Ajaj, R. and others",
    collaboration = "The DEAP",
    title = "{Search for dark matter with a 231-day exposure of liquid argon using DEAP-3600 at SNOLAB}",
    eprint = "",
    archivePrefix = "",
    primaryClass = "",
    doi = "https://doi.org/10.1103/PhysRevD.100.022004",
    journal = "Phys. Rev. D.",
    volume = "100",
    number = " 022004",
    pages = "",
    year = "2019",
}

@article{SNO+geo,
    author = "Abreu, M. and others",
    collaboration = "The SNO+",
    title = "{Measurement of reactor antineutrino oscillations with 1.46 ktonne-years of data at SNO+}",
    eprint = "",
    archivePrefix = "",
    primaryClass = "",
    doi = "https://doi.org/10.48550/arXiv.2511.11856",
    journal = "arXiv: High Energy Physics - Experiment",
    volume = "2511.11856 ",
    number = "",
    pages = "",
    year = "2025"
}

@article{BNL30ton,
    author = " Andrede, S. and others",
    title = "{Design, construction, and operation of a 30-ton Water-based Liquid scintillator detector at Brookhaven National Laboratory}",
    eprint = "",
    archivePrefix = "",
    primaryClass = "",
    doi = "https://doi.org/10.48550/arXiv.2603.20019
",
    journal = "arXiv: Instrumentation and Detectors",
    volume = "2603.20019",
    number = "",
    pages = "",
    year = "2026",
}

@article{Button,
    author = " Bae, J. and others",
    collaboration = "The Button",
    title = "{The BUTTON-30 detector at Boulby}",
    eprint = "",
    archivePrefix = "",
    primaryClass = "",
    doi = "10.1088/1748-0221/21/03/P03008",
    journal = "JINST",
    volume = "21",
    number = "",
    pages = "P03008",
    year = "2026",
}

@article{EosWater,
    author = " Arora, S. and others",
    collaboration = "The Eos",
    title = "{Performance of the Eos detector with water}",
    eprint = "",
    archivePrefix = "",
    primaryClass = "",
    doi = "https://doi.org/10.48550/arXiv.2606.10234
",
    journal = "arXiv: High Energy Physics - Experiment",
    volume = "2606.10234",
    number = "",
    pages = "",
    year = "2026",
}

@article{LBNE,
    author = " Montanari, D. and others",
    collaboration = "",
    title = "{Performance and results of the LBNE 35 ton membrane cryostat
prototype}",
    eprint = "",
    archivePrefix = "",
    primaryClass = "",
    doi = "https://doi.org/10.1016/j.phpro.2015.06.092",
    journal = "Physics Procedia",
    volume = "67",
    number = "",
    pages = "308 – 313",
    year = "2015",
}

@article{EosDesign,
    author = " Anderson, T. and others",
    collaboration = "The Eos",
    title = "{EOS: a demonstrator of hybrid optical detector technology}",
    eprint = "",
    archivePrefix = "",
    primaryClass = "",
    doi = "https://doi.org/10.48550/arXiv.2211.11969",
    journal = "arXiv: Instrumentation and Detectors",
    volume = "2211.11969",
    number = "",
    pages = "",
    year = "2023",
}

@article{Braidwood,
    author = "Bolton, T",
    collaboration = "",
    title = "{The Braidwood Reactor Anitneutrino Experiment}",
    eprint = "",
    archivePrefix = "",
    primaryClass = "",
    doi = "https://doi.org/10.1016/j.nuclphysbps.2005.05.041",
    journal = "Nuclear Physics B - Proceedings Supplements",
    volume = "149",
    number = "",
    pages = "166 – 169",
    year = "2005",
}

@article{COHERENT,
    author = "Bolton, T",
    collaboration = "The COHERENT",
    title = "{Simulating the neutrino flux from the Spallation Neutron Source for the COHERENT experiment}",
    eprint = "",
    archivePrefix = "",
    primaryClass = "",
    doi = "https://doi.org/10.1103/PhysRevD.106.032003",
    journal = "Phys. Rev. D.",
    volume = "106",
    number = "",
    pages = "032003",
    year = "2022",
}

@article{LBNECDR,
    author = "",
    collaboration = "The LBNE",
    title = "{The Long Baseline Neutrino Experiment (LBNE) Water Cherenkov Detector (WCD) Conceptual Design Report (CDR)}",
    eprint = "",
    archivePrefix = "",
    primaryClass = "",
    doi = "https://doi.org/10.48550/arXiv.1204.2295",
    journal = "arXiv: Instrumentation and Detectors",
    volume = "1204.2295",
    number = "",
    pages = "",
    year = "2012",
}

@article{RHIC,
    author = "Muller, B. and Nagle, J",
    collaboration = "",
    title = "{Results from the Relativistic Heavy Ion Collider}",
    eprint = "",
    archivePrefix = "",
    primaryClass = "",
    doi = "https://doi.org/10.1146/annurev.nucl.56.080805.140556",
    journal = "Ann. Rev. Nucl. Part. Sci.",
    volume = "56",
    number = "",
    pages = "93-135",
    year = "2006",
}

\end{document}